\documentclass[11pt, a4paper]{article}
\usepackage[utf8]{inputenc}
\usepackage[T1]{fontenc}
\usepackage[english]{babel}
\usepackage{csquotes}
\usepackage[backend=biber, natbib=true, style=apa]{biblatex}
\usepackage{amsmath, amsfonts, amssymb}
\usepackage{graphicx} 
\usepackage{caption}
\usepackage[belowskip=2ex]{subcaption}
\usepackage[margin=1in]{geometry}
\usepackage{verbatim}
\usepackage{float}
\usepackage{comment}
\usepackage{multirow, rotating}
\usepackage{tabularx, booktabs}
\usepackage{longtable}
\usepackage{hyperref}
\usepackage[bottom]{footmisc}
\usepackage{pdflscape}
\usepackage{afterpage}
\usepackage{makecell}
\usepackage{xcolor}
\usepackage{listings}
\usepackage{titling}
\usepackage{colortbl}
\usepackage{appendix}
\usepackage{tikz}
\usetikzlibrary{decorations.pathreplacing}

\title{GDP nowcasting with artificial neural networks: How much does long-term memory matter?}

\author{
Kristóf Németh\thanks{
Corresponding author. 
Budapest University of Technology and Economics, 
Faculty of Electrical Engineering and Informatics, 
Magyar Tudósok krt. 2, 1117 Budapest, Hungary; 
E-mail: \href{mailto:manzotta@gmail.com}{manzotta@gmail.com}} \and 
Dániel Hadházi\thanks{
Budapest University of Technology and Economics, 
Department of Measurement and Information Systems, 
Magyar Tudósok krt. 2, 1117 Budapest, Hungary; 
E-mail: \href{mailto:hadhazi@mit.bme.hu}{hadhazi@mit.bme.hu}}
}
\date{}

\date{}

\begin{document}

\maketitle

\begin{abstract}
We apply artificial neural networks (ANNs) to nowcast quarterly GDP growth for the U.S. economy. 
Using the monthly FRED-MD database, we compare the nowcasting performance of five different ANN architectures: the multilayer perceptron (MLP), the one-dimensional convolutional neural network (1D CNN), the Elman recurrent neural network (RNN), the long short-term memory network (LSTM), and the gated recurrent unit (GRU). 
The empirical analysis presents results from two distinctively different evaluation periods. 
The first (2012:Q1 -- 2019:Q4) is characterized by balanced economic growth, while the second (2012:Q1 -- 2024:Q2) also includes periods of the COVID-19 recession. 
During the first evaluation period, longer input sequences slightly improve nowcasting performance for some ANNs, but the best accuracy is still achieved with 8-month-long input sequences at the end of the nowcasting window. 
Results from the second test period depict the role of long-term memory even more clearly. 
The MLP, the 1D CNN, and the Elman RNN work best with 8-month-long input sequences at each step of the nowcasting window. 
The relatively weak performance of the gated RNNs also suggests that architectural features enabling long-term memory do not result in more accurate nowcasts for GDP growth.  
The combined results indicate that the 1D CNN seems to represent a \textit{``sweet spot''} between the simple time-agnostic MLP and the more complex (gated) RNNs. 
The network generates nearly as accurate nowcasts as the best competitor for the first test period, while it achieves the overall best accuracy during the second evaluation period. 
Consequently, as a first in the literature, we propose the application of the 1D CNN for economic nowcasting. 
\\

Journal of Economic Literature (JEL) codes: C45, C51, C53 

\end{abstract}

\section{Introduction}  \label{sec:intro}

Gross Domestic Product (GDP) is arguably the most important flow-type monetary measure of the economic activity of a country or region. 
GDP is often calculated on an annual basis, but in most advanced economies it is also measured at a quarterly frequency. 
Considering its relevance, the availability and reliability of these quarterly data are of great significance. 
However, due to the time required for data collection and statistical data processing, statistical offices are delayed in releasing GDP data for the current quarter. 
The first measurements are usually available  only a few months after the end of the current quarter, which makes it difficult to assess economic activity in real-time (i.e., synchronously). 
Therefore, many decision makers (such as central banks, budget offices, etc.) rely on the current quarter forecast, that is, the nowcast of GDP growth for more informed policy making. 
This underscores the need for accurate nowcasts. 

In nowcasting current quarter GDP growth, one typically tries to extract the informational content of higher frequency monthly data to track the real-time development of economic activity. 
Such an analysis faces the following three problems \citep{giannone2008nowcasting}: (i) Combining monthly predictor variables and a quarterly target variable. 
(ii) Handling a large number of potential regressors. 
(iii) Monthly data are released at different times (unsynchronized) within a quarter. 
This leads to the so-called \textit{``ragged edge''} problem, where some monthly features may have missing values, usually at the end of the sample period \citep{wallis1986forecasting}.

In our study, we apply different artificial neural networks (ANNs) to nowcast the quarterly U.S. GDP growth of the current quarter. 
Using the monthly FRED-MD database, we measure the nowcasting performance of five different ANN architectures relative to a naive constant growth model and a benchmark DFM specification: 
The multilayer perceptron (MLP), the one-dimensional convolutional neural network (1D CNN), the Elman recurrent neural network (RNN), the long short-term memory network (LSTM), and the gated recurrent unit (GRU). 
These architectures are similar in that they all try to meet the above-mentioned challenges of nowcasting. 
However, the models have a different structure. 
First, the MLP's relatively simple architecture is not explicitly designed for sequence modeling. 
While we can form ``training sequences'' by hand and add them as input, the MLP does not recognize the temporal ordering in the data. 
This is not to say that the MLP cannot be used effectively in applications related to time series analysis, but in this regard, the MLP is the exception. 
Compared to the other ANNs investigated in this paper, the MLP is a simple, time-agnostic architecture, which is not tolerant to shifts (translations) in time \citep{ekman2021learning, goodfellow-et-al_2016}. 
The 1D CNN and all other RNNs investigated in this paper are specifically designed for sequence modeling. 
They are capable of learning order dependence in sequential prediction problems. 
In other words, they are trained with \textit{real} input sequences. 
Despite this similarity, even the 1D CNN and the RNNs differ substantially in how they handle the sequential data.\footnote{We will describe these differences in more detail in Section \ref{sec:models_and_methods}.}
From the perspective of the empirical analysis, the question comes down to what architectural features are advantageous in terms of GDP nowcasting. 

The first main contribution of this paper is that it proposes the utilization of a neural network architecture that has yet to be applied for economic nowcasting, namely the one-dimensional convolutional neural network (hereinafter, 1D CNN). 
Our results suggest that the 1D CNN could be a highly suitable architecture for GDP nowcasting as it generates accurate nowcasts for both of our evaluation periods, i.e., both during a balanced growth period and in times of high economic turbulence. 
While this ANN architecture has been neglected compared to the MLP and the more advanced gated RNNs, we would argue that it represents a \textit{``sweet spot''} for this type of analysis in many aspects. 
In contrast to the MLP, the 1D CNN has an intrinsic architectural constraint that usefully limits the hypothesis space, i.e., the range of mappings the network can learn during the training process. 
During training, the network learns the coefficients of different finite impulse response filters, so it can only learn a composition of time-invariant (translation-invariant) mappings. 
Intuitively, the network performs a series of hierarchical pattern recognition tasks where the result is time-invariant to the relative position of those patterns \citep{goodfellow-et-al_2016}. 
It is worth noting that the vast majority of the traditional time series approaches (e.g., the dynamic factor model) are also built around this assumption; namely, the conditional distribution of the target variable only depends on time through the regressors (predictors). 
From this point of view, it seems especially advantageous to narrow the hypothesis space to this type of mappings. 
Unlike the MLP, the 1D CNN is trained with input sequences, reflecting the architecture's temporal ``awareness''. 
Training of the 1D CNN is also helped by the perception of the temporal ordering in input data, ultimately leading to less trainable parameters (weights). 

During the empirical analysis, we also investigate the nowcasting performance of three different recurrent neural networks (RNNs): 
The Elman RNN, also referred to as basic or \textit{vanilla} RNN, and two types of its more advanced variants augmented with a so-called \textit{gating} mechanism: 
the long short-term memory network (LSTM) and the gated recurrent unit (GRU). 
Similarly to the 1D CNN, RNNs can learn order dependence in input sequences and are restricted to learning time-invariant (translation-invariant) mappings between their inputs and outputs. 
However, compared to the 1D CNN, they handle sequential data in a more general, more complex way. 
Through their memory cells (basically, neurons forming a recurrent layer), they can learn a much wider range of time-invariant mappings rather than just a composition of finite impulse response filters. 
Hence, the hypothesis space of an RNN can be much larger than that of an 1D CNN containing the same number of parameters (weights).  

The major novelty of gated RNNs is that they introduce a so-called \textit{gating mechanism} (practically, element-wise multiplication) that enables the network to develop long-term memory. 
This feature helps to overcome the vanishing gradient problem of the vanilla RNNs, making them a much more viable option for modeling time series with path-dependent behavior, i.e., with long-term memory \citep{chung2014empirical, hochreiter1997long}. 
Financial time series, such as stock prices or even more foreign exchange rates, generally show characteristics of path dependency (e.g., a slowly decaying autocovariance function). 
For those series, it can easily happen that the current value of the exchange rate is affected (supported) by a distant technical level whose effect, however, is no longer reflected in any nearby values. 
As a result, this old technical level should be remembered by the network in a predictive exercise. 
Since gated RNNs can learn those long-term dependencies between input and output, they can be highly effective in such predictive exercises. 
Formation of long-term memory comes at a cost, however: 
They contain much more trainable weights than an Elman RNN with the same number of neurons in their (hidden) state vector. 
Consequently, they usually need a large number of training samples to obtain a good level of generalization capability \citep{ekman2021learning, goodfellow-et-al_2016}. 

The second main contribution of the paper is that, first in the literature, it presents the results of a comprehensive nowcasting \textit{competition} between different ANN architectures. 
The competitor ANN architectures are then trained with different training configurations. 
These training configurations differ (i) in the length of the training sequences and (ii) in information sets available intra-quarterly. 
The nowcasting competition presented in this paper has three important aspects and purposes. 
First, it aims to identify the most suitable ANN architectures for GDP nowcasting. 
The second research question we try to answer is how important long-term memory is in GDP nowcasting. 
So, we investigate how the different competitor models' nowcasting performance changes with the length of the input sequences. 
Finally, we also try to measure how the nowcasting accuracy of the competitors changes with consecutive intra-quarterly data releases. 
Hence, we suppose a three-step nowcasting window along which nowcasts are conducted and evaluated. 
We refer to the subsequent steps of that nowcasting window as n\textit{owcasting scenarios}. 
For example, the 1-month nowcasting scenario refers to the first step of the nowcasting window where the information set contains monthly regressor data until the first of the current quarter. 
We believe that the results of the empirical analysis might be indicative of some aspects of the underlying data-generating process. 
If, for example, the data-generating process of GDP growth was highly path-dependent (i.e., non-ergodic), then that would \textit{likely} speak for the use of those gated RNNs with long input sequences.\footnote{Input sequence is a more general term than training sequence because input sequences can also be used for validation and inference (prediction). 
Nonetheless, we will use these terms interchangeably since we always use sequences of the same length for validation and prediction as for training.  
} 

We present the results from two evaluation (test) periods during the empirical analysis. 
While those evaluation periods overlap in time, they are very different in the characteristics of the target series, i.e., GDP growth. 
The first evaluation period ranges from 2012:Q1 to 2019:Q4, so it ends before the economic consequences of the COVID-19 pandemic hit the U.S. economy. 
The results for this period indicate that longer input sequences lead to slightly more accurate nowcasts for some competitor ANNs. 
However, the best accuracy is still achieved with 8-month-long sequences at the end of the nowcasting window. 
Given its best training configuration, the GRU generates a relative RMSE of $0.589$ and $0.938$ towards the naive model and the benchmark DFM, respectively. 
During the first evaluation period, the 1D CNN yields its most accurate nowcasts with 36 long input sequences in the 3-month nowcasting scenario. 
That is indicated by a relative RMSE of $0.627$ and $0.999$ against the naive model and the benchmark DFM. 
It is important to note that longer input sequences do not provide the 1D CNN with significantly more accurate nowcasts than the configuration, which is trained with 8 long sequences at any step of the nowcasting window. 
Our second evaluation period is from 2012:Q1 to 2024:Q2, so it also includes periods of the COVID-19 crisis. 
Under such circumstances, shorter input sequences lead to more accurate nowcasts for all competitor models. 
In the second evaluation period, the best accuracy is achieved by the 1D CNN trained with 8 long sequences. 
At the end of the nowcasting window, the 1D CNN generates a relative RMSE of 0.199 and 0.523 towards the naive model and the benchmark DFM, respectively. 
Differences in accuracy are statistically significant relative to both benchmark models on a 10\% level, according to the \textcite{diebold1995comparing} test. 
We can see that architectural complexity clearly hinders the gated RNNs' nowcasting performance in the second evaluation period. 
This time, they produce less accurate nowcasts than even the basic Elman RNN. 
Combined results from the two evaluation periods show that longer input sequences generally do not improve our models' predictive accuracy. 
That is especially interesting to see in the case of the LSTM and the GRU: 
While their architectural design enables them to handle long input sequences, forming a long-term memory, even these networks tend to have better accuracy with shorter sequences. 
Considering the results from the two test periods as a whole, architectural features supporting the learning of long-term dependencies do not play an important role in GDP nowcasting. 
Instead, they exhibit a poor trade-off between complexity and generalization capability for this specific predictive analysis. 

The remainder of the paper is structured as follows: 
Section \ref{sec:literature} briefly summarizes the related literature on the topic. 
Section \ref{sec:models_and_methods} presents the different models applied in the empirical analysis, i.e., the five different neural network architectures mentioned above.  
Section \ref{sec:empirical_analysis} describes the data behind the analysis and exposes the design of the nowcasting exercise. 
Then, it presents the results, including feature importance interpretations. 
Finally, Section \ref{sec:conclusion} concludes. 



\section{Related works}     \label{sec:literature}

\textcite{stock2002forecasting} and \textcite{stock2002macroeconomic} use principal components and diffusion indexes to forecast macroeconomic time series. 
They show that when the data follow an approximate dynamic factor model with many predictors and time series observations, a few indexes can summarize the informational content of the predictors. 
Forecasts based on these indexes outperformed univariate AR models, small vector autoregressions, and leading indicator models. 
Following these papers, \textcite{giannone2008nowcasting} combine a dynamic factor model (DFM) and the Kalman smoother to nowcast the quarterly GDP growth for the U.S. economy. 
They propose a two-step estimator for the common factors which combines principal components and Kalman filtering techniques: 
In the first step, the parameters of the model are estimated from an OLS regression on principal components extracted from a balanced panel. 
In the second step, the common factors are extracted by applying the Kalman smoother on the entire data set.\footnote{Consistency properties of the two-step estimator are described in \textcite{doz2011two}}. 
Finally, they define the nowcast as the linear projection of quarterly GDP on the extracted common factors (``bridging with factors''). 
The bridge equation can be given as a mixed-frequency measurement equation which applies approximate temporal aggreagation \citep{mariano2003new}. 
The estimation procedure described in \textcite{giannone2008nowcasting} addresses all three problems of nowcasting mentioned before: 
(i) it can handle a large number of potential regressors, (ii) it can cope with staggered data-release dates, and (iii) it deals with the mixed frequency issue of combining monthly predictor variables and quarterly GDP. 

After the seminal work of \textcite{giannone2008nowcasting}, DFMs have been found particularly successful at addressing many of the data issues inherent in nowcasting, and have been applied in many empirical studies. 
\textcite{marcellino2010factor} compare the nowcasting performance of three different factor estimation methods (one of them is the same as in \textcite{giannone2008nowcasting}) and combine the DFM with mixed-frequency data sampling \citep{ghysels2007midas}. 
They find that choice of the factor estimation technique has no substantial impact on the nowcast performance. 
On the other hand, mixed-data sampling based on common factors results in more accurate nowcasts for the German GDP growth than factor models based on time-aggreagted data. 
\textcite{matheson2014new} develop monthly indicators for tracking short-run trends in real GDP growth and conduct nowcasting analysis for 32 advanced and emerging-market economies. 
\textcite{chernis2017dynamic} perform a similar analysis for the Canadian GDP growth. 
\textcite{botha2021nowcasting} apply a suite of statistical models to nowcast South African GDP growth from various model classes. 
Besides classical time series approaches (e.g., ARIMA, DFM), they also adopt different machine learning techniques (e.g., Lasso model, Elastic net). 
While the indicator model (a combination of individual OLS models) wins the nowcasting competition, combined results from the real-time analysis also show that both model combination and selection provide more accurate nowcasts for GDP growth. 
The real-time analysis presented by \textcite{richardson2021nowcasting} focuses more on machine learning (ML) algorithms. 
They apply numerous ML models (SVM, LSboost, Elastic net, among others) to nowcast the New Zealand GDP growth and compare the results to an AR(1) benchmark model. 
Results of \textcite{richardson2021nowcasting} show that all competitor ML algorithms can produce more accurate nowcasts than those of the AR and dynamic factor models. 
\\

Although ANNs have been the catalyst for numerous advances in various fields and disciplines in recent years, their impact on macroeconomics has been relatively muted. 
This can partly be explained by the fact that the length of the aggregated macro series does not allow for their effective use in many cases. 
Along with that, early applications have already shown great potential for larger economies, where monthly and quarterly series provide a sufficient number of training samples (training sequences). 
\textcite{tkacz2001neural} uses an MLP to forecast the Canadian GDP growth rate between 1989 and 1992 by applying lagged GDP growth and several other financial variables. 
He finds that the MLP yields statistically lower forecast errors for the
year-over-year growth rate of real GDP relative to linear and univariate models. However, when forecasting quarterly GDP growth, the MLP cannot outperform a naive constant growth model. 
While \textcite{tkacz2001neural} uses quarterly data to forecast Canadian GDP by using ANNs, \textcite{heravi2004linear} apply ANNs to forecast monthly industrial production between 1978 and 1995 for the three largest European economies, with data provided by Eurostat. 
They find that linear models generally produce more accurate out-of-sample forecasts than neural networks in terms of RMSE evaluation. 
In contrast, neural networks dominate linear models in predicting the direction of change. 
The underlying ANN architecture is also the MLP. 
Besides these studies, ANNs have been used in financial forecasting as well. 
For instance, \textcite{kuan1995forecasting} investigate the forecasting ability of feedforward (MLP) and recurrent neural networks (Elman RNN) for daily exchange rates of five major currencies between 1980 and 1985. 
The study shows that ANNs have significant market timing ability (i.e., forecasts of the direction of future price changes) and significantly lower out-of-sample MSPE (relative to the random walk model) in two out of the five currencies investigated: 
Namely, the Japanese yen and the British pound. 
More recently, \textcite{torres2018applying} apply recurrent neural networks to daily data of several crypto-currencies, exchange rates, commodities, and stocks between 2013 and 2017. 

\textcite{loermann2019nowcasting}, one of the works most closely related to our study, shows that the MLP can outperform the constant growth benchmark model and the DFM in terms of now- and forecasting accuracy, while it generates at least as good now- and forecasts as the Survey of Professional Forecasters. 
They use an expanding estimation window with one quarter of an increment. Along the expanding estimation window, they iteratively re-optimize the range of monthly predictors and the architecture of the MLP itself. 
This is done by a combined filter and wrapper approach presented by \cite{crone2010feature}. 
The evaluation period ranges from 1999:Q3 to 2018:Q3, where the performance advantage of the MLP over a naive constant growth benchmark model is significant on a 1\% level and measured by a 0.398 relative RMSE.  
Compared to the DFM, the MLP generates a 0.834 relative RMSE, and its advantage is significant on a 10\% significance level based on \textcite{diebold1995comparing}. 
Figure 2 in \textcite{loermann2019nowcasting} also suggests that the significant performance advantage of the MLP is largely related to the prediction of the Great Recession: 
From 2008:Q4 to 2009:Q4, the DFM collects more than half of its performance deficit in terms of CSSED.\footnote{CSSED stands for the cumulative sum of squared forecast error differences, which is computed as defined in \textcite[10]{loermann2019nowcasting}.} 
As stated in the paper, automated filters and wrappers are only available for the MLP, so we chose not to use them for two reasons. 
First, we did not want to provide such a one-sided advantage to the MLP in our empirical analysis, which is centered around a nowcasting competition between different ANN architectures. 
Second, we do not re-optimize the specification of the competitor ANN architectures along our rolling estimation window. 
Using the definitions of \textcite{crone2010feature}, we do not use any wrappers. 
At the same time, we apply an automated feature selection procedure, similar to a filter, which works the same for all competitor ANN architectures and provides the most relevant low-dimensional representation of the features for every predictive step along the rolling window.\footnote{This automated feature selection procedure is described in detail in Section \ref{sub-sec:feature_selection}.} 

Compared to the MLP, \textcite{hopp2021economic} proposes a more complex recurrent network architecture, namely the long short-term memory network. 
The paper shows that the LSTM generates more accurate nowcasts compared to the DFM for three different target series: global merchandise export values and volumes, and global services exports. 
In addition to better empirical performance for the three target series, LSTMs provide advantages over DFMs by being able to handle large numbers of features without computational bottlenecks and the ability to use any mixture of frequencies in features or target. 
Disadvantages include the random aspects of the training process (e.g., sampling for mini-batches, weight initialization) and the lack of interpretability in their coefficients \citep{hopp2021economic}. 

The possible performance advantage of nonlinear models was already foreseen in \textcite{stock2002macroeconomic}. 
As the discussion points out, indexes based on principal components or common factors are linear functions of the data, so there is a possibility that further forecasting gains can be realized using nonlinear models \citep[154]{stock2002macroeconomic}. 
This valuable insight is particularly relevant to our study. 
When we try to identify the most suitable ANN architecture for GDP nowcasting, we also look for the potential sources of a good (bad) nowcasting performance. 
Specifically, we are interested in whether more complex architectures enabling long-term memory result in more accurate nowcasts for GDP growth. 


\section{Models and methods} \label{sec:models_and_methods}

Artificial neural networks are flexible nonlinear estimators which represent the set of alternatives to the classical time series approaches. 
Similarly to most machine learning algorithms, ANNs can be used for classification problems, predictive exercises where the target variable is categorical, and also for regression function estimation. 
One of the biggest strengths of ANNs is that they can handle the curse of dimensionality in an \textit{automated} way, much more naturally than traditional time series models.
The latter set of models require the preparatory use of a dimension reduction method (e.g., PCA) or an unobserved component model for feature extraction, where the number of potential explanatory variables is large. 
For example, PCA can give us an optimal lower dimensional representation of the data in the sense that it changes the basis to the direction of maximum variance. 
However, we cannot be sure if there is no more relevant lower-dimensional representation of the features for the prediction of our target variable. 
Similarly, when we extract common factors from the input data, the procedure should be driven heavily by our a priori knowledge (choice): 
We should have a relatively concrete idea about the potentially relevant explanatory variables (features). 
After the set of observable indicators have been defined, method of explanatory factor analysis extracts common factors in such a way that they can explain the co-movements, i.e., the covariance structure of our features best. 
Here we see again that feature extraction is somewhat separated from the essential modeling task. 
Hence, it is not surprising that DFM-based GDP nowcasting is referred to as a two-staged estimation procedure in the literature \citep{giannone2008nowcasting}. 
By contrast, ANNs, can, in principle at least, learn the relevant feature representation by themselves.   

The demand to connect the feature extraction with the actual predictive modeling task arises naturally and justifiably. 
Here is where artificial neural networks perform really well because they are inherently capable of learning the relevant representation of the features for the prediction of the target variable.\footnote{We should note that this kind of \textit{connected} feature extraction (with some limitations) can also be conducted with classical time series approaches. However, ANNs have an advantage in representation learning.}

This is not to say however, that ANNs are generally superior modeling frameworks compared to traditional linear models. 
While ANNs handle the \textit{curse of dimensionality} more easily, they can perform worse in other areas. 
For example, they require a much more sophisticated data pre-processing step where we must deal with missing observations as well. 
One possible solution to fill up missing values is applying univariate ARMA(p,q) forecasts for each single input series. 
In this case, the lag lengths of the individual ARMA(p,q) models can be selected based on the value of Akaike or Schwarz information criterion \citep{greene2003econometric}, \parencite{loermann2019nowcasting}. 
Although missing data issue for ANNs can be solved in many ways, state space models handle this problem by design. 

If we think about the three big challenges of nowcasting mentioned in the introduction, we can say that mixed-frequency data can be treated relatively well in both modeling frameworks. 
Compared with the DFM, proper construction of our training and test samples with mixed-frequency data may require some additional effort. 

More generally, we should also mention that even the smallest ANNs have much more parameters than a typical statistical model. 
The large number of trainable (\textit{free}) parameters results in a relatively large hypothesis space.
From the bias-variance trade-off follows that with a larger hypothesis space, it is more likely to find a hypothesis which fits the training data very well \citep{hastie2009elements}. 
This phenomenon is called overfitting, and it happens when a model learns the fine details and noise in the training data rather than the actual \textbf{systematic} relationship between the regressors and the target variable(s). 
Overfitting can substantially weaken a statistical model's generalization capability, and ANNs are substantially more prone to overfitting compared to the traditional time series models \citep{ekman2021learning}. 

So far, we tried to summarize the most important characteristics of ANNs as a general modeling framework. 
In the following sections, we investigate those ANN architectures adopted in the empirical analysis in more detail.

\subsection{Multilayer perceptron}

The multilayer perceptron (MLP), often referred to as fully connected neural network, is the straightforward extension of a single Rosenblatt perceptron.\footnote{The Rosenblatt perceptron is a special type of neuron (just like the Adaline, for example), but hereafter, we will use the more generic name \textit{neuron}}.  
With nonlinear activation (squashing) functions, the MLP can perform nonlinear classification and regression tasks. 
It is one of the simplest types of a \textit{feedforward} neural network. 
Figure \ref{fig:mlp_01} below shows the architecture of an MLP with an input layer, three hidden layers, and an output layer. 
As we see, the information in the model flows in only one direction: 
From the direction of the input layer, through the hidden layers, and to the output layer, i.e., forward. 
In other words, the computational graph of an MLP does not contain any cycles or loops. 

\begin{figure}[H]
\begin{center}
\includegraphics[width = \textwidth]{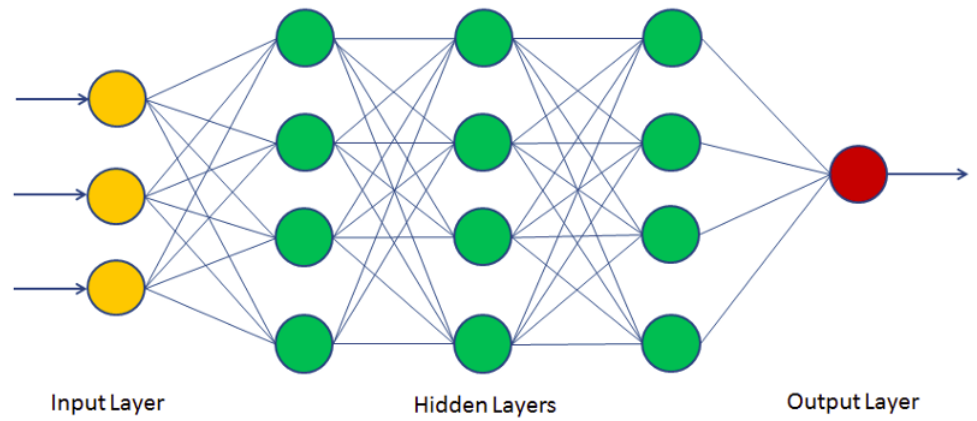}
\caption{The architecture of the multilayer perceptron.}
\label{fig:mlp_01}
\end{center}
\end{figure}

According to Figure \ref{fig:mlp_01}, the MLP's input-output mapping -- if we assume the same activation function $f(.)$ in each layer -- is as follows:
\begin{align}   \label{eq:mlp_mapping}
\mathbf{y} = f \left( \mathbf{w}^{l} f \left( \mathbf{w}^{l-1} f \left( \mathbf{w}^{l-2} \dots f \left( \mathbf{w}^{1} \mathbf{x} + \mathbf{b}^{1} \right) + \mathbf{b}^{l-2} \right) + \mathbf{b}^{l-1} \right) + \mathbf{b}^{l} \right)
\end{align}
where $\mathbf{w}^{l}$ is the matrix combining the weight vectors of the neurons in the $l$-th layer, $\mathbf{b}^{l}$ is the bias vector in the $l$-th layer, and $f(.)$ denotes the activation function (e.g., sigmoid) applied to the outputs of the neurons belonging to the same layer. 
An MLP, even in its simplest form when it contains one hidden layer, implements a nonlinear mapping on its parameters. 
As Equation \ref{eq:mlp_mapping} suggests each hidden layer of an MLP can learn an arbitrary affine transformations of its inputs.\footnote{An affine transformation is the combination of a linear transformation with translations. 
Translation can be learned through the bias.)} 
After performing an affine transformation on the inputs, the resulting weighted sum is fed into a nonlinear activation or squashing function ($f(.)$), which then computes the output of the given layer. 

\textcite{hornik1989multilayer} shows that when the activation function is nonlinear, the MLP, even with a single hidden layer, can be proven to be a universal function approximator. 
Based on the universal approximator property, the MLP can also be used in time series-related applications. 
However, its architecture lacks any explicit temporal aspect. 
Since a hidden layer can learn any arbitrary affine transformation of its inputs, the size of the hypothesis space can be too large for our predictive analysis. 
Along with that MLPs can theoretically learn time-invariant mappings between their inputs and outputs, their underlying (hidden) architecture is not invariant to shifts (translations) in time. 
A more fitting feedforward architecture for time series prediction can be the time-delay neural network, or in other words, the one-dimensional convolutional neural network (1D CNN).

\subsection{1D Convolutional neural network}    \label{sec:1d_cnn}

Convolutional neural networks (CNNs), similarly to MLPs, are also feedforward neural networks that typically consist of convolutional, pooling and fully connected layers. 
Without exaggeration, we can say that 2 dimensional CNNs have become the standard for various computer vision tasks in the last decade. 
As the field of application for 2D CNNs has constantly broadened, 1D CNNs have also received growing attention. 
1D CNNs have recently achieved state-of-the-art performance levels in several applications related to time series analysis, such as biomedical data classification, anomaly detection, and structural health monitoring \citep{kiranyaz20211d}. 
We have not yet seen any application in the literature that is specifically related to economic nowcasting or forecasting. 
Nonetheless, in the following, we try to argue that GDP nowcasting might have exactly those characteristics that can render the 1D CNN a suitable competitor model. 

In GDP nowcasting, we supposedly want to find a time-invariant mapping between our regressors (features) and our target variable. 
Intuitively, this means that the same regressor vector should have the same response in the target variable, even if it is shifted in time -- assuming steady state. 
While MLPs can also learn time-invariant mappings, their architecture does not support this type of \textit{restricted} learning by design. 
In contrast, 1D CNNs learn the coefficients of multiple finite impulse response filters in their convolutional layers. 
Consequently, the network is forced to learn a composition of time-invariant mappings in its convolutional layers. 
In other words, the network's architecture intrinsically restricts the hypothesis space to the composition of time-invariant mappings, which we supposedly want to find in the nowcasting of GDP growth. 
Figure \ref{fig:conv1d_01} below shows a simple example of single channel 1D convolution, where length of the input sequence is 7 and the length (size) of the convolution filter (kernel) is 3. 
According to the definition of the discrete convolution, the length of the resulting sequence is 5. 

\begin{figure}[H]
\begin{center}
\includegraphics[width = \textwidth]{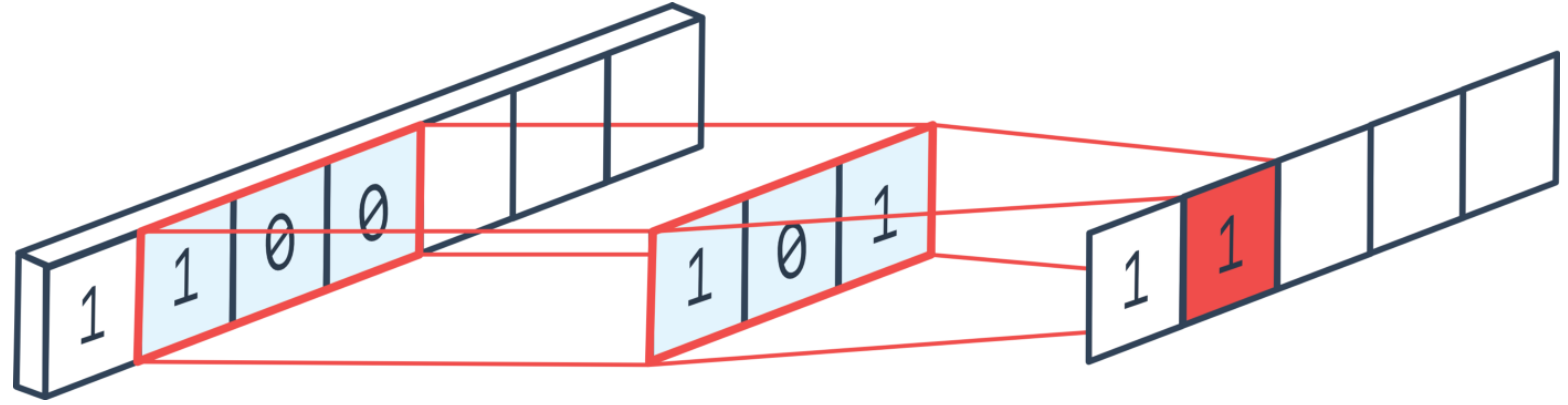}
\caption{1D convolution with a single input and output channel. 
Source: \textcite{ganesh2019}.}
\label{fig:conv1d_01}
\end{center}
\end{figure}

As Figure \ref{fig:conv1d_01} suggests, a 1D CNN practically realizes finite impulse response filters on the input sequences, where filter coefficients (weights) are estimated during the training process. 
If we think about 1D CNN's relation to traditional time series analysis, filtering is the first thing that should come into our minds. 
In traditional time series analysis, we often apply different frequency filters to reveal some hidden characteristics of the target series. 
For example, the Hodrick -- Prescott filter is a popular frequency filter used to extract the unobserved trend and cycle components from a time series \citep{Hodrick1997PostwarUB}. 
In terms of its characteristics, it is a low-pass pass filter, which attenuates the high-frequency components in the frequency spectrum of the target series. 
Naturally, we can think of any other frequency filter, whether low-pass, high-pass, or band-pass. 
While these filters are designed in the frequency domain, they are realized (applied) in the time domain. 
As multiplication in the frequency domain is equivalent to convolution in the time domain, we can see the 1D CNN's strong relation to the traditional filtering techniques. 
There is one key difference between them, however. 
When we apply a 1D CNN, we do not design our filters based on some a priori knowledge or assumption, such as higher frequencies should be associated with an unobserved cyclical component. 
Instead, we estimate the filter coefficients (weights) as an inherent part of the predictive analysis. 
By applying multiple finite impulse response filters, the network tries to create (learn) those input representations relevant to the prediction of the target variable. 
Since kernels within a convolutional layer are relatively short, they typically act as either a low-pass or a high-pass filter, depending on the learned weights of the kernel. 
Nonetheless, by combining multiple kernels or layers, the network can also approximate the behavior of band-pass filters. 

Based on this simple example, we can also see that unlike MLPs, 1D CNNs are trained with sequential data where the network receives a fixed size input (training) sequence for each observation of the target variable. 
Obviously, in a more realistic setting we can have multiple input series (channels) and multiple convolution filters as well. 
In this case, we will have an output sequence (channel) for each convolving filter. 
Figure \ref{fig:conv1d_02} below illustrates a case where we have 4 input channels (i.e., the input depth is 4), and we have 4 different convolution filters of length 5. 
Accordingly, we have 4 output channels (i.e., the output depth is 4). 
More generally, in a convolutional layer we will have a separate filter corresponding to each output channel of the feature map. 

\begin{figure}[H]
\begin{center}
\includegraphics[width = \textwidth]{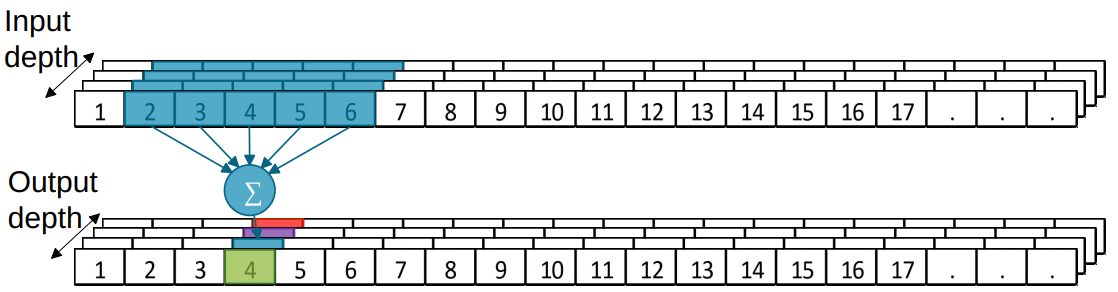}
\caption{1D convolution with multiple input and output channels.}
\label{fig:conv1d_02}
\end{center}
\end{figure}

As we have seen, a 1D convolutional layer applies an 1D convolution over an input sequence (signal) composed of several input channels (features or planes). 
In a more formal manner, the output vector (tensor) of the $j$-th output channel or feature map ($c_{out, j}$) in a 1D convolutional layer with input size $(C_{in}, L_{in})$ and output size $(C_{out}, L_{out})$ can be described as follows: 
\begin{align}   \label{eq:1dcnn_mapping}
c_{out, j} = b_j + \sum_{i=0}^{C_{in}-1} f_{j, i} \star c_{in, i} 
\end{align}
where $f_{j, i}$ denotes the convolution kernel (filter) corresponding to the $j$-th output channel (feature map) which slides over the $i$-th input channel. 
If $k$ defines the filter size (length) within a convolution layer, $f_{j, i}$ can be represented by a $k \times 1$ row vector (1D tensor), containing $k$ trainable weigths (parameters). 
Continuing the interpretation of Equation \ref{eq:1dcnn_mapping}, $\star$ is the convolution operator\footnote{Rigorously speaking, it is the valid cross-correlation operator because filters (kernels) in a convolutional layer are not reversed.}, $C$ denotes the number of channels, and $L$ stands for the sequence length. 
From the definition of convolution (cross-correlation), it follows that $L_{out} = L_{in} - k + 1$.\footnote{This is only true, if we assume the default values for the stride, padding and dilatation parameters. 
The more general formula can be found at: \url{https://pytorch.org/docs/stable/generated/torch.nn.Conv1d.html}}. 

Based on Equation \ref{eq:1dcnn_mapping}, in each 1D convolutional layer the networks learns the coefficients of different FIR filters. 
Not counting the biases, we will have $C_{out} \times C_{in} \times k$ trainable weights (parameters) in a given layer. 
Let us consider $f_{j, i}$ again, i.e., the convolution filter corresponding to the $j$-th output channel (feature map) which slides over the $i$-th input in channel.  
Since $k$ defines the length of the convolution filters, it also controls how many terms of the input sequence are considered in the convolution while computing a single value in the output channel. 
Given a filter of size $k$, the possible time delays ($d$) are $d = 0, 1, \dots, k-1$. 
Since the $f_{j, i}$ slides over the input sequence (series) in $L_{out}$ steps, the following relationship will hold for every time delay $d$ in every possible sliding step $\tau$: 
\begin{align}   \label{eq:1dcnn_weight_sharing}
w_{j, i}(d) = \frac{\partial c_{out, j}(t)}{\partial c_{in, i}(t-d)} = \frac{\partial c_{out, j}(t-\tau)}{\partial c_{in, i}(t-d-\tau)}
\end{align} 
where $c_{out, j}(t)$ stands for the value of the $j$-th output channel (feature map) in time $t$, and the corresponding time-related indices ranges as follows: $t = \left[ 0, 1, \dots, ( L_{out} - 1 ) \right]$, $d = \left[ 0, 1, \dots, k-1 \right]$, and $\tau = \left[ 0, 1, \dots, ( L_{out} - 1 - t ) \right]$. 

Intuitively, Equation \ref{eq:1dcnn_weight_sharing} shows that between given pairs of output and input channels (e.g., for a given $(j, i)$ pair), the weights belonging to the same time delay ($d$) are the same.
Consequently, we have locally shared weights within a convolutional layer \citep{ekman2021learning}.\footnote{Based on this, it is not surprising that the 1D CNN is often called as Time delay neural network (TDNN).} 
By contrast, in the case of the MLP, Equation \ref{eq:1dcnn_weight_sharing} obviously does not hold. 
As the MLP learns any arbitrary affine transformation within its hidden layers, it will also learn different weights for the same time delay. 
More precisely, it has no architectural restriction that would guarantee to learn shared weights for the same delay (shift) in time. 
While theoretically possible for the MLP to learn the same mapping as a given 1D CNN, that would require many more weights to train. 
It is worth noting that weight sharing has a potential drawback as well. 
With locally shared weights, the 1D CNN is restricted to learn only time-invariant mappings within its convolutional layers. 
The results of the empirical analysis will indicate whether that is really a drawback or not in this specific application.

\subsection{Recurrent neural networks}

\begin{figure}[H]
\begin{center}
\includegraphics[width = \textwidth]{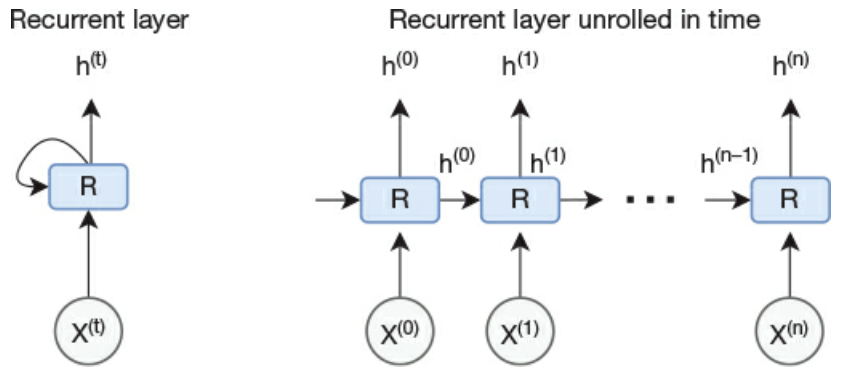}
\caption{Recurrent layer drawn with one node representing an entire layer (left). 
Recurrent layer unrolled in time (right). 
Source: \textcite{ekman2021learning} p. 247.}
\label{fig:rnn_unrolled}
\end{center}
\end{figure}

This section presents the other three ANN architectures investigated in the empirical analysis: the Elman recurrent neural network, the long short-term memory network (LSTM), and the gated recurrent unit (GRU). 
As opposed to the unidirectional relationship between inputs and outputs, previously seen in feedforward networks (MLP, 1D CNN), recurrent neural networks (RNNs) introduce a feedback loop, where layer outputs can be fed back into the network \citep{dematos1996feedforward}. 
Those neurons that create a cycle in the computational graph allow the network to exhibit dynamic behavior. 
They form the network's internal (hidden) state or, more intuitively, its memory. 
This architecture makes RNNs well-suited to applications with a temporal aspect or flow, such as natural language processing, speech recognition, or, in our case, time series analysis. 
Figure \ref{fig:rnn_unrolled} shows how a recurrent layer can be unrolled in time. 
We create a realization of the recurrent layer for each timestep in the input sequence. 
This way, we practically convert the recurrent layer into a number of feedforward layers \citep{ekman2021learning}. 

It should be emphasized that Figure \ref{fig:rnn_unrolled} depicts a schematic representation of a recurrent layer, where the rectangular node, denoted by $R$, represents a full layer of recurrent neurons.\footnote{The number of recurrent neurons is often referred as hidden size in Deep learning frameworks, e.g., in Pytorch.} 
Once we have the network unrolled in time, we can backpropagate the error in the same way as we do for a feedforward network. 
In cases with long input sequences, i.e, with many timesteps, it can be computationally more expensive though. 
This time, the backpropagation algorithm will produce one update value for each timestep, but when we later want to update the weight, there is only one weight to update. 
In other words, we have weight sharing between layers. 
The algorithm is known as \textit{backpropagation through time} (BPTT). 
\textcite{werbos1990backpropagation} has written a detailed description about the algorithm, which also contains references to papers in which the algorithm was first used. 

Similarly to the 1D convolutional layer, we also have weight sharing for the recurrent layer. 
This time, however, weight sharing does not work within a layer but between the consecutive layers of the unrolled network. 
As we have seen for the 1D CNN, weight sharing is beneficial because we have fewer weights to train. 
At the same time, the potential drawback of weight sharing is that the network is restricted to learning only a certain type of time-invariant mappings within its convolutional layers.\footnote{Results of the empirical analysis will show, whether that is really a drawback or not in this application.} 
As for the RNN, weight sharing provides a similar benefit of requiring fewer weights to train. 
Unfortunately, it surely has a negative implication this time, which is related to the operation of the BPTT algorithm. 

In line with Figure \ref{fig:rnn_unrolled}, let us consider a simple case where a whole recurrent layer, represented by the rectangular node R, consists of a single neuron. 
Thus, we have a single weight for this recurrent neuron. 
What happens if this weight happens to be less than one, and we multiply it by itself as many times as many timesteps we have in the unrolled network (T)? 
It is easy to see that it will approach 0 (i.e., a vanishing gradient). 
Alternatively, in the other case, if the weight happens to be greater than one and we multiply it by itself T times, it will approach infinity (i.e., an exploding gradient). 
Not to mention that, the vanishing gradient problem described above is usually further worsened by saturated activation functions. 

Due to vanishing gradients, RNNs tend to have a very ``short'' memory, limiting their usefulness in many applications \citep{ekman2021learning}. 
Especially in those that require to learn long-term dependencies between inputs and outputs. 
Even with these considerations, we included one specific variant of the basic RNN in our nowcasting contest: the Elman network or Elman RNN.  
In an Elman network, each recurrent layer applies the following mapping for each element in the input sequence \citep{elman1990finding}:
\begin{align}   \label{eq:rnn_hidden_state}
h_t = tanh \left( x_{t} W_{i, h}^{T} + b_{i, h}  + h_{t-1} W_{h, h}^{T} + b_{h, h} \right),
\end{align}
where $h_{t}$ is the hidden state at time $t$, $x_t$ is the input at time $t$, and $h_{t-1}$ is the hidden state of the previous layer at time $t-1$ or the initial hidden state at time $0$. 
Based on the above, it is relatively easy to see that RNNs, similarly to 1D CNNs, will learn time-invariant mappings in their recurrent layers. 
Providing that the initialization of the hidden state is the same for all training sequences: e.g., $h_0$ consists of zeros. 

After the short description of the Elman RNN, we present how to overcome the main limitations (vanishing gradients) associated with the basic RNN by using more advanced units when building the network. 
Introducing our last two competitor ANN architectures, we shift our focus on more advanced RNN variants that implement a so-called gating mechanism, namely the long short-term memory network and the gated recurrent unit \citep{chung2014empirical, hochreiter1997long}. 

Long short-term memory (LSTM) network is a special form of gated RNNs which was introduced by \textcite{hochreiter1997long}. 
A common way of drawing LSTM was introduced in a popular blog post that explains how LSTM works \citep{olah2015lstm}. 
Our summary on LSTMs is based on \textcite{ekman2021learning}, \textcite{goodfellow-et-al_2016} and \textcite{olah2015lstm}. 
Figure \ref{fig:lstm_unrolled} below illustrates an LSTM layer unrolled in time for three timesteps. 
For each timestep $t$, the corresponding cell receives $c_{t-1}$ and $h_{t-1}$ from the previous timestep (cell) and $x_t$ from the current timestep and outputs new values for the cell state ($c$) and the hidden state ($h$). 
Accordingly, $h_t$ and $c_t$ denote the \textit{hidden state} (also referred to as the \textit{output} of the LSTM cell) and the \textit{cell state} (also known as the \textit{internal state}) at time step $t$, respectively.

\begin{figure}[H]
\begin{center}
\includegraphics[width = \textwidth]{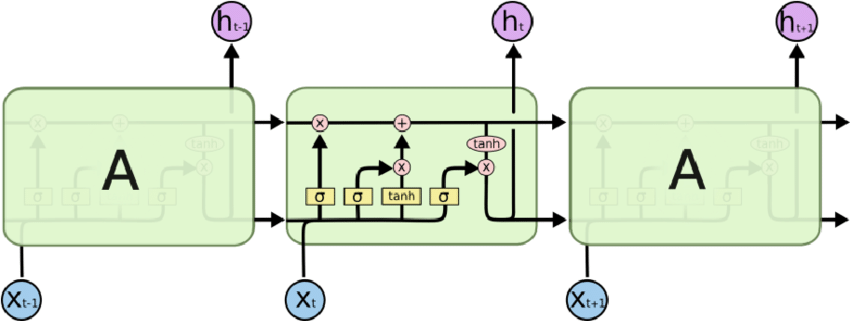}
\caption{LSTM layer unrolled in time. 
Source: \textcite{olah2015lstm}.}
\label{fig:lstm_unrolled}
\end{center}
\end{figure}

As Figure \ref{fig:lstm_unrolled} illustrates, LSTM introduces a memory cell and three gates: an input gate, an output, and a forget gate. 
Essentially, the architecture then allows gradients to flow unchanged through the so-called \textit{gradient highway}, mitigating the vanishing gradient problem of standard RNNs and rendering them more suitable for applications where long memory matters.
Considering our empirical analysis, we are interested to see whether allowing for long memory does lead to better nowcasts. 

Gated recurrent unit (hereafter, GRU) represents the other gated recurrent network adopted in our nowcasting analysis. 
Considering its architecture, GRU is very similar to an LSTM but belongs to a newer generation of gated RNNs \citep{chung2014empirical}. 
The GRU removes the cell state (denoted by $c_t$ in the LSTM) and uses its hidden state (denoted by $h_t$, similar to the LSTM) even for transferring information. 
This way, it only has two gates, a reset gate and an update gate. 
Thereby, GRUs has fewer parameters and tensor operations compared to LSTMs \citep{chung2014empirical}. 
Consequently, they are a little speedier to train, and a bit less prone to overfitting. 
Researchers and engineers usually try both to determine which one works better for their specific use case. 
So we do the same during the empirical analysis. 
In line with the above, all three RNNs described above are trained with sequential data, similarly to the 1D CNN. 
\\

We have seen that gated RNNs provide a solution to the vanishing gradient problem through different realizations of the gating mechanism. 
This way, they can explore long-term dependencies in training sequences and connect information from the past to the present. 
Considering our empirical analysis, it is an essential question if there are such training sequences whose much older values have a significant impact on current-quarter GDP growth while that impact is no longer reflected in any more recent values of those sequences. 
To put this into a more formal manner, let us consider the conditional probability distribution of our target variable. 
Suppose that $t$ denotes months and $q=3t$ stands for quarters. 
In that case, the conditional probability distribution of quarterly GDP growth can be written as $\mathbb{P}(y_q | \Omega_{t})$, where the information set $\Omega_{t}$ contains monthly regressor (indicator) series of length $l$ with measurements available until the $t$-th month: $\Omega_{t} = \left[ \mathbf{x_{t}}, \mathbf{x_{t-1}}, \dots, \mathbf{x_{t-l+1}} \right]$. 
If the series of quarterly GDP growth has no long-term memory, i.e., it is ergodic, then Equation \ref{eq:target_conditional_dist} is expected to hold for a given length of training sequences ($l$) and for all $t^{*} < t - l$ time period -- at least with good approximation:  
\begin{align}   \label{eq:target_conditional_dist}
\mathbb{P}(y_t|\Omega_{t^{*}}, \Omega_{t}) = \mathbb{P}(y_t|\Omega_{t}),
\end{align}
This question is of essential relevance to our predictive analysis because the choice of the fitting neural network architecture largely depends on what we think about the data-generating process of GDP growth. 
In other words, in the context of system theory, what do we think about the impulse response function of the target variable? 
It is worth noting that both the MLP and the 1D CNN build around the assumption of fixed finite support for their target variable, in our case, GDP growth.\footnote{The support of a real-valued function $f(.)$ is the subset of the function domain containing the elements which are not mapped to zero. In our case, finite support means that the impulse response is bounded in time.} 
While gated RNNs still assume a finite impulse response for the target variable, they allow for a much longer support. 
Furthermore, through their gating mechanism, they can dynamically adjust for the length of the impulse response. 
With this type of flexibility, however, comes complexity (much more trainable parameters) as well. 

We should also note that gated RNNs are even more complex than those standard RNNs: 
While in standard RNNs cells are represented by a single neuron, LSTM and GRU cells contain several neurons inside, requiring at least a matrix for every gate. 
As a result, they are more prone to overfitting and highly sensitive to input noise. 
Moreover, due to their feedback loop, the stability analysis of (gated) RNNs is also more complicated than the feedforward networks. 
Even the initialization of such recurrent networks can be complicated in many cases \citep{ekman2021learning}. 
On the other hand, if we think about the economy as a system with a long memory, then gated RNNs might provide more accurate nowcasts than any previously presented architectures, despite the aforementioned difficulties. 

Next, in the empirical analysis, we let the data speak and evaluate the nowcasting accuracy of the different ANN architectures presented above. 
Accordingly, our competitor models will be the MLP, the 1D CNN, the Elman RNN, the LSTM, and the GRU. 
We believe that the results of this comprehensive nowcasting competition will identify the suitable architecture(s) for this predicitve analysis and also reveal some aspects of the underlying data-generating-process of GDP growth. 


\section{Empirical analysis} \label{sec:empirical_analysis} 

In this section, we first describe the data and the applied feature selection procedure. 
Then, we define the nowcasting exercise. 
After that, we present the results of the empirical analysis. 
Finally, we study the contribution of the various monthly indicators to the nowcasts.

\subsection{Data and feature selection}     \label{sub-sec:feature_selection}

\begin{figure}[H]
\begin{center}
\includegraphics[width = \textwidth]{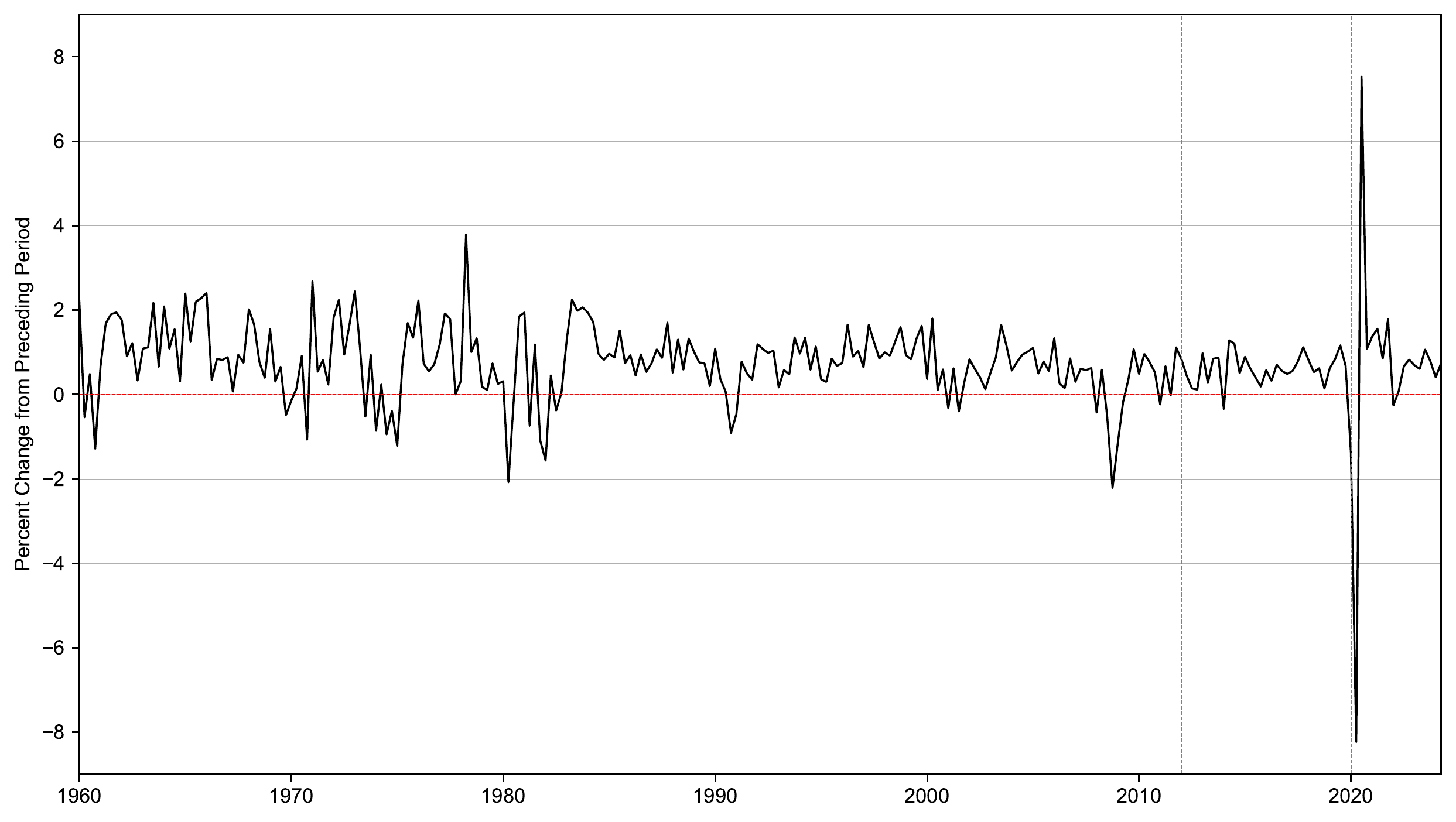}
\caption{U.S. Gross Domestic Product, percent change from preceding period (quarter). 
Vertical dashed lines indicate the two evaluation periods. 
Source: Federal Reserve Economic Data (FRED).}
\label{fig:gdp_growth}
\end{center}
\end{figure}

The target variable of the empirical analysis is the percentage change (calculated as log-difference) in the U.S. Gross Domestic Product (GDPC1) relative to the preceding period.\footnote{The series for the quarterly U.S. GDP can be downloaded from: \url{https://fred.stlouisfed.org/series/GDPC1}.} 
The base series for the U.S. GDP is seasonally adjusted, measured at a quarterly frequency, and available from 1947:Q1 to 2024:Q2 by the time the empirical analysis was closed. 
Since a significant proportion of the monthly regressor data is only available from 1960, we adjusted the full sample from 1960:Q1 to 2024:Q2, containing 258 target measurements. 
Figure \ref{fig:gdp_growth} plots the target series for quarterly GDP growth over the full sample period. 
Then, Table \ref{tab:target_descriptives} presents some main descriptive statistics for the target series and reports the test statistics of the KPSS and the ADF stationary test. 
Those statistics are reported for the full sample period and two sub-samples corresponding to our evaluation periods. 

\begin{table}[H]
\centering
\caption{Main descriptive statistics of GDP growth and the test statistics of the KPSS and the ADF stationary test.} 
\centering
\def\arraystretch{1.0}
\begin{tabular}{ p{3.6cm} | c | c | c }
{} & 1960:Q1 -- 2024:Q2 & 2012:Q1 -- 2019:Q4 & 2012:Q1 -- 2024:Q2 \\
\hline \hline
$\bar{y}_t$                                        &            $0.740$ &            $0.618$ &            $0.598$ \\
$\sigma(y_t)$                                      &            $1.068$ &            $0.359$ &            $1.684$ \\
$100 \cdot \left( \sigma(y_t) / \bar{y}_t \right)$ &          $144.331$ &           $58.121$ &          $281.576$ \\
\midrule
KPSS                                               &       $0.475^{**}$ &            $0.164$ &            $0.137$ \\
ADF                                                &    $-15.534^{***}$ &     $-5.739^{***}$ &     $-9.193^{***}$ \\
\bottomrule 
\end{tabular}
\caption*{\begin{footnotesize}
\textbf{Notes:} For the KPSS test, the null hypothesis is that the series is stationary, and the alternative hypothesis is that the series has a unit root, i.e., it is non-stationary. 
Oppositely, the null hypothesis of the ADF test is that the series has a unit root. 
The stars denote statistical significance at 10\%($*$), 5\%($**$) and 1\%($***$) level of the corresponding test. 
\end{footnotesize}}
\label{tab:target_descriptives}
\end{table}

Table \ref{tab:target_descriptives} shows that average quarterly growth slightly decreased by the end of the sample. 
There are more noticeable changes in the volatility, which cannot be viewed as stable, even approximately. 
The relative volatility is only one-third of its full sample value during the first evaluation period but double that during the second test period. 
Thus, both of our evaluation periods have their unique characteristics.  
Since the first evaluation period displays balanced, largely stable growth, predicting the subtle changes in economic activity is decisive for nowcasting performance. 
From the perspective of the bias-variance trade-off, characteristics of the first evaluation period favor those models (estimators) with the lowest bias and is more tolerant of models with a higher variance \citep{hastie2009elements}. 
By contrast, nowcasting performance in the second evaluation period depends largely on the timely prediction of the COVID crisis. Consequently, idiosyncrasies of the second test period favor models (estimators) with the lowest variance. 
Based on the KPSS and the ADF statistics, quarterly GDP growth can be considered stationary over the two distinct evaluation periods. 
Considering the full sample period, however, the stationarity of the target series seems questionable, as the KPSS and ADF tests result in contradictory conclusions. 

Figure \ref{fig:target_acf_pacf} characterizes the persistence of the target series as it plots the ACF and PACF functions for the same sample periods. 
Then, inspired by \textcite{botha2021nowcasting}, Figure \ref{fig:target_seasonality} shows the values of GDP growth from each respective quarter for the relevant sample periods. 
The main conclusions from these figures are:
\begin{itemize}
    \item [-] GDP growth cannot be described by its past observations. 
    Thus, similar to \textcite{giannone2008nowcasting},  we do not include any lagged values of the target series in the regressor vector; 
    \item [-] The initial seasonal adjustment removed all the predictable patterns from the target series.
\end{itemize}

In our empirical analysis, we try to extract the information content of several monthly macroeconomic indicators (regressors or features) whose intra-quarterly development may be related to current economic growth. 
The data for the input features comes from the FRED-MD database, the monthly database for Macroeconomic Research of the Federal Reserve Bank of St. Louis, which is described extensively in \textcite{mccracken2016fred}.\footnote{Monthly vintages of the FRED-MD database are openly available and can be downloaded from \url{https://research.stlouisfed.org/econ/mccracken/fred-databases/}}. 
The database contains a total of 134 monthly indicators (features or regressors) which are grouped into eight categories:\footnote{Newer vintages of the FRED-MD database contain slightly fewer features (typically around 125), but most of them are available across all vintages used in the empirical analysis (see later in Section \ref{sec:nowcasting_exercise}).} 
\begin{enumerate}
    \item Output and income (Table \ref{tab:fred-md_group_01})
    \item Labour market (Table \ref{tab:fred-md_group_02})
    \item Housing (Table \ref{tab:fred-md_group_03})
    \item Consumption, orders and inventories (Table \ref{tab:fred-md_group_04})
    \item Money and credit (Table \ref{tab:fred-md_group_05})
    \item Interest and exchange rates (Table \ref{tab:fred-md_group_06})
    \item Prices (Table \ref{tab:fred-md_group_07})
    \item Stock market (Table \ref{tab:fred-md_group_08})
\end{enumerate}

The complete list of the data and the series-related data transformations are described in detail in Appendix \ref{app:data}. 
Monthly indicator data are available from 1960:M1 to 2024:M6, so we have 774 observations for the potential regressors. 
For the empirical analysis, all monthly indicators (features) are detrended using the corresponding data transformations. 
This is important because all of our competitor models, except for the MLP, are designed to estimate time-invariant prediction rules. 
Consequently, when both the target variable and the regressors (features) are stationary, we can generally expect better predictive performance from these models compared to when modeling non-stationary series.\footnote{Figure \ref{fig:ipdcongd} shows the raw levels of Industrial Production: Durable Consumer Goods (IPDCONGD, upper panel) and the result of the adequate data transformation (lower panel). 
Similarly, Figure \ref{fig:features_dfm} displays the result of the data transformation for those monthly indicators also used in the benchmark DFM.} 
As the combined results of Table \ref{tab:stationarity_regressors_raw} and Table \ref{tab:stationarity_regressors_transformed} suggest, most of the monthly indicators are transformed to be approximately stationary. 
Overall, we can say that the statistical properties of the target series and the regressors align well with the requirements of the adopted ANN architectures. 

For the training of the different ANNs, we use all the monthly indicator series included in the FRED-MD database. 
At the same time, we introduce a bottleneck (encoder) layer, as the first layer in each network, to reduce the dimensionality of the feature space. 
The resulting low-dimensional embedding (state) vector contains a linear combination of the original variables, so in this respect, it performs a transformation on the data similar to principal component analysis (PCA). 
In the case of PCA, however, the weights of the resulting linear combination are determined to minimize the quadratic error between the variable reconstructed from the resulting low-dimensional representation and the original. 
By contrast, in the solution we propose, the weights of the linear combination are trained in the same integrated training process.
Thus they are determined to minimize the prediction error. 
Considering the relatively small number of available training samples, this type of dimension reduction should significantly help each ANN's generalization capability. 
Besides dimension reduction, we also apply regularization to the weight vector for the bottleneck layer ($\mathbf{w}_b$). 
Equation \ref{eq:bottleneck_regularization} describes the cost function augmented with the an L1 (Lasso) regularization term \citep{jiang2016variable}: 
\begin{align}   \label{eq:bottleneck_regularization}
C(\mathbf{w}) = \frac{1}{N} \sum_{i=1}^{N} L \left( y_{i}, f(\mathbf{w}, \mathbf{x_i}) \right) + \lambda \lVert \mathbf{w}_b \rVert_{1} 
\end{align}
where $L()$ is the loss (criterion) function selected for training, $\lVert \mathbf{w}_b \rVert_{1}$ denotes the L1 norm of $\mathbf{w}_b$ and $\lambda$ is a hyperparameter which determines how severe the penalty is.\footnote{We will specify the loss function in Section \ref{sec:nowcasting_exercise}, during the description of the training process.} 
Intuitively, a higher value of $\lambda$ induces a bigger penalty to the L1 norm thus it leads to a more sparse weight vector. 
As Equation \ref{eq:bottleneck_regularization} suggests, we do not use regularization in any other layer of the different networks. 
This L1 (Lasso) regularization leads to a sparse weight vector in the bottleneck (encoder) layer and ultimately results in feature selection by giving zeros to many of the input features \citep{jiang2016variable}. 
Figure \ref{fig:weights} below shows the histograms of the estimated parameters (weights and biases) for the bottleneck layer ($\mathbf{w}_b^{*}$) without regularization (\ref{fig:weights_default}) and with L1 regularization(\ref{fig:weights_l1}). 

\begin{figure}[H]
\centering
   \begin{subfigure}[t]{0.45\linewidth}
   \centering
   \includegraphics[width=\linewidth]{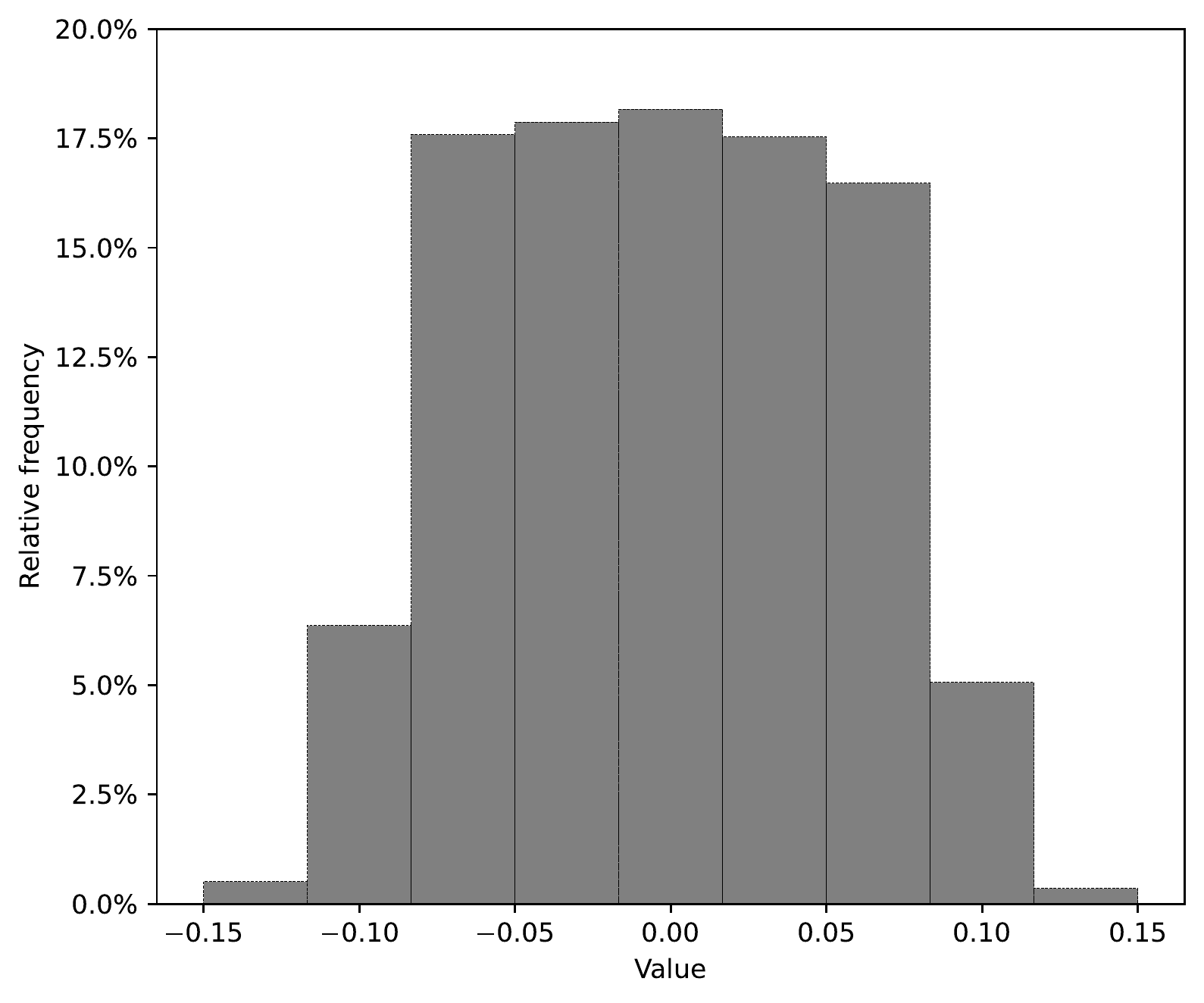}
   \caption{Histogram of $\mathbf{w}_b^{*}$ without regularization.}
   \label{fig:weights_default} 
\end{subfigure}
\hfill
\begin{subfigure}[t]{0.45\linewidth}
   \centering
   \includegraphics[width=\linewidth]{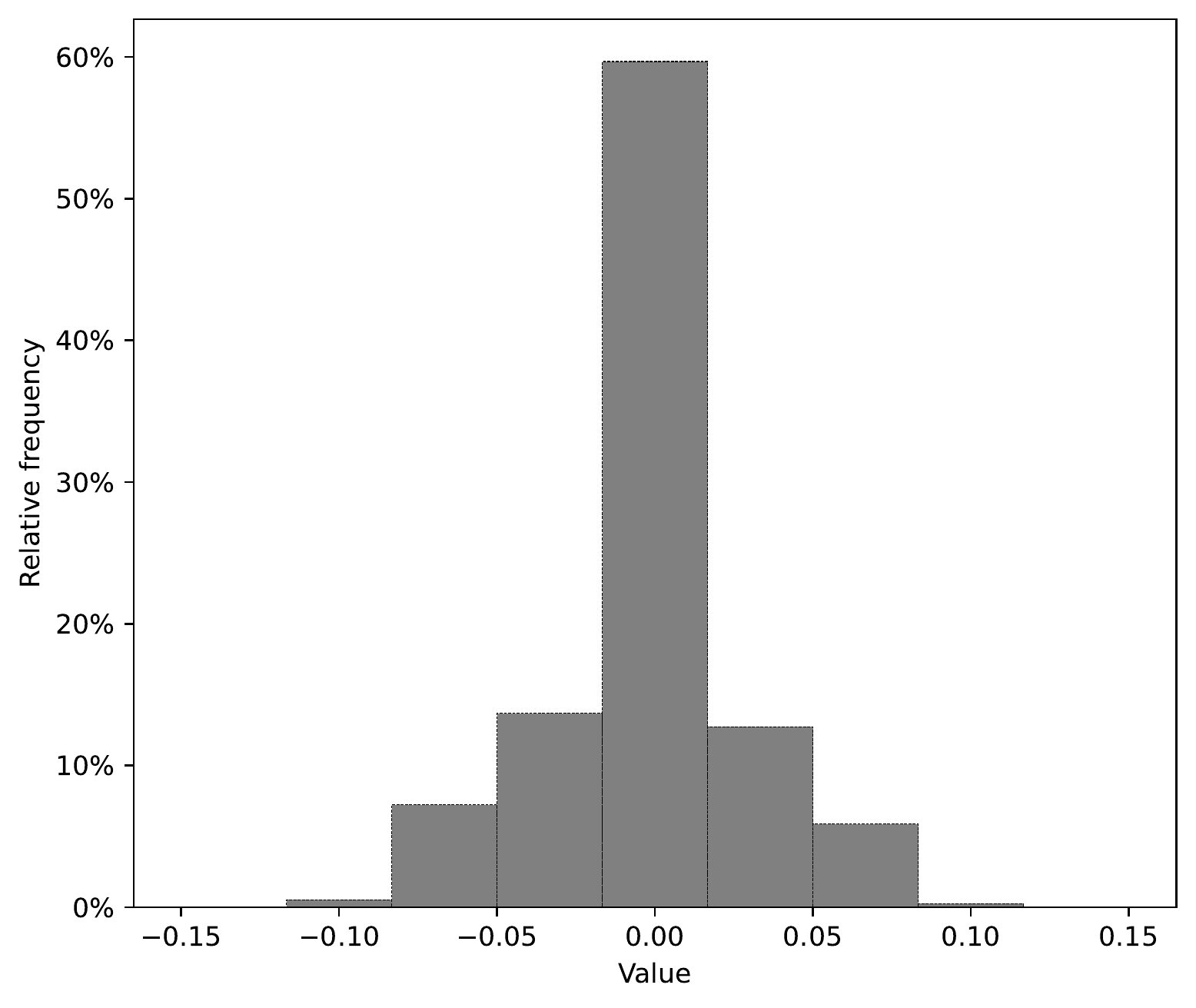}
   \caption{Histogram of $\mathbf{w}_b^{*}$ with L1 regularization.}
   \label{fig:weights_l1}
\end{subfigure}
\centering
\caption{Empirical distribution of the estimated parameters (weights and biases) for the bottleneck (encoder) layer in the 1D CNN. Source: Own editing based on FRED-MD.}
\label{fig:weights}
\end{figure}

\subsection{The nowcasting exercise}    \label{sec:nowcasting_exercise}

Our full dataset ranges from 1960:Q1 to 2024:Q2, so that it contains feature observations for every month and target observations for every quarter (i.e., for every third month). 
We assume that quarterly GDP measurements are available for months M3, M6, M9, and M12. 
In other words, we assume them to be dated by the last month of the corresponding quarter.\footnote{FRED data releases associate these quarterly target measurements with timestamps M1, M4, M7, and M10. However, using these timestamps would be counter-intuitive in an empirical analysis focused on nowcasting. 
Following the literature, we assume that quarterly GDP measurements are published at the end of the current quarter. 
That corresponds with the reality since the first measurements (first estimates) are available about a month after the end of the current quarter.} 
We evaluate the results from two evaluation (test) periods during the empirical analysis. 
The first evaluation period ranges from 2012:Q1 to 2019:Q4, so it ends before the COVID-19 pandemic would hit the US economy. 
It contains 32 observations for the quarterly GDP growth and 96 measurements for the monthly indicators. 
In comparison, the second evaluation period lasts until 2024:Q2, so it also includes those periods of the COVID crisis. 
Specifically, it ranges from 2012:Q1 to 2024:Q2, containing 50 observations for the target variable and 150 measurements for the regressors. 
We use the real-time monthly vintages of the FRED-MD database according to those test periods. 
Thus, the first FRED-MD vintage used in the empirical analysis is released for 2012:M1. 
Depending on the information set based on which nowcasts are generated, we use every third FRED-MD vintage until the end of the evaluation period.\footnote{Table \ref{tab:fred_md_timestamps} in Appendix \ref{app:results} reports the monthly timestamps of those FRED-MD vintages corresponding to each intra-quarterly information set.}

During the empirical analysis, we distinguish three different nowcasting scenarios depending on the underlying information set ($\mathbf{\Omega}$) available for our models. 
The \textit{1-month} nowcasting scenario presumes that monthly indicator data is available up through the first month within a quarter ($m_1$). 
In line with the above, we let a quarter $q$ be dated by its last month. 
For example, the first quarter of 2012 (2012:Q1) is associated with the monthly timestamp 2012:M3. 
If $q$ stands for quarters and $t$ denotes months, then the adequate relation between quarterly and monthly time indices can be expressed as $t = 3q$. 
Hence, the information set corresponding to the \textit{1-month} nowcasting scenario, i.e., $\Omega_{m_{1}}$, consists of input sequences where target observations ($y_{q}$) are associated with regressor vectors that contain monthly indicator data until the first intra-quarterly month: $m_1 = 3(q-1)+1$. 
Formally, $\Omega_{m_{1}} = \bigcup\displaylimits_{q} \{ \left( \varphi_{m_{1}}( \mathbf{x} ), y_{q} \right) \}$, where $\left( \varphi_{m_{1}}( \mathbf{x} ), y_{q} \right)$ is an input sequence composed by a quarterly target measurement ($y_{q}$) and its associated monthly regressor vector $\varphi_{m_{1}}( \mathbf{x} )$. 
The regressor vector $\varphi( \mathbf{x}_{m_{1}} )$ is defined as $\varphi_{m_{1}}( \mathbf{x} ) = \left[ \mathbf{x}_{m_{1}-l+1}, \dots, \mathbf{x}_{m_{1}} \right]$, where $l$ stands for the length (also measured in months) of the input sequences. 
Similarly, the second \textit{'2-month'} information set or nowcasting scenario ($\Omega_{m_{2}}$) is related to those FRED-MD vintages released during the second intra-quarterly month. 
Following the above definitions, $\Omega_{m_{2}} = \bigcup\displaylimits_{q} \{ \left( \varphi_{m_{2}}( \mathbf{x} ), y_{q} \right) \}$, where $m_2 = 3(q-1) + 2$ is the second intra-quarterly month. 
Finally, the third \textit{'3-month'} information set or nowcasting scenario ($\Omega_{m_{3}}$) corresponds to those FRED-MD vintages released in the last month of a given quarter: 
$\Omega_{m_{3}} = \bigcup\displaylimits_{q} \{ \left( \varphi_{m_{3}}( \mathbf{x} ), y_{q} \right) \}$, where $m_3 = 3(q-1) + 3$ stands for the last month of the current quarter. 
Together, the distinguished information sets form a three-step nowcasting window along which density nowcasts are generated by the different models and algorithms. 
Intuitively, we expect to see the most accurate nowcasts at end of the nowcasting window ($\mathbf{\Omega} = \Omega_{m_{3}}$), where all intra-quarterly indicator data is available for prediction.\footnote{Figure \ref{fig:visu_regressor_vectors} in Appendix \ref{app:results} illustrates the formation of input sequences (regressor vectors) corresponding to each intra-quarterly information set.}

For each information set within a given quarter, the nowcast is defined as the expected value of GDP growth conditional on the available information set \citep{giannone2008nowcasting}: 
\begin{align}   \label{eq:nowcast_defintion}
\hat{y}_{q|m_{j}} = \mathbb{E}(y_{q} | \Omega_{m_{j}}, \mathcal{M}_c), \qquad m_j \in \{ m_1, m_2, m_3 \}
\end{align}
where $\mathcal{M}_c$ denotes one of the competitor ANNs, i.e., $\mathcal{M}_c \in \{$ MLP, 1D CNN, RNN, LSTM, GRU $\}$. 
The underlying model determines what type of functions $f(.)$, mapping the regressors into a predicted value for the target variable, can be learned during the training (estimation) process. 
Thus, the nowcast for a given quarter $q$ is generated as follows:
\begin{align}  \label{eq:nowcast_generated}
\hat{y}_{q} = \hat{f}_{c, j} \left( \varphi_{m_{j}} (\mathbf{x}) \right)
\end{align}
where $\hat{f}_{c, j}$ represents that specific mapping (function) which is learned by the $c$-th competitor ANN at the $j$-th step of the nowcasting window (in the $j$-th nowcasting scenario). 
As Equation \ref{eq:nowcast_generated} suggests, similarly to \textcite{giannone2008nowcasting}, we do not include any lagged values of the target variable in the regressor vector. 
Regressor vectors are constructed by sampling the monthly features with a fixed-length rolling window, determined by $l$.  
Since each regressor vector is related to a given quarterly target observation, successive training (input) sequences differ in their last 3 monthly values. 
Consequently, they overlap in $l-3$ monthly values. 

All of the competitor ANNs are trained (estimated) based on a rolling training (estimation) window, where the initial training window (set) ranges from 1960:Q1 to 2019:Q4. 
By forming 8-month-long training sequences, we lose two target measurements, for which we cannot form valid regressor vectors. 
In this case, the rolling window will include 198 training sequences. 
The corresponding initial validation window (set) used for early stopping is from 2010:Q1 to 2011:Q4. 
Following the recommendation of the literature, we keep a small fraction of the data aside from training and use it for early stopping: 
Having 8 validation sequences for early stopping is very close to that ratio proposed by \textcite{amari1997asymptotic}. 
Accordingly, the first nowcast is generated for 2012:Q1. 
The number of increments between consecutive rolling windows, i.e., the step size, is 1 quarter (3 months). 
Based on Equation \ref{eq:nowcast_generated}, we generate then the series of nowcasts for the first evaluation period where $q \in \{ $2012:Q1$, \dots, $2019:Q4$ \}$. 
We do the same for the second test period, where $q \in \{ $2012:Q1$, \dots, $2024:Q2$ \}$. 
As the loss (criterion) function, we apply the mean squared error in line with the definition of the nowcast in Equation \ref{eq:nowcast_defintion}. 
During training, we update (optimize) the weights of the competitor ANNs using the Adam optimizer \citep{kingma2014adam}. 
When ANNs are trained with backpropagation, the resulting coefficients correspond to a local minimum of the loss function. 
Because there are multiple local minima, different random initializations of the backpropagation algorithm result in different weights and, therefore, different nowcasts.  
\textcite{kourentzes2014neural} suggests retraining the network for multiple random initializations, generating a distribution of nowcasts. 
The mean of a kernel density estimate of the different nowcasts shows superiority in terms of nowcasting accuracy when compared to a single nowcast applying only one trained network. 
This so-called ensemble operator approach is also applied here for 30 different random initializations. 

The predictive accuracy of the competitor ANNs is evaluated relative to a naive constant growth model and a dynamic factor model (DFM). 
The specification and the detailed description of our benchmark DFM can be found in \textcite{fulton2024dfm}.
Based on \textcite{fulton2024dfm}, our benchmark DFM includes the following four monthly indicators in its measurement vector: 
\begin{itemize}
\item [-] Industrial production: IPMANSICS (instead of IPMAN), 
\item [-] Real personal income excluding transfer receipts: W875RX1 (included in FRED-MD), 
\item [-] Manufacturing and trade sales: CMRMTSPLx (instead of CMRMTSPL), 
\item [-] Employees on non-farm payrolls: PAYEMS (included in FRED-MD). 
\end{itemize}

The FRED-MD database includes two adjusted series for IPMAN and CMRMTSPL: IPMANSICS and CMRMTSPLx, respectively \citep{mccracken2016fred}. 
Although they use a different classification system (SIC for IPMANSICS and CMRMTSPLx, while NAICS for IPMAN and CMRMTSPL), overall dynamics of the two variants are very similar, and the adjusted series included in FRED-MD are almost fully correlated with those used by \textcite{fulton2024dfm}. 
The corresponding data transformations (i.e., taking log differences) proposed by \textcite{mccracken2016fred} are the same as those used by \textcite{fulton2024dfm}. 
While this subset of features might seem rather small, it is worth noting that state-of-the-art nowcasting performance can be achieved even with a single highly correlated regressor and an adequately selected (constructed) nowcasting framework, as shown by \citep{labonne2020capturing}. 
After estimating the DFM, we apply a second, separate state space model to bind the values of the extracted monthly factor to quarterly GDP growth. 
Alternatively, one could adopt a fully integrated state space approach, resulting in a mixed-frequency measurement vector in the DFM's specification. 
However, since this would require modifying the specification given by \textcite{fulton2024dfm}, we chose the two-step estimation procedure, which is slightly faster to execute and yields very similar results \citep{marcellino2010factor}. 
Figure \ref{fig:nowcast_benchmarks} below illustrates the nowcasting accuracy of our benchmark models during the two evaluation periods. 

\begin{figure}[H]
\centering
     \begin{subfigure}[b]{0.85\textwidth}
         \centering
         \includegraphics[width=\textwidth]{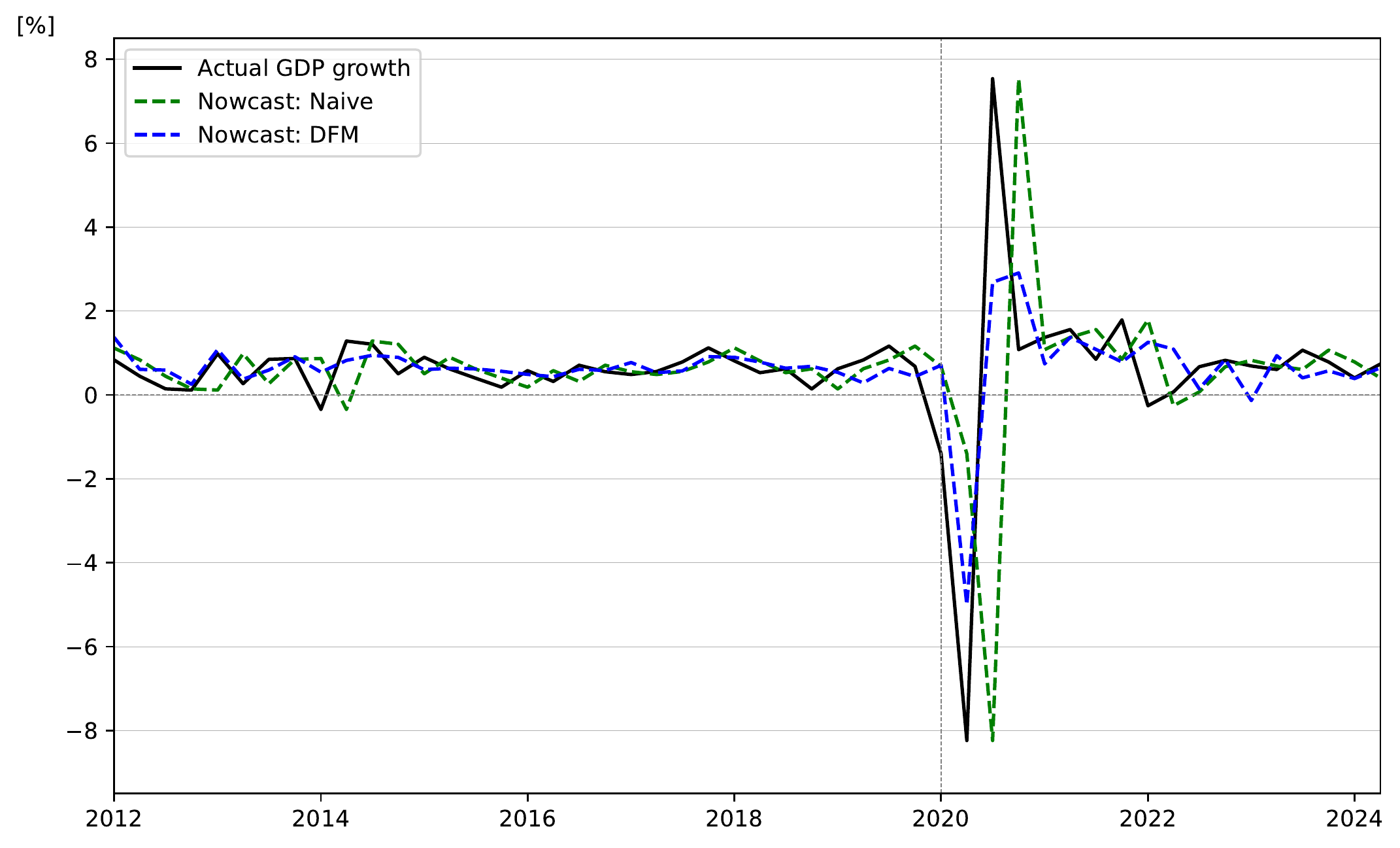}
         \caption{Realised GDP growth (Q-on-Q percentage change, $y_{q}$) against the nowcasts generated by the naive constant growth model ($\hat{y}_{b_{1}, q}$) and the benchmark DFM ($\hat{y}_{b_{2}, q}$). 
         The first evaluation period ranges from 2012:Q1 to 2019:Q4. 
         The second evaluation period is from 2012:Q1 to 2024:Q2.}
         \label{fig:nowcast_benchmarks_predicted}
     \end{subfigure}
     \hfill
     \hfill
     \begin{subfigure}[b]{0.85\textwidth}
         \centering
         \includegraphics[width=\textwidth]{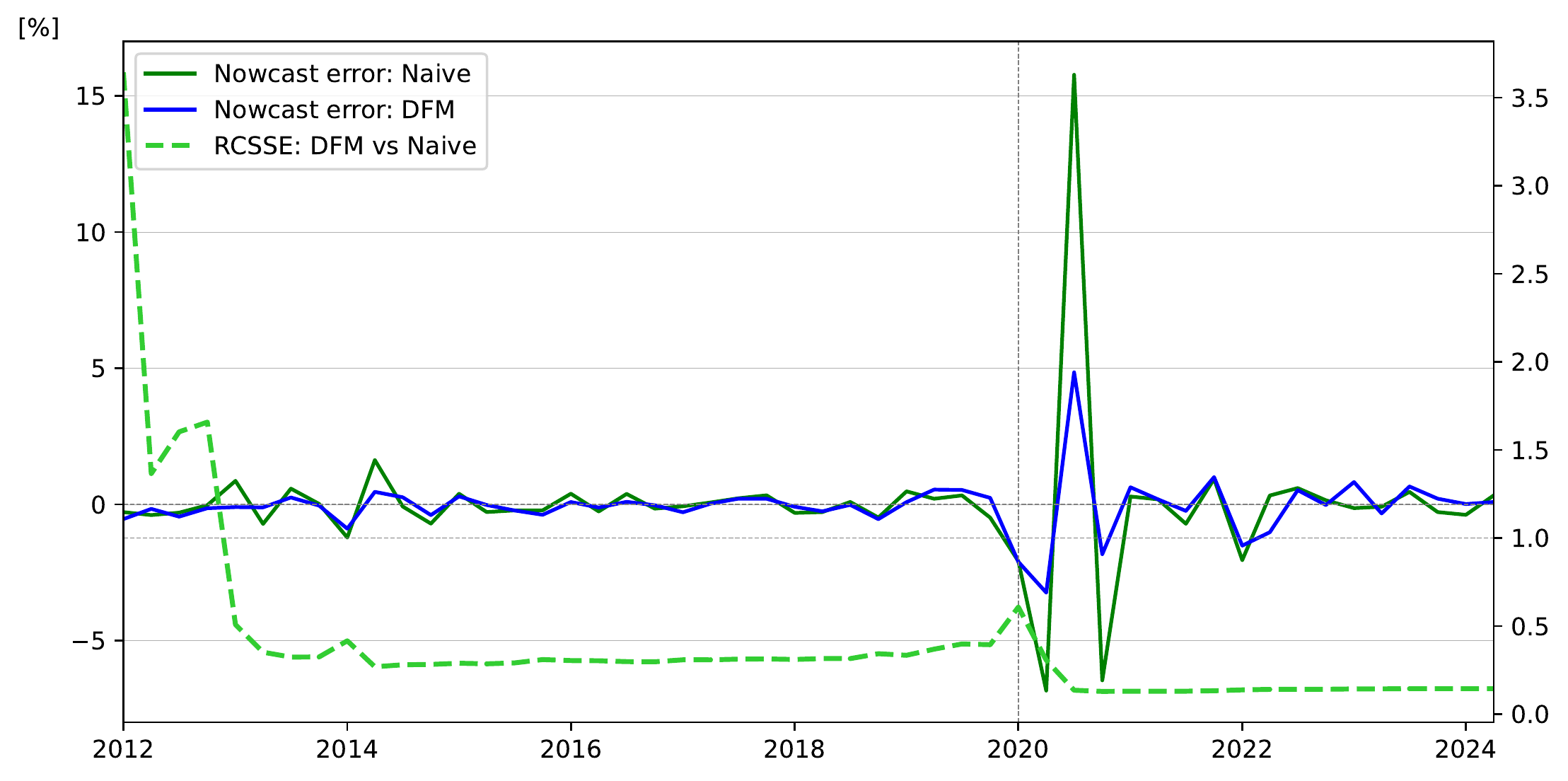}
         \caption{Nowcast errors ($\hat{e}_{b_{j}, q} = y_{Q} - \hat{y}_{b_{j}, q}$, left axis) and the relative cumulative sum of the squared nowcast errors (RCSSE) for the DFM. 
         Against the naive benchmark model, the RCSSE (right axis) for the DFM is computed as $RCSSE_{DFM} = \sum_{q} \hat{e}_{b_{2}, q}^{2} / \sum_{q} \hat{e}_{b_{1}, q}^{2}$, where $\hat{e}^{2}_{b_{j}, q}$ denotes the squared nowcast error of the $j$-th benchmark model.}
         \label{fig:nowcast_benchmarks_error}
     \end{subfigure}     
    
\caption{Nowcasting performance of the benchmark DFM. Source: Own editing based on FRED-MD.}
\label{fig:nowcast_benchmarks}

\end{figure}

Figure \ref{fig:nowcast_benchmarks_predicted} shows that even the naive benchmark model fits the data surprisingly well during the first evaluation period, i.e., in periods of quasi-stable economic growth. 
However, it fails when large shocks (positive or negative) hit the economy, as that is apparent during the second test period. 
The economic impact of the COVID-19 pandemic resulted in a V-shaped recession: a sharp downturn in 2020:Q2 followed by a quick rebound in 2020:Q3. 
While signs of the economic downturn could have been seen already in the macro indicators for April, the naive benchmark model fails to predict the forthcoming recession in Q2. 
Incorporating the additional information content from higher frequency data, the DFM performs much better in those critical quarters and generates a relative RMSE of $0.380$ towards the naive benchmark during the second test period. 
Albeit the DFM also beats the naive benchmark by a statistically significant margin in the first evaluation period, it yields virtually the same accuracy from 2014 to 2019 as the naive model.  
This finding is apparent in Figure \ref{fig:nowcast_benchmarks_error}, where we display the DFM's relative cumulative sum of squared nowcast errors against the naive benchmark. 
Notwithstanding that our restricted evaluation period does not include those periods of the COVID crisis, it also has its own unique characteristics. 
Table \ref{tab:eval_benchmarks} summarizes the above and reports the evaluation metrics for our benchmark models. 
It might be interesting to see Table 1 and Table 2 together, as the combined results point out the strong negative relationship between the volatility of GDP growth and the nowcasting performance of the naive benchmark model.  

\begin{table}[H]
\centering
\caption{Accuracy of nowcasts generated by the naive constant growth model and the benchmark DFM: RMSE and MAE evaluation.}
\def\arraystretch{1.0}
\begin{tabular}{l | l l | l l | l l}
\toprule
{} & \multicolumn{2}{c|}{$\Omega_{m_{1}}$} & \multicolumn{2}{c|}{$\Omega_{m_{2}}$} & \multicolumn{2}{c}{$\Omega_{m_{3}}$} \\
{} & RMSE & MAE & RMSE & MAE & RMSE & MAE \\
\toprule
\multicolumn{7}{c}{2012:Q1 -- 2019:Q4} \\
\midrule
Naive                &      $0.515$ &       $0.390$ &       $0.515$ &       $0.390$ &       $0.515$ &        $0.390$ \\
DFM (absolute) &      $0.364$ &       $0.262$ &       $0.345$ &       $0.263$ &       $0.323$ &        $0.252$ \\
DFM (relative) &  $0.707^{*}$ &  $0.673^{**}$ &  $0.670^{**}$ &  $0.674^{**}$ &  $0.628^{**}$ &  $0.647^{***}$ \\
\hline \hline
\multicolumn{7}{c}{2012:Q1 -- 2024:Q2} \\
\midrule
Naive                &      $2.672$ &      $1.011$ &      $2.672$ &       $1.011$ &      $2.672$ &       $1.011$ \\
DFM (absolute) &      $1.329$ &      $0.642$ &      $0.955$ &       $0.520$ &      $1.015$ &       $0.547$ \\
DFM (relative) &  $0.497$ &  $0.635^{*}$ &  $0.357^{*}$ &  $0.515^{**}$ &  $0.380^{*}$ &  $0.541^{**}$ \\
\bottomrule
\end{tabular}
\caption*{\begin{footnotesize}
\textbf{Notes:} This table reports the RMSE and MAE values for our benchmark models. 
Absolute RMSEs and MAEs are reported for the naive constant growth model and the benchmark DFM. 
For the DFM, relative RMSE and MAE values are measured against the performance of the naive constant growth model. 
A value below one indicates that the competitor model beats the naive benchmark model. 
The stars denote statistical significance at the 10\%($^{*}$), 5\%($^{**}$) and 1\%($^{***}$) level of the one-sided \textcite{diebold1995comparing} test. 
Columns are related to the individual nowcasting scenarios. 
For example, $\Omega_{m_{1}}$ refers to the 1-month nowcasting scenario, i.e., to the first step of the nowcasting window. 
\end{footnotesize}}
\label{tab:eval_benchmarks}
\end{table}

\subsection{Results}

In the following, we present the results of the empirical analysis. 
As the COVID-19 pandemic hit the US economy in 2020, we decided to evaluate our models on two different evaluation periods: 
The first evaluation period ranges from 2012:Q1 to 2019:Q4, so it does not reflect the economic impact of the COVID-19 pandemic. 
In comparison, the second evaluation period is from 2012:Q1 to 2024:Q2, containing those periods of the COVID crisis as well. 
We believe that the results from these two evaluation periods show a more balanced and well-rounded picture of the different models' generalization capability. 
These two test periods have very different characteristics. 
The first can be described as a period of balanced economic growth. 
In such circumstances, even the naive benchmark model is expected to be a strong contender for our competitor models. 
Consequently, competitor ANNs should accurately predict the subtle changes in economic growth to beat our benchmark models. 
In contrast, the second evaluation period would favor the models that can timely capture the sharp V-shaped recovery of 2020:Q2 and 2020:Q3. 
We believe that results from these two evaluation periods will provide a robust assessment of the generalization capability of the different ANN architectures, showing their strengths and weaknesses. 

Table \ref{tab:eval_no_covid} presents the results of the nowcasting competition from the first evaluation (test) period, which ranges from 2012:Q1 to 2019:Q4. 
We will see how the length of the training sequences affects the competitor models' nowcasting accuracy in the different nowcasting scenarios. 
Accordingly, Table \ref{tab:eval_no_covid} reports the results for four different sequence length configurations: $l = [8, 18, 36, 48]$. 
We also investigated the nowcasting performance with longer sequences. However, training sequences longer than 48 months did not yield more accurate nowcasts for any competitor model in RMSE or MAE terms. 
This applies to both evaluation periods. 
In the inference phase, we use input sequences (i.e., test sequences) of the same length as we used by forming the training sequences in the training phase. 
Given that test and training sequences are the same length, we use the more generic term \textit{input sequences} for both when describing the results. 
Table \ref{tab:eval_no_covid} reports the relative RMSEs generated by the competitor ANNs, while the results of the MAE evaluation are provided in Table \ref{tab:eval_no_covid_mae} in Appendix \ref{app:results}. 

\begin{table}[H]
\centering
\caption{Nowcasting performance of the competitor models relative to a naive constant growth model for GDP and a benchmark DFM specification: RMSE evaluation. Evaluation period: 2012:Q1 -- 2019:Q4.} 
\centering
\def\arraystretch{1.0}
\begin{tabular}{l | l l | l l | l l}
& \multicolumn{2}{c |}{$\Omega_{m_{1}}$} & \multicolumn{2}{c |}{$\Omega_{m_{2}}$} & \multicolumn{2}{c}{$\Omega_{m_{3}}$} \\
& Naive & DFM & Naive & DFM & Naive & DFM \\
\hline\hline

MLP ($l$ = 8)     &   $\mathbf{0.707^{*}}$ &  $\mathbf{1.000}$ &  $\mathbf{0.670^{**}}$ &  $\mathbf{1.000}$ &  $\mathbf{0.655^{**}}$ &  $\mathbf{1.043}$ \\
MLP ($l$ = 18)    &   $0.716^{*}$ &  $1.014$ &   $0.698^{*}$ &  $1.041$ &   $0.695^{*}$ &  $1.107$ \\
MLP ($l$ = 36)    &   $0.735^{*}$ &  $1.040$ &   $0.704^{*}$ &  $1.051$ &   $0.716^{*}$ &  $1.140$ \\
MLP ($l$ = 48)    &   $0.762^{*}$ &  $1.079$ &   $0.739^{*}$ &  $1.103$ &   $0.738^{*}$ &  $1.175$ \\
\midrule

1D CNN ($l$ = 8)  &   $0.745^{*}$ &  $1.055$ &   $0.674^{*}$ &  $1.005$ &  $0.637^{**}$ &  $1.015$ \\
1D CNN ($l$ = 18) &   $0.723^{*}$ &  $1.023$ &  $0.657^{**}$ &  $0.980$ &   $0.651^{*}$ &  $1.037$ \\
1D CNN ($l$ = 36) &   $\mathbf{0.697^{*}}$ &  $\mathbf{0.987}$ &  $\mathbf{0.620^{**}}$ &  $\mathbf{0.926}$ &  $\mathbf{0.627^{**}}$ &  $\mathbf{0.999}$ \\
1D CNN ($l$ = 48) &   $0.762^{*}$ &  $1.078$ &   $0.688^{*}$ &  $1.026$ &  $0.657^{**}$ &  $1.046$ \\
\midrule

RNN ($l$ = 8)     &   $0.710^{*}$ &  $1.005$ &  $0.667^{**}$ &  $0.995$ &  $\mathbf{0.597^{**}}$ &  $\mathbf{0.950}$ \\
RNN ($l$ = 18)    &   $0.701^{*}$ &  $0.992$ &  $0.666^{**}$ &  $0.994$ &  $0.606^{**}$ &  $0.965$ \\
RNN ($l$ = 36)    &   $0.727^{*}$ &  $1.029$ &  $\mathbf{0.663^{**}}$ &  $\mathbf{0.989}$ &  $0.612^{**}$ &  $0.974$ \\
RNN ($l$ = 48)    &  $\mathbf{0.694^{**}}$ &  $\mathbf{0.982}$ &   $0.667^{*}$ &  $0.995$ &  $0.617^{**}$ &  $0.984$ \\
\midrule

LSTM ($l$ = 8)    &   $0.713^{*}$ &  $1.009$ &   $0.663^{*}$ &  $0.988$ &  $\mathbf{0.608^{**}}$ &  $\mathbf{0.968}$ \\
LSTM ($l$ = 18)   &   $0.718^{*}$ &  $1.016$ &  $0.662^{**}$ &  $0.987$ &  $0.615^{**}$ &  $0.979$ \\
LSTM ($l$ = 36)   &  $\mathbf{0.698^{**}}$ &  $\mathbf{0.988}$ &   $\mathbf{0.655^{*}}$ &  $\mathbf{0.978}$ &  $0.631^{**}$ &  $1.005$ \\
LSTM ($l$ = 48)   &   $0.707^{*}$ &  $1.000$ &   $0.656^{*}$ &  $0.979$ &  $0.623^{**}$ &  $0.993$ \\
\midrule

GRU ($l$ = 8)     &  $0.703^{**}$ &  $0.994$ &  $\mathbf{0.633^{**}}$ &  $\mathbf{0.945}$ &  $\underline{\mathbf{0.589^{**}}}$ &  $\underline{\mathbf{0.938}}$ \\
GRU ($l$ = 18)    &  $\mathbf{0.697^{**}}$ &  $\mathbf{0.986}$ &  $0.644^{**}$ &  $0.961$ &  $0.589^{**}$ &  $0.938$ \\
GRU ($l$ = 36)    &   $0.710^{*}$ &  $1.005$ &   $0.661^{*}$ &  $0.986$ &  $0.590^{**}$ &  $0.940$ \\
GRU ($l$ = 48)    &   $0.709^{*}$ &  $1.004$ &  $0.652^{**}$ &  $0.973$ &  $0.594^{**}$ &  $0.946$ \\
\bottomrule
\end{tabular}
\caption*{\begin{footnotesize}
\textbf{Notes:} This table reports the relative RMSE of GDP growth for the competitor ANNs relative to a naive constant growth model and a benchmark DFM specification. 
A value below one indicates that the competitor model beats the naive benchmark model. 
The stars denote statistical significance at 10\%($*$), 5\%($**$) and 1\%($***$) level of the one-sided \textcite{diebold1995comparing} test. 
Columns are related to the individual nowcasting scenarios: e.g. $\Omega_{m_{1}}$ refers to the 1-month nowcasting scenario. 
Within each nowcasting scenario, best (lowest) relative RMSE values are highlighted in bold for the given competitor model. 
The overall best relative RMSE values are highlighted in bold and underlined. 
\end{footnotesize}}
\label{tab:eval_no_covid}
\end{table}

\newpage

Based on Table \ref{tab:eval_no_covid}, it is apparent that all competitor ANNs can beat the naive benchmark model in terms of nowcasting accuracy. 
The performance advantage of the ANNs is statistically significant with any length of input sequences at each step of the nowcasting window. 
Except for the MLP, the competitors can also outperform the benchmark DFM with properly selected regressor vectors. 
However, the margins are small and not statistically significant according to the \textcite{diebold1995comparing} test. 

Focusing on the nowcasting performance of the different ANN architectures, we see that the MLP generates the most accurate nowcasts with 8-month-long input sequences. 
With 8 long sequences, the MLP outperforms the naive benchmark model at each step of the nowcasting window. 
Table \ref{tab:eval_no_covid} shows that the performance advantage is only significant on a 10\% level in the 1-month nowcasting scenario, but it strengthens in significance in the 2-month and 3-month nowcasting scenarios. 
At the end of the nowcasting window, the MLP yields a relative RMSE of $0.655$, beating the naive benchmark on a 5\% level. 
This accuracy results in a relative RMSE of $1.034$ against the DFM, indicating almost identical nowcasting performance. 
According to Table \ref{tab:eval_no_covid}, the MLP yields basically the same accuracy as the DFM in the 1-month and 2-month nowcasting scenarios as well. 
While this also holds for the different sequence length configurations, relative RMSE values slightly increase as we take longer input sequences. 
In general, the MLP performs well compared to both benchmark models, but its (absolute) accuracy somewhat lags behind that of the other four competitor ANNs explicitly designed for sequence modeling. 
It is also worth noting that longer input sequences yield gradually less accurate nowcasts for the MLP. 
We assume that the MLP's performance is somewhat hindered by its simple time-agnostic architecture during the first evaluation period, where nowcasting performance largely depends on identifying the more subtle movements in economic activity. 

Compared to the MLP, the 1D CNN produces slightly more accurate nowcasts during the first evaluation period. 
With 8-month-long input sequences, the 1D CNN generates a relative RMSE of $0.637$ against the naive benchmark model, where the difference is statistically significant on a 5\% level. 
This performance advantage increases with 36 long input sequences, indicated by a relative RMSE of $0.627$ towards the naive benchmark. 
Compared with the DFM, the 1D CNN generates almost identically accurate nowcasts for all of the training configurations.  
Unlike the MLP, however, we also see some relative RMSEs below one, with 18 and 36 long input sequences. 
Compared to the MLP, the 1D CNN capitalizes a bit more on the longer input sequences, but there is only a slight improvement in nowcasting accuracy. 
It is easy to see that random aspects of the training process (e.g., sampling for mini-batches, weight initialization) can easily turn this difference around. 
Table \ref{tab:eval_no_covid_shortest} in Appendix \ref{app:results} measures each competitor's nowcasting performance relative to the training configuration, which uses the shortest, 8-month-long input sequences. 
Results in Table \ref{tab:eval_no_covid_shortest} indicate that, while longer input sequences may result in slightly more accurate nowcasts for some ANNs, the performance gain associated with longer regressor vectors is not significant in any training configurations. 
If we evaluate the nowcasting performance of the 1D CNN along the nowcasting window, we see the same intuitive results as for the MLP. 
The performance advantage over the naive benchmark strengthens in significance. 
In terms of statistical significance, the relative performance gain over the naive benchmark model does not improve with longer input sequences but becomes worse above 36 months. 
Table \ref{tab:eval_no_covid} also shows that the nowcasting accuracy of the 1D CNN gradually improves with consecutive intra-quarterly data releases. 
With 36-month-long sequences, the 1D CNN yields an overall best relative RMSE of $0.926$ towards the DFM in the nowcasting scenario. 
In general, the 1D CNN yields strong nowcasting performance across all the training configurations, beating the benchmark model on a 5\% significance level regarding both RMSE and MAE evaluation. 

Results in Table \ref{tab:eval_no_covid} indicate that the Elman (basic or ``vanilla'') RNN also reaches its full potential with 8 long input sequences. 
With input sequences this long, the basic RNN significantly outperforms the naive benchmark model and generates equally accurate nowcasts as the benchmark DFM. 
With 8 long input sequences, the Elman RNN generates a relative RMSE of $0.597$ against the naive benchmark model at the end of the nowcasting window. 
The performance advantage is statistically significant on a 5\% level. 
For the same training configuration, the relative RMSE is $0.950$ compared to the DFM, which does not mean a statistically significant margin. 
From Table \ref{tab:eval_no_covid}, it can be observed that longer input sequences yield slightly more accurate nowcasts in the 2-month and 3-month nowcasting scenarios. 
We also see that longer input sequences only result in significantly more accurate nowcasts towards the naive benchmark model at the first step of the nowcasting window. 
Intuitively, the best accuracy is achieved at the end of the nowcasting window. 
As Table \ref{tab:eval_no_covid} shows, the LSTM has almost identical nowcasting performance as the basic RNN during the first evaluation period. 
With 8 long input sequences, in the 3-month nowcasting scenario, the LSTM produces a relative RMSE of $0.608$ and $0.968$ against the naive benchmark model and the DFM, respectively. 
While accuracy slightly improves with longer sequences in the 2-month nowcasting scenario, differences are not significant this time either. 
Generally, relative RMSEs are virtually the same towards both benchmark models as in the case of the basic RNN. 

\begin{figure}[H]
     \centering
     \begin{subfigure}[b]{0.85\textwidth}
         \centering
         \includegraphics[width=\textwidth]{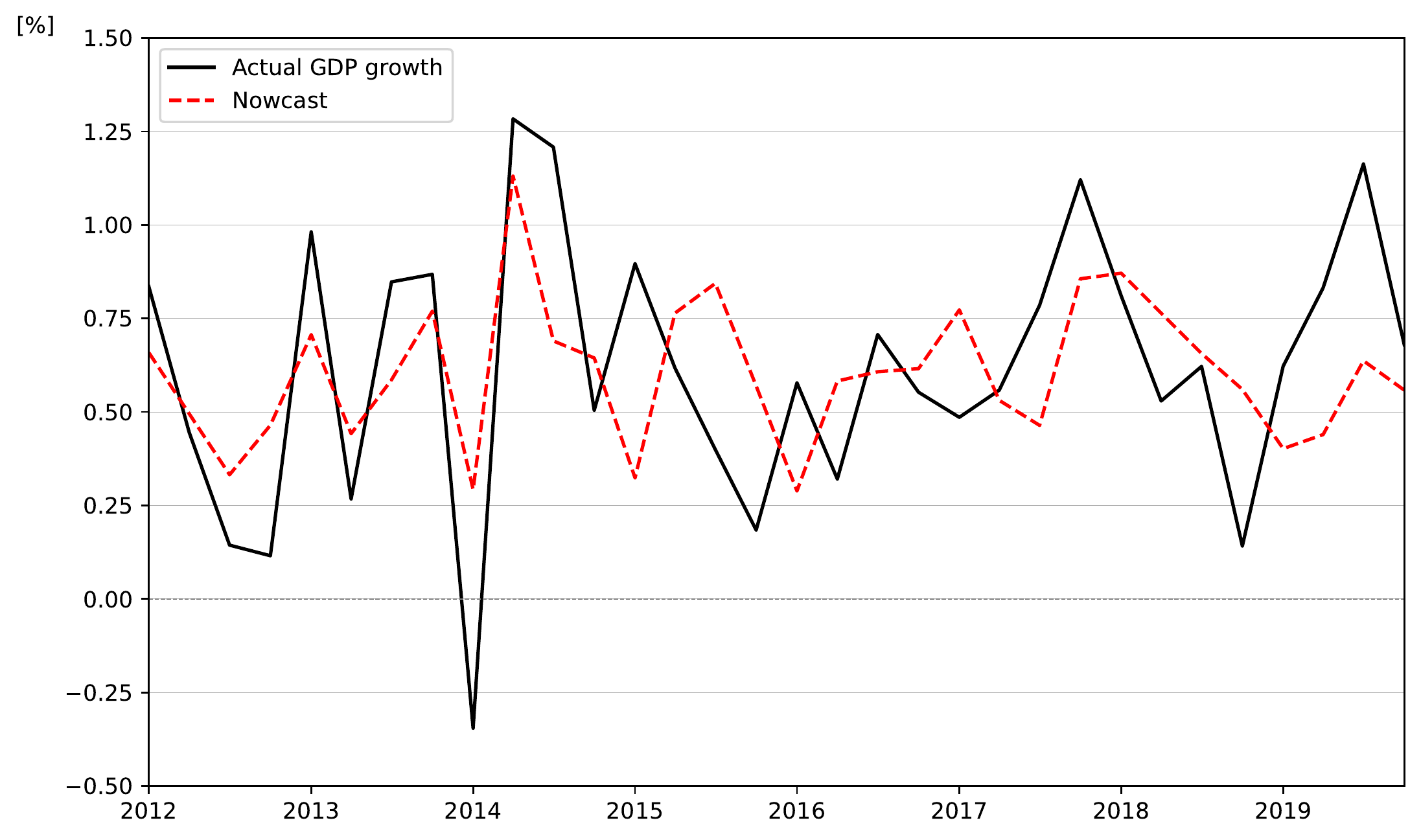}
         \caption{The actual quarterly GDP growth ($y_{q}$) versus the nowcasts generated by the best-performing competitor model ($\hat{y}_{c, q}$). 
         Best accuracy is achieved by the GRU, with 8 long input sequences ($l = 8$) in the 3-month nowcasting scenario / at the end of the nowcasting window ($v = 3t$). 
         The evaluation period ranges from 2012:Q1 to 2019:Q4.}
         \label{fig:predicted_vs_actuals_test1}
     \end{subfigure}
     \hfill
     \begin{subfigure}[b]{0.85\textwidth}
         \centering
         \includegraphics[width=\textwidth]{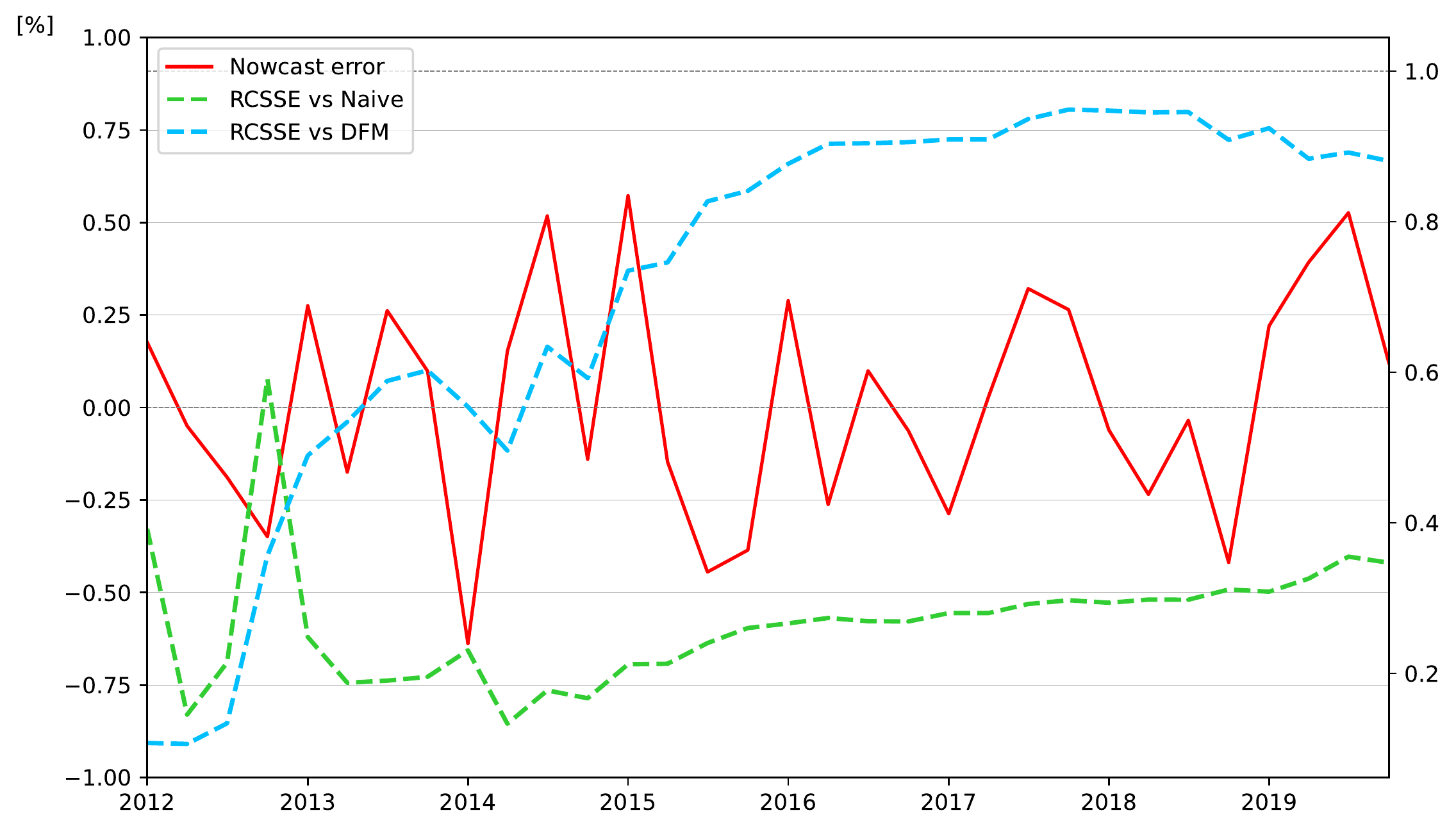}
         \caption{Nowcast errors ($\hat{e}_{c, q} = y_{q} - \hat{y}_{c, q}$, left axis) and the relative cumulative sum of squared nowcast errors (RCSSE) for the best-performing competitor model. 
         The RCSSE (right axis) towards a given benchmark ($b_j$) is computed as $RCSSE_{c_{j}} = \sum_{q} \hat{e}_{c, q}^{2} / \sum_{q} \hat{e}_{b_{j}, q}^{2}$, where $\hat{e}^{2}_{c, q}$ denotes the squared nowcast error of the competitor model ($c$).}
         \label{fig:residuals_test1}
     \end{subfigure}     
    
    \caption{Nowcasting performance of the best-performing competitor model (GRU, $l=8, v=3t-1$) during the first evaluation period. 
    Source: Own editing based on FRED-MD.}
    \label{fig:bestmdl_no_covid}

\end{figure}

Results for the other gated RNN show a largely similar picture, with some subtle differences. 
With 8 long input sequences, the GRU yields a relative RMSE of $0.562$ towards the naive benchmark model at the end of the nowcasting window. 
The difference is significant on a 5\% level. 
Compared to the DFM, the relative RMSE is $0.938$. 
While these results are the best among the ANNs for the first evaluation period, even the GRU cannot outperform the DFM by a statistically significant margin. 
We should also see that the results are very similar to those of the other competitor ANNs. 
As Table \ref{tab:eval_no_covid} shows, the GRU delivers almost the same nowcasting performance as the basic RNN during the first evaluation period. 
Even with these considerations, the GRU achieves the overall best accuracy during the first evaluation period. 
Figure \ref{fig:predicted_vs_actuals_test1} below plots the nowcasts generated by the best-performing GRU configuration against the actual GDP growth during the first evaluation period. 
Then Figure \ref{fig:residuals_test1} shows the series of the nowcast errors and how the cumulative sum of the squared nowcast errors evolved relative to the benchmark model. 

Table \ref{tab:eval_no_covid} shows that the two gated RNNs, the LSTM and the GRU, generate almost identically accurate nowcasts as the basic Elman network during the first evaluation period. 
Whilst the LSTM performs slightly worse, the GRU performs slightly better in the different nowcasting scenarios. 
In addition, as Table \ref{tab:eval_no_covid} indicates, the GRU achieves the best accuracy with 8 long input sequences. 
These results overall suggest that architectural features enabling for long-term memory does not improve the nowcasting accuracy during the first evaluation period. 
In our interpretation, the slight advantage of the GRU can be much more attributed to the characteristics of the target series and, in this context, to the bias-variance trade-off. 
From this perspective, the GRU seems to have just the right amount of free (trainable) parameters (weights) considering the characteristics of the first evaluation period. 
On the one hand, the GRU has three times more parameters than an Elman RNN with a same-sized hidden state. 
Having many more parameters means the model will have a much higher chance of fitting the training data well, i.e., with less bias. 
On the other hand, the GRU is a more efficient and parsimonious architecture than the LSTM, having only three-quarters as many parameters as an evenly sized LSTM. 
Since the first evaluation period is characterized by stable, balanced economic growth, nowcasting accuracy will be determined by predicting the more subtle changes in the target series. 
In such circumstances, having a low bias is particularly important. 
Consequently, a mapping with more degrees of freedom is generally preferred. 
From this point of view, it is not surprising that the characteristics of the first evaluation period favor those models with more trainable weights, the GRU in our case. 

Overall, results from the first evaluation period indicate that the competitor ANNs generally produce the most accurate nowcasts with only 8-month-long input sequences. 
The superiority of the 8-month-long regressor vectors is most evident at the last step of the nowcasting window ($\mathbf{\Omega} = \Omega_3$). 
While longer input sequences may result in slightly more accurate nowcasts for some training configurations, the performance gain stemming from longer input sequences is not significant in any case (as shown in Table \ref{tab:eval_no_covid_shortest}). 
Regarding the competitor models' nowcasting performance across the different nowcasting scenarios, the additional informational content of the second-month data releases often results in significantly more accurate nowcasts. 
Comparing the 3-month and the 2-month nowcasting scenarios, we see that although additional informational content of the third-monthly data releases can improve the absolute accuracy of the nowcasts, the performance gain relative to the naive benchmark model does not change in terms of statistical significance. 

After we have seen the results for the first evaluation period, Table \ref{tab:eval_with_covid} reports the outcome of the RMSE evaluation for the second evaluation period, which ranges from 2012:Q1 to 2024:Q2. 
As before, the results of the MAE evaluation can be found in Table \ref{tab:eval_with_covid_mae} in Appendix \ref{app:results}. 

\begin{table}[H]
\centering
\caption{Nowcasting performance of the competitor models relative to a naive constant growth model for GDP and a benchmark DFM specification: RMSE. Evaluation period: 2012:Q1 -- 2024:Q2.} 
\centering
\def\arraystretch{1.0}
\begin{tabular}{l | l l | l l | l l}
& \multicolumn{2}{c |}{$\Omega_{m_{1}}$} & \multicolumn{2}{c |}{$\Omega_{m_{2}}$} & \multicolumn{2}{c}{$\Omega_{m_{3}}$} \\
& Naive & DFM & Naive & DFM & Naive & DFM \\
\hline\hline

MLP ($l$ = 8)     &  $\mathbf{0.509}$ &  $\mathbf{1.023}$ &  $\mathbf{0.333}$ &  $\mathbf{0.932}$ &  $\mathbf{0.289^{*}}$ &  $\mathbf{0.762}$ \\
MLP ($l$ = 18)    &  $0.549$ &  $1.105$ &  $0.378$ &  $1.058$ &  $0.369$ &  $0.972$ \\
MLP ($l$ = 36)    &  $0.602$ &  $1.210$ &  $0.418^{*}$ &  $1.171$ &  $0.469$ &  $1.233$ \\
MLP ($l$ = 48)    &  $0.605$ &  $1.217$ &  $0.457$ &  $1.279$ &  $0.479$ &  $1.260$ \\
\midrule

1D CNN ($l$ = 8)  &  $\mathbf{0.486}$ &  $\mathbf{0.977}$ &  $\mathbf{0.384}$ &  $\mathbf{1.076}$ &  $\underline{\mathbf{0.199^{*}}}$ &  $\underline{\mathbf{0.523^{*}}}$ \\
1D CNN ($l$ = 18) &  $0.590$ &  $1.186$ &  $0.443$ &  $1.239$ &  $0.227^{*}$ &  $0.597^{*}$ \\
1D CNN ($l$ = 36) &  $0.658$ &  $1.323$ &  $0.567$ &  $1.588$ &  $0.448$ &  $1.179$ \\
1D CNN ($l$ = 48) &  $0.648$ &  $1.303$ &  $0.587$ &  $1.641$ &  $0.464$ &  $1.223$ \\
\midrule

RNN ($l$ = 8)     &  $\mathbf{0.484}$ &  $\mathbf{0.974}$ &  $\mathbf{0.507}$ &  $\mathbf{1.420}$ &  $\mathbf{0.325}$ &  $\mathbf{0.854}$ \\
RNN ($l$ = 18)    &  $0.483$ &  $0.971$ &  $0.593$ &  $1.659$ &  $0.334^{*}$ &  $0.880$ \\
RNN ($l$ = 36)    &  $0.516$ &  $1.037$ &  $0.587$ &  $1.642$ &  $0.434^{*}$ &  $1.141$ \\
RNN ($l$ = 48)    &  $0.515$ &  $1.035$ &  $0.573$ &  $1.602$ &  $0.451$ &  $1.188$ \\
\midrule

LSTM ($l$ = 8)    &  $\mathbf{0.515}$ &  $\mathbf{1.036}$ &  $\mathbf{0.470}$ &  $\textbf{1.315}$ &  $0.558$ &  $1.469$ \\
LSTM ($l$ = 18)   &  $0.530$ &  $1.067$ &  $0.479$ &  $1.339$ &  $\mathbf{0.532}$ &  $\mathbf{1.399}$ \\
LSTM ($l$ = 36)   &  $0.564$ &  $1.134$ &  $0.498$ &  $1.393$ &  $0.558$ &  $1.469$ \\
LSTM ($l$ = 48)   &  $0.552$ &  $1.109$ &  $0.491$ &  $1.375$ &  $0.570$ &  $1.501$ \\
\midrule

GRU ($l$ = 8)     &  $\mathbf{0.517}$ &  $\mathbf{1.040}$ &  $\mathbf{0.448}$ &  $\mathbf{1.254}$ &  $0.568$ &  $1.496$ \\
GRU ($l$ = 18)    &  $0.516$ &  $1.038$ &  $0.465$ &  $1.302$ &  $\mathbf{0.546}$ &  $\mathbf{1.438}$ \\
GRU ($l$ = 36)    &  $0.563$ &  $1.132$ &  $0.505$ &  $1.412$ &  $0.634$ &  $1.669$ \\
GRU ($l$ = 48)    &  $0.541$ &  $1.089$ &  $0.451$ &  $1.262$ &  $0.622$ &  $1.637$ \\
\bottomrule

\end{tabular}
\caption*{\begin{footnotesize}
\textbf{Notes:} This table reports the relative RMSE of GDP growth for the competitor ANNs relative to a naive constant growth model and a benchmark DFM specification. 
A value below one indicates that the competitor model beats the naive benchmark model. 
The stars denote statistical significance at 10\%($*$), 5\%($**$) and 1\%($***$) level of the one-sided \textcite{diebold1995comparing} test. 
Columns are related to the individual nowcasting scenarios: e.g. $\Omega_{m_{1}}$ refers to the 1-month nowcasting scenario. 
Within each nowcasting scenario, best (lowest) relative RMSE values are highlighted in bold for the given competitor model. 
The overall best relative RMSE values are highlighted in bold and underlined. 
\end{footnotesize}}
\label{tab:eval_with_covid}
\end{table}

\newpage

Table \ref{tab:eval_with_covid} shows that all competitor ANNs achieve much better accuracy than the naive benchmark model, which looks intuitive knowing the characteristics of the second evaluation period. 
The detection of the sharp V-shaped COVID is clearly beyond the capabilities of a model assuming constant growth. 
Controversially though, we see a somewhat opposite picture regarding statistical significance of the ANNs' performance advantage: 
While we see much better (lower) relative RMSE values, none of the competitor models can beat the naive benchmark model on a 5\% significance level. 
The reason for this counter-intuitive phenomenon can be traced back to the special characteristics of the evaluation period. 
The sharp V-shaped recession in 2020:Q2 and Q3 is indicated by extreme values (\textit{outliers}) in the target series.
The naive benchmark model, which assumes constant growth, performs particularly badly in those periods, generating huge swings for the Diebold -- Mariano loss differential as well. 
Recalling the formula of the \textcite{diebold1995comparing} test statistic, these extreme values affect the variance of the loss differential (series) much more than its mean. 
Along with these considerations, we believe that the results in Table \ref{tab:eval_with_covid} provide some valuable insights into the generalization capabilities of the different competitor models and the role of long-term memory. 

Results from the second evaluation period indicate that the MLP's simple time-agnostic architecture performs better compared to both benchmark models.  
The results in Table \ref{tab:eval_with_covid} show that the best-performing configuration for the MLP is again the one trained with the shortest training sequences ($l = 8$). 
In the 3-month nowcasting scenario, the MLP yields a relative RMSE of $0.289$ versus the naive benchmark model where the performance advantage is significant on a 5\% level. 
For this configuration, the relative RMSE towards the DFM is $0.762$, which, while statistically not significant, indicates a clear performance advantage. 
The MLP also looks more competitive this time compared to the other ANN architectures designed for sequence modeling. 
With its best training configuration, it easily beats the two gated RNNs and outperforms even the Elman RNN. 
This significant performance gain during the second evaluation period corresponds to the fact that the MLP detects the COVID crisis almost as well as our overall best-performing model, the 1D CNN.\footnote{Similarly to that for the first evaluation period, Table \ref{tab:eval_with_covid_shortest} in Appendix \ref{app:results} reports the relative RMSE and MAE values for each competitor ANN relative to the training configuration using the shortest input sequences.} 

\begin{figure}[H]
     \centering
     \begin{subfigure}[b]{0.85\textwidth}
         \centering
         \includegraphics[width=\textwidth]{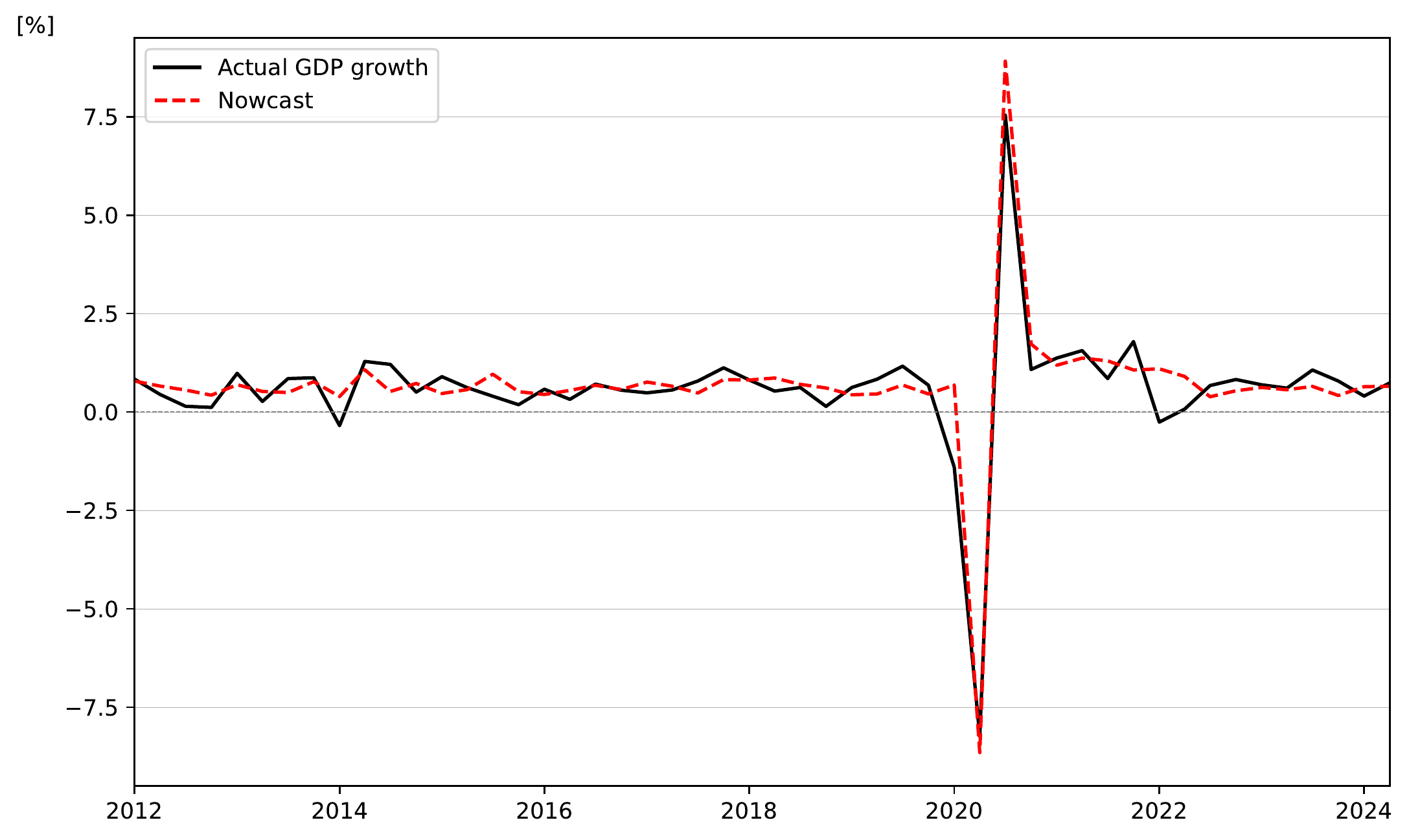}
         \caption{The actual quarterly GDP growth ($y_{q}$) versus the nowcasts generated by the best-performing competitor model ($\hat{y}_{c, q}$). 
         Best accuracy is achieved by the 1D CNN, with 8 long input sequences ($l = 8$) in the 3-month nowcasting scenario ($v = 3t$). 
         The evaluation period ranges from 2012:Q1 to 2024:Q2.} 
         \label{fig:predicted_vs_actuals_test2}
     \end{subfigure}
     \hfill
     \begin{subfigure}[b]{0.85\textwidth}
         \centering
         \includegraphics[width=\textwidth]{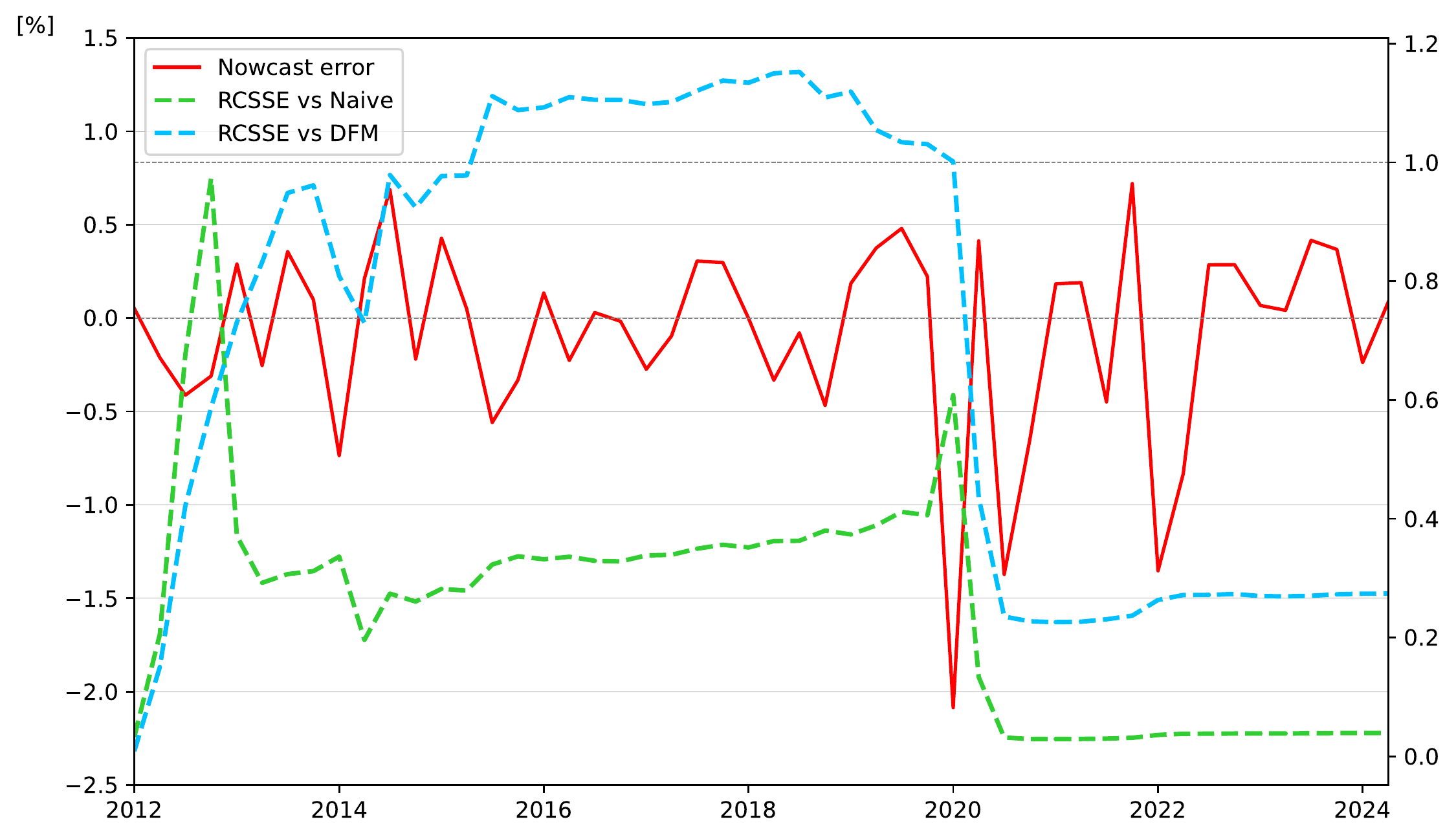}
         \caption{Nowcast errors ($\hat{e}_{c, q} = y_{q} - \hat{y}_{c, q}$, left axis) and the relative cumulative sum of squared nowcast errors (RCSSE) for the best-performing competitor model. 
         The RCSSE (right axis) is computed as $RCSSE_c = \sum_{q} \hat{e}_{c, q}^{2} / \sum_{q} \hat{e}_{b, q}^{2}$, where $\hat{e}^{2}_{c, q}$ denotes the squared nowcast error of the competitor model $c$.}
         \label{fig:residuals_test2}
     \end{subfigure}     
    
    \caption{Nowcasting performance of the best-performing competitor model (1D CNN, $l=8, v=3t$) during the second evaluation period. 
    Source: Own editing based on FRED-MD.}
    \label{fig:bestmdl_with_covid}

\end{figure}

While the 1D CNN also achieved good accuracy during the first evaluation period, it performs even better this time. 
Compared to the results from the first evaluation period, the 1D CNN clearly favors the shortest possible input sequences during the second test period. 
The network performs best in all three nowcasting scenarios with 8-month-long input sequences. 
Concretely, with 8 long input sequences in the 3-month nowcasting scenario, the 1D CNN generates a relative RMSE of $0.199$ and $0.523$ against the naive benchmark model and the DFM, respectively. 
With this performance, the 1D CNN achieves the best accuracy among the competitor ANNs and significantly beats both of our benchmark models by a statistically significant margin. 
As Table \ref{tab:eval_with_covid} shows, that is quite a unique feature that no other competitor ANN could achieve. 
Only the 1D CNN can replicate this level of accuracy with its second-best training configuration, using 18 long input sequences. 
Figure \ref{fig:predicted_vs_actuals_test2} below plots the nowcasts generated by the best-performing competitor model (1D CNN, $l = 8$, $v = 3t$) against the actual GDP growth. 
Then Figure \ref{fig:residuals_test2} shows how the series of the nowcast errors and the cumulative nowcast errors have evolved during the second test period. 

Results in Table \ref{tab:eval_with_covid} indicate that the Elman RNN with 8  and 18 long input sequences generates fairly accurate nowcasts as well. 
While the Elman network slightly lags behind the best MLP and 1D CNN configurations in terms of nowcasting accuracy, it still manages to beat both benchmark models in some training configurations. 
At the end of the nowcasting window, the Elman RNN generates a relative RMSE of $0.325$ beating the naive benchmark on a 5\% level. 
The relative RMSE value towards the benchmark DFM is $0.854$, which is associated with a better prediction of the sharp recession and recovery caused by the COVID pandemic. 
Similar to the 1D CNN, the vanilla RNN also delivers near its peak accuracy in the training configuration with input sequences of length 18. 
With input sequences longer than 18 months, the RNN's nowcasting accuracy declines, and the network generates less accurate nowcasts than the DFM. 
Along with that, results for the Elman RNN indicate a relatively good nowcasting performance during the second test period. 

Table \ref{tab:eval_with_covid} also shows that the basic RNN generates more accurate nowcasts than its architecturally more advanced variants, especially at the end of the nowcasting window. 
While the two gated RNNs achieved roughly the same accuracy as the vanilla RNN during the first evaluation period, they clearly have difficulty here with the detection of the COVID crisis. 
As we see, the gated RNNs generate relative RMSEs lower than one against the naive benchmark model, but the differences are not significant even in their most favorable training configurations. 
In addition, both the LSTM and the GRU fail to beat the benchmark DFM in any training configuration during the second evaluation period. 
Whereas the GRU had a slight edge over the competitors during the first evaluation period, even the basic RNN shows significantly better generalization capability this time. 
Remembering that the GRU (LSTM) has three (four) times more trainable parameters than an Elman RNN with a same-sized hidden state, architectural complexity clearly comes at a high cost here. 
The gating mechanism implemented in the LSTM and the GRU weakens the networks' generalization capability rather than helping with the detection the COVID recession. 
In the context of the bias-variance trade-off, that many more parameters introduced by the gating mechanism increase the variance of these estimators too much.  

Overall, results from the second evaluation period show that nowcasting accuracy declines with longer input sequences for all competitor models. 
The weak performance of the gated RNNs especially suggests that long-term memory does not help predict sharp changes in economic growth, like the V-shaped recession and recovery of 2020:Q2 and Q3. 
In general, architectural complexity and longer input sequences seem to hinder the detection of structural breaks and affect the competitor models' generalization capability badly during the second evaluation period. 

As we proposed, the 1D CNN might represent a \textit{``sweet spot''} in terms of architectural design between the simple time-agnostic MLP and the more complex (gated) recurrent networks. 
Based on the combined results of Table \ref{tab:eval_no_covid} and Table \ref{tab:eval_with_covid}, the 1D CNN seems to be a very suitable ANN architecture for GDP nowcasting:
It achieves nearly the same accuracy as the best competitor during the first evaluation period, while it generates the most accurate nowcasts for the second evaluation period.

\subsection{Contribution of features to the fitted values}  \label{sec:features_contribution}

In this section, we study the contribution of the various monthly indicators to the nowcasts. 
We restrict the analysis to our best-performing 1D CNN ($\mathbf{\Omega} = \Omega_{m_{3}}, l=8$) model, relying on the SHAP (SHapley Additive exPlanations) approach. 

The SHAP approach, presented by \textcite{lundberg2017unified}, offers a unified framework for interpreting predictions, which is also applicable to complex models, such as neural networks. 
SHAP has its roots in cooperative game theory and gives explanations of model outputs following the idea captured by Shapley values \citep{shapley1953value}. 
The approach casts the output generation of a model as a game where features act as players contributing to the output value. 
Computed in different coalitions of players, i.e., features, Shapley values per feature give the contribution of a feature value to the actual prediction (the generated nowcast) over the average prediction. 
SHAP values by \textcite{lundberg2017unified} represent a new class of linear feature importance measures, which are essentially an optimized, applied version of classic Shapley values designed to make feature attribution feasible even for complex machine learning models. 
SHAP values approximate the contribution of each value within a given input sequence to the output of the model. 
They explain how to get from the base value $\mathbb{E}\left[ \hat{f}(\mathbf{x}) \right]$ that would be predicted if we did not know any element of the regressor vector to the model's actual prediction $\hat{f}(\mathbf{x})$. 
Thus, the estimated SHAP value of a given feature $i$, denoted by $\phi_{i}$ is defined as follows:
\begin{align}
\hat{f}(\mathbf{x}) \approx \sum_{i} \phi_{i}(\mathbf{x}) + \mathbb{E} \left[ \hat{f} (\mathbf{x}) \right],     \label{eq:shap_additivity} \\
\phi_{i}(\mathbf{x}) = w_{i} (x_{i} - \mathbb{E} \left[ x_{i} \right])  \label{eq:shap_def_linear},
\end{align}
where $\mathbb{E} \left[ \hat{f}(\mathbf{x}) \right]$ is the expected (mean) response of the trained network given all input sequences (regressor vectors) contained in the training set, and $w_{i}$ represents the contribution weight associated with the $i$-th feature. 
For linear models, $w_{i}$ can be obtained directly from the model's weight coefficients. 
For complex nonlinear models, such as neural networks, $w_{i}$ has a more complex interpretation and must be computed (approximated) as a composite of several components. 
Accordingly, \textcite{lundberg2017unified} also present different model-type-specific approximation methods, which provide the tools for the efficient computation of those measures. 
We used Deep SHAP from these methods to compute the expected contribution of the different features to the nowcasts generated by the 1D CNN. 
SHAP also allows the assessment of the directional impact of features per data point on the prediction. 
However, in contrast to the absolute importance, the directionalities do not lend themselves to straightforward interpretations in all cases. 
Since the 1D CNN uses input sequences of length $l$ both during training and inference, estimated SHAP values for a given feature will also form sequences of the same length. 
It can easily happen that estimated SHAP values for a given feature indicate a positive contribution to the model's prediction at certain timesteps while negative at others. 
So, to make the (visual) interpretation feasible, we aggregated the SHAP values related to a given input sequence across the different timesteps. 
By taking the mean, we can express how much an input sequence, as a whole, affects the nowcast in a given quarter. 

\begin{figure}[H]
     
     \centering
     \begin{subfigure}[b]{0.9\textwidth}
         \centering
         \includegraphics[width=\textwidth]{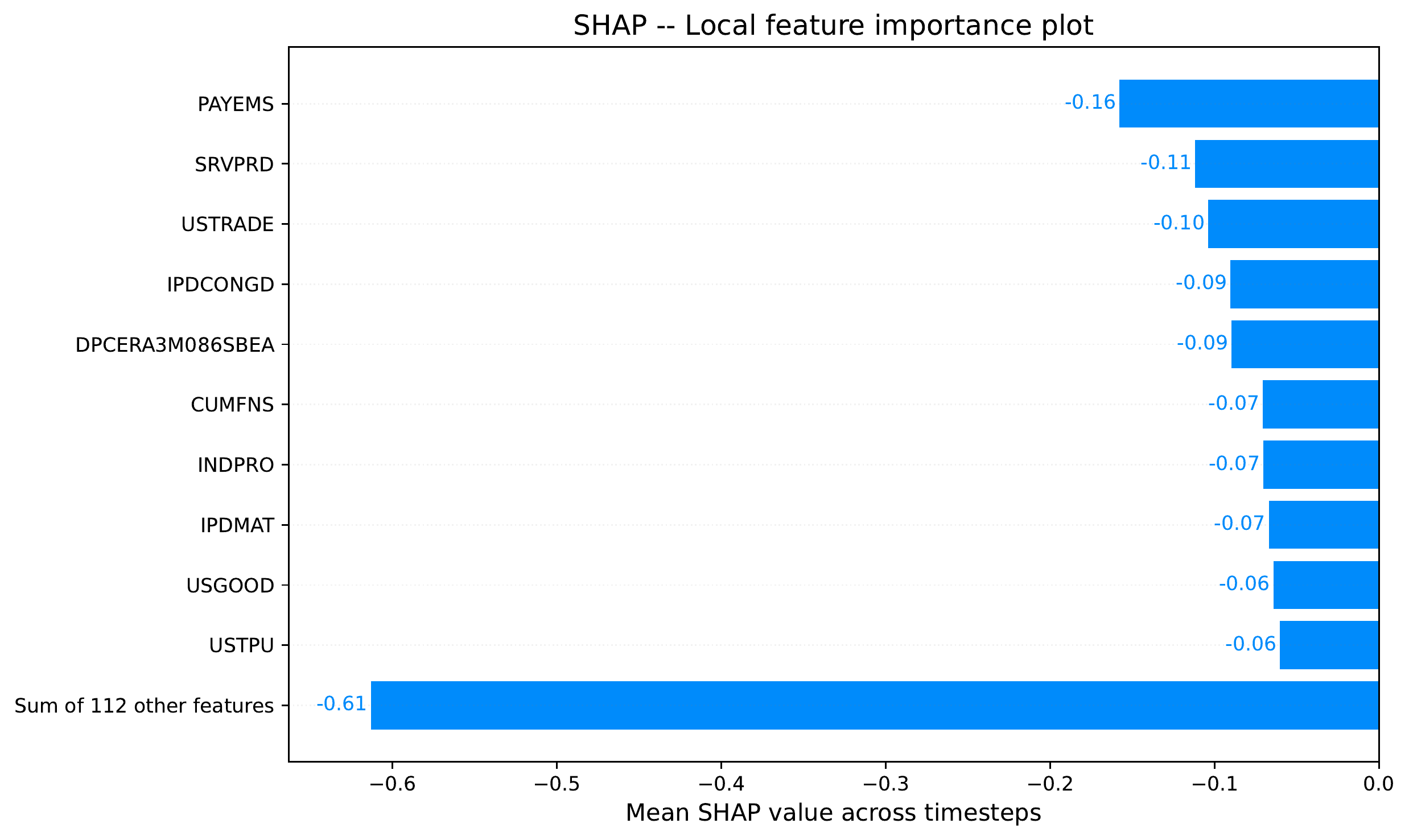}
         \caption{Local importance plot corresponding to test sequences for 2020:Q2, i.e. the sharp economic downturn caused by the COVID-19 pandemic.}
         \label{fig:shap_local_recession_others}
     \end{subfigure}
     \hfill
     \begin{subfigure}[b]{0.9\textwidth}
         \centering
         \includegraphics[width=\textwidth]{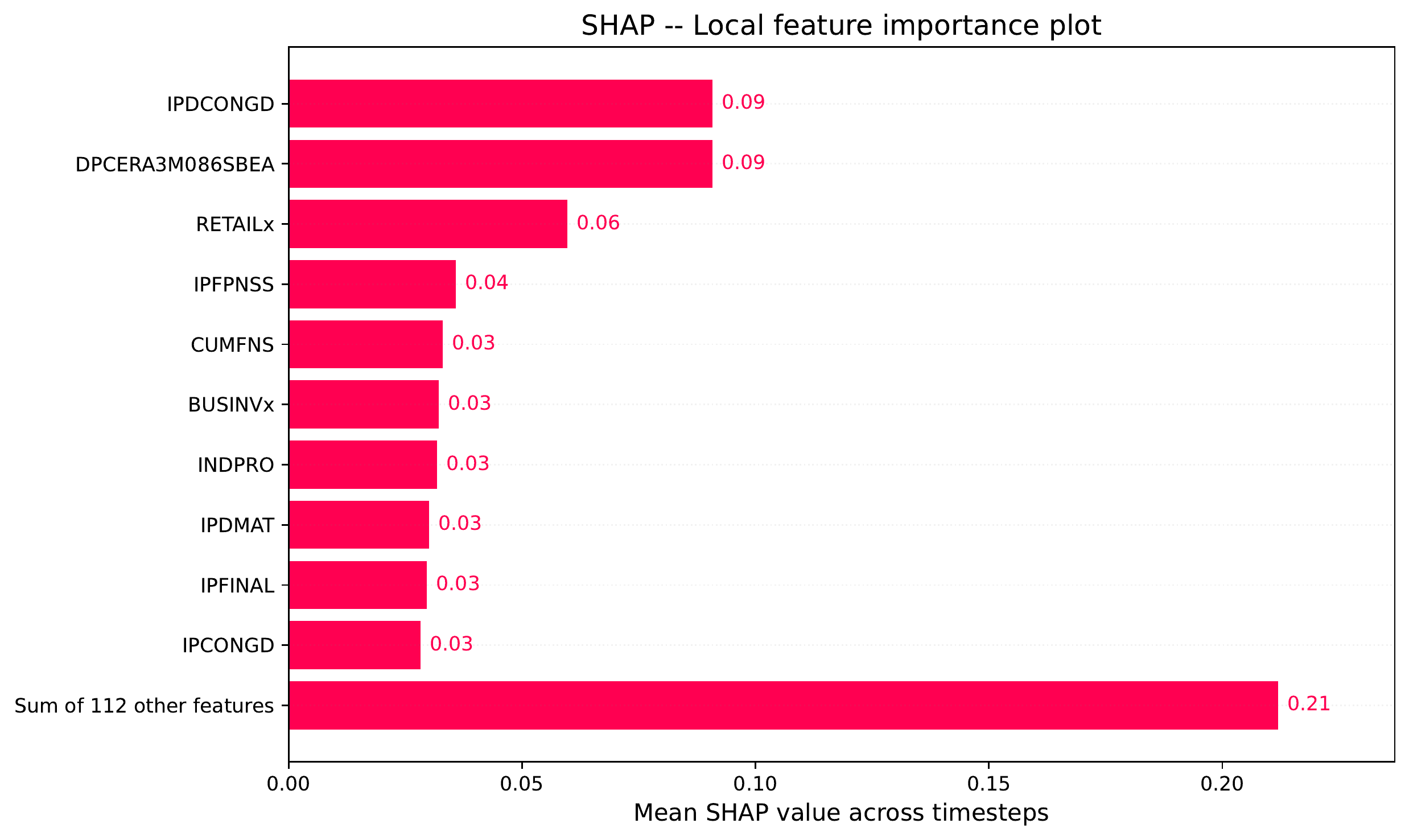}
         \caption{Local importance plot corresponding to test sequences for 2020:Q3, i.e. the sharp rebound after the COVID recession.}
         \label{fig:shap_local_recovery_others}
     \end{subfigure} 
    
\caption{Local feature importance plots for 2020:Q2 and 2020:Q3. Source: Own calculation based on FRED-MD.}    
\label{fig:shap_local_importance}

\end{figure}

In Figure \ref{fig:shap_local_importance}, we present the local feature importance plots for the two quarters of the V-shaped COVID recession, i.e., for 2020:Q2 and 2020:Q3, respectively. 
For both test sequences, the bars display the aggregated (mean) SHAP values across the different timesteps of the ten most important features. 

\newpage

Figure \ref{fig:shap_local_recession_others} shows that the 1D CNN forms its prediction mainly based on two groups of features when the pandemic-driven recession hits the U.S. economy in 2020:Q2. 
Five of the most important input sequences ending in June 2020:M6 are related to the labor market (Group 2 in Table \ref{tab:fred-md_group_02}), while four of them are included in the category called Output and income (Group 1 in Table \ref{tab:fred-md_group_01}). 
The three most important contributors are all related to the labor market. 
More concretely, \textit{PAYEMS} stands for the number of nonfarm employees in the U.S., who account for approximately 80 percent of the workers (ca. 158 million persons in June 2024.) 
During the COVID-19 recession, 21.9 million nonfarm employment jobs were lost, which is the largest share of jobs lost during a recession on record (dating back to 1939). 
The unemployment rate also increased to a record high 14.7 percent by April 2020 (2020:M4).\footnote{The detailed description of the labor market statistics, e.g., the definition of the service-providing industries can be found at: \url{https://www.bls.gov/}.} 
As Figure \ref{fig:shap_local_recession_others} indicates, the other 112 features also have a fairly  large contribution to the nowcast generated for 2020:Q2. 
Since the model's output relies on most of the features even in this highly turbulent period, we can have a first positive impression about the model's stability. 

Next, Figure \ref{fig:shap_local_recovery_others} displays the local feature importance plot for the sharp economic rebound in 2020:Q3. 
First, we can see that five of the ten most important contributors displayed in Figure \ref{fig:shap_local_recession_others} occur here in Figure \ref{fig:shap_local_recovery_others} as well. 
As Figure \ref{fig:shap_local_recovery_others} illustrates, there is one marked change in the model's focus: 
Monthly indicators related to the category Output and income (Group 1) remain just as important as before but we cannot see any features related to the Labor market among the ten most important contributors. 
Their place is taken by those monthly indicators listed in the FRED-MD category \textit{Consumption, orders and inventories} (Group 4, in Table \ref{tab:fred-md_group_04}). 
This change in the model's behavior aligns well with the general intuition that the labor market tends to react slower to economic shocks than production or consumption. 
There are multiple factors and theories that account for the lagged response of the labor market: wage stickiness, adjustment costs, and labor market regulations, among others. 
While the U.S. jobs recovery was faster following the COVID-19 recession than other recent recessions, it still took 28 months to reach its pre-crisis (2019:Q4) level. 
Accordingly, the recovery was reflected more apparently in indicators related to other sectors, especially production and consumption. 
For example, as Figure \ref{fig:ipdcongd} shows, the production of durable consumer goods (\textit{IPDCONGD}) rebounded immediately to its pre-crisis level during the third quarter. 
Figure \ref{fig:shap_local_recession_others} and \ref{fig:shap_local_recovery_others} indicate that our model captures this type of asymmetry across the COVID downturn and upturn \citep{ferraro2018asymmetric}, and provides an intuitive interpretation of what factors drove them. 
It is also worth mentioning that the other 112 features still have a fairly large contribution to the model's prediction, even if it is smaller than that for the previous quarter. 

\begin{figure}[H]
\begin{center}
\includegraphics[width = \textwidth]{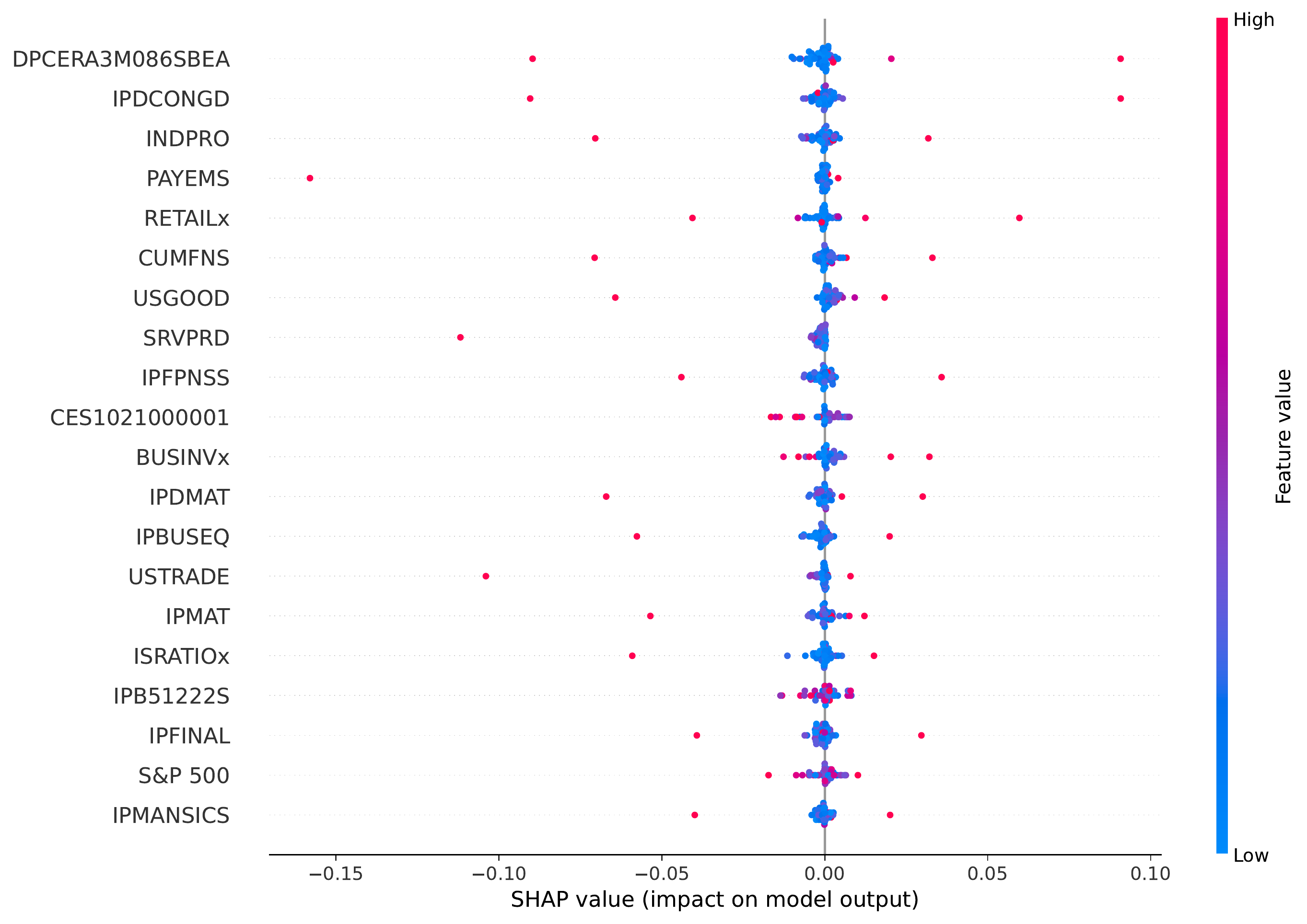}
\caption{SHAP beeswarm plot. Source: Own calculations based on FRED-MD.}
\label{fig:shap_beeswarm}
\end{center}
\end{figure}

Figure \ref{fig:shap_beeswarm} displays a so-called beeswarm plot, illustrating the estimated SHAP values of the 20 most important features from 2012:Q1 to 2024:Q2, i.e., for the second evaluation period. 
The 20 most important contributors were selected based on their mean absolute SHAP over that period. 
To generate the plot, we also had to aggregate the input sequences contained in the test set (i.e., in the second evaluation period) across different timesteps. 
For the $i$-th feature, we first demeaned the input sequence by calculating the average feature value, $\mathbb{E}_{\tau} (x_{\tau}^{i})$, based on the training set. 
Then, we computed the mean absolute value of the demeaned series. 
The procedure can be formalized as follows: 
\begin{align}
\alpha_{q}^{i} = \mathbb{E}_{t} \left(|x_{t}^{i} - \mathbb{E}_{\tau} (x_{\tau}^{i})| \right),  \label{eq:shap_aggregate_sequences} 
\end{align}
where $x_{t}^{i}$ denotes monthly values of the $i$-th feature, $t \in \{ m_{3}(q)-l+1, \dots, m_{3}(q) \} $, and $\alpha_{q}^{i}$ is the aggregated feature value for feature $i$, corresponding to the $q$-th quarterly target observation from the test set. 
Unlike the temporal aggregation of SHAP values, taking the mean absolute values appeared to be the only meaningful choice. 
This type of aggregation ensures that input sequences containing extreme values for one or more timesteps will be associated with a high feature value (indicated by red dots in Figure \ref{fig:shap_beeswarm}). 
If we think of the test sequence for 2020:Q3, simple averaging of the regressor vector would cancel out the extreme lows and extreme highs, resulting in unrealistic low feature values. 

Figure \ref{fig:shap_beeswarm} reinforces our earlier observations regarding the stability of the trained network. 
Except for those special quarters of the V-shaped COVID recession, aggregated SHAP values are centered around zero, indicating the balanced and robust behavior of the model. 
Figure \ref{fig:shap_beeswarm} shows that our model learns a stable mapping between the target variable and the regressors. 
This mapping leverages the informational content of numerous regressors, so it does not depend on any specific feature to a large extent. 
During the COVID crisis, however, the model effectively adapts to the unique circumstances reflected in extremely high feature values. 

\section{Conclusion}    \label{sec:conclusion}

Our study applies different ANN architectures to nowcast GDP growth for the U.S. economy. 
The main research question of the paper is how the length of input sequences (regressor vectors) affects each ANN's nowcasting performance. 
Previously, we have seen the results from two distinctively different evaluation periods. 
The first (2012:Q1 -- 2019:Q4) is characterized by balanced economic growth, while the second (2012:Q1 -- 2024:Q2) also includes periods of the COVID-19 crisis. 

Results from the first evaluation period show that all competitor ANNs beat the naive constant growth model in terms of nowcasting accuracy. 
Intuitively, the performance advantage of the competitors generally strengthens in significance with subsequent intra-quarterly data releases. 
Towards the benchmark DFM, relative RMSEs (and MAEs) are close to one, signaling a similar nowcasting performance. 
As the results in Table \ref{tab:eval_no_covid} indicate, longer input sequences result in slightly more accurate nowcasts for some competitor ANNs. 
However, the best accuracy is still achieved with 8-month-long input sequences at the end of the nowcasting window.
Furthermore, longer input sequences do not lead to significantly more accurate nowcasts compared to those generated with 8-month-long sequences for any competitors. 
During the second evaluation period, where timely detection of the V-shaped COVID recession (i.e., a structural break) is apparently decisive for nowcasting performance, all ANNs work best with the shortest input sequences. 
Here, suitable ANN architectures with properly selected regressor vectors can outperform the benchmark DFM at the last step of the nowcasting window.  

Combined results from the two evaluation periods suggest that architectural features enabling long-term memory, i.e., the effective use of long input sequences, may not play an important role in GDP nowcasting. 
Instead, they exhibit a poor trade-off between complexity and generalization capability for our predictive analysis. 
On the other hand, the 1D CNN represents a ``sweet spot'' in terms of architectural complexity, showing consistently good nowcasting performance in both test periods. 
With 8-month-long input sequences, the 1D CNN beats the naive benchmark model during the first evaluation period, indicated by a relative RMSE of $0.637$. 
The difference is statistically significant on a 5\% level. 
Here, the network performs basically on par with the DFM: the relative RMSE towards the DFM is $1.043$. 
During the second evaluation period, the 1D CNN achieves the overall best accuracy with the same length on input sequences ($l=8$), generating a relative RMSE of $0.199$ and $0.523$ against the naive constant growth model and the benchmark DFM, respectively. 
At the last step of the nowcasting window, the 1D CNN outperforms both of our benchmark models by a statistically significant margin. 
Based on SHAP values, we investigated the contribution of the features to the generated nowcasts. 
The analysis shows that our trained network forms its predictions based on the information content of the whole feature set, indicating the stability of the learned mapping. 
While generating highly accurate nowcasts for both the downturn and recovery during the COVID crisis, the model also provides an intuitive interpretation of what factors drove them. 

Finally, we should emphasize that the paper presents the results for different behavioral models and not for an exact system identification procedure, where we know the underlying data-generating process. 
Accordingly, empirical results mainly speak about the different ANN architectures' compatibility with this predictive analysis rather than the characteristics of GDP growth as a time series. 
Nonetheless, the results of the empirical analysis suggest that long-term memory does not play an important role in GDP nowcasting. 

\newpage

\section*{Acknowledgements}  \label{sec:acknowledgements}

We would like to thank Róbert Lieli, Professor at the Central European University, for his invaluable feedback and support. 
He has helped and inspired our work in many ways. 
We thank Ádám Reiff and the Budapest School for Central Bank Studies for allowing us to attend Professor Massimiliano Marcellino's course related to this topic. 
We also thank Ádám Czelleng and Máté Kővágó for their feedback and support.

\medskip
\printbibliography

@article{amari1997asymptotic,
  title={Asymptotic statistical theory of overtraining and cross-validation},
  author={Amari, Shun-ichi and Murata, Noboru and Muller, K-R and Finke, Michael and Yang, Howard Hua},
  journal={IEEE transactions on neural networks},
  volume={8},
  number={5},
  pages={985--996},
  year={1997},
  publisher={IEEE}
}

@article{botha2021nowcasting,
  title={Nowcasting South African gross domestic product using a suite of statistical models},
  author={Botha, Byron and Olds, Tim and Reid, Geordie and Steenkamp, Daan and van Jaarsveld, Rossouw},
  journal={South African Journal of Economics},
  volume={89},
  number={4},
  pages={526--554},
  year={2021},
  publisher={Wiley Online Library}
}

@article{chernis2017dynamic,
  title={A dynamic factor model for nowcasting Canadian GDP growth},
  author={Chernis, Tony and Sekkel, Rodrigo},
  journal={Empirical Economics},
  volume={53},
  number={1},
  pages={217--234},
  year={2017},
  publisher={Springer}
}

@article{chung2014empirical,
  title={Empirical evaluation of gated recurrent neural networks on sequence modeling},
  author={Chung, Junyoung and Gulcehre, Caglar and Cho, KyungHyun and Bengio, Yoshua},
  journal={arXiv preprint arXiv:1412.3555},
  year={2014}
}

@article{crone2010feature,
  title={Feature selection for time series prediction -- A combined filter and wrapper approach for neural networks},
  author={Crone, Sven F and Kourentzes, Nikolaos},
  journal={Neurocomputing},
  volume={73},
  number={10-12},
  pages={1923--1936},
  year={2010},
  publisher={Elsevier}
}

@article{dematos1996feedforward,
  title={Feedforward versus recurrent neural networks for forecasting monthly japanese yen exchange rates},
  author={Dematos, Giovani and Boyd, Milton S and Kermanshahi, Bahman and Kohzadi, Nowrouz and Kaastra, Iebeling},
  journal={Financial Engineering and the Japanese Markets},
  volume={3},
  number={1},
  pages={59--75},
  year={1996},
  publisher={Springer}
}

@article{diebold1995comparing,
  title={Comparing Predictive Accuracy},
  author={Diebold, Francis X and Mariano, Roberto S},
  journal={Journal of Business \& Economic Statistics},
  pages={253--263},
  year={1995},
  publisher={JSTOR}
}

@article{doz2011two,
  title={A two-step estimator for large approximate dynamic factor models based on Kalman filtering},
  author={Doz, Catherine and Giannone, Domenico and Reichlin, Lucrezia},
  journal={Journal of Econometrics},
  volume={164},
  number={1},
  pages={188--205},
  year={2011},
  publisher={Elsevier}
}

@article{elman1990finding,
  title={Finding structure in time},
  author={Elman, Jeffrey L},
  journal={Cognitive science},
  volume={14},
  number={2},
  pages={179--211},
  year={1990},
  publisher={Wiley Online Library}
}

@article{ferraro2018asymmetric,
  title={The asymmetric cyclical behavior of the US labor market},
  author={Ferraro, Domenico},
  journal={Review of Economic Dynamics},
  volume={30},
  pages={145--162},
  year={2018},
  publisher={Elsevier}
}

@article{ghysels2007midas,
  title={MIDAS regressions: Further results and new directions},
  author={Ghysels, Eric and Sinko, Arthur and Valkanov, Rossen},
  journal={Econometric reviews},
  volume={26},
  number={1},
  pages={53--90},
  year={2007},
  publisher={Taylor \& Francis}
}

@article{giannone2008nowcasting,
  title={Nowcasting: The real-time informational content of macroeconomic data},
  author={Giannone, Domenico and Reichlin, Lucrezia and Small, David},
  journal={Journal of monetary economics},
  volume={55},
  number={4},
  pages={665--676},
  year={2008},
  publisher={Elsevier}
}

@article{heravi2004linear,
  title={Linear versus neural network forecasts for European industrial production series},
  author={Heravi, Saeed and Osborn, Denise R and Birchenhall, CR},
  journal={International Journal of Forecasting},
  volume={20},
  number={3},
  pages={435--446},
  year={2004},
  publisher={Elsevier}
}

@article{hochreiter1997long,
  title={Long short-term memory},
  author={Hochreiter, Sepp and Schmidhuber, J{\"u}rgen},
  journal={Neural computation},
  volume={9},
  number={8},
  pages={1735--1780},
  year={1997},
  publisher={MIT Press}
}

@article{Hodrick1997PostwarUB,
  title={Postwar U.S. Business Cycles: An Empirical Investigation},
  author={Robert J. Hodrick and Edward C. Prescott},
  journal={Journal of Money, Credit and Banking},
  year={1997},
  volume={29},
  pages={1-16},
  url={https://api.semanticscholar.org/CorpusID:154995815}
}

@article{hopp2021economic,
  title={Economic nowcasting with long short-term memory artificial neural networks (LSTM)},
  author={Hopp, Daniel},
  journal={arXiv preprint arXiv:2106.08901},
  year={2021}
}

@article{hornik1989multilayer,
  title={Multilayer feedforward networks are universal approximators},
  author={Hornik, Kurt and Stinchcombe, Maxwell and White, Halbert},
  journal={Neural networks},
  volume={2},
  number={5},
  pages={359--366},
  year={1989},
  publisher={Elsevier}
}

@article{jiang2016variable,
  title={Variable selection with prior information for generalized linear models via the prior LASSO method},
  author={Jiang, Yuan and He, Yunxiao and Zhang, Heping},
  journal={Journal of the American Statistical Association},
  volume={111},
  number={513},
  pages={355--376},
  year={2016},
  publisher={Taylor \& Francis}
}

@article{kingma2014adam,
  title={Adam: A method for stochastic optimization},
  author={Kingma, Diederik P and Ba, Jimmy},
  journal={arXiv preprint arXiv:1412.6980},
  year={2014}
}

@article{kiranyaz20211d,
  title={1D convolutional neural networks and applications: A survey},
  author={Kiranyaz, Serkan and Avci, Onur and Abdeljaber, Osama and Ince, Turker and Gabbouj, Moncef and Inman, Daniel J},
  journal={Mechanical systems and signal processing},
  volume={151},
  pages={107398},
  year={2021},
  publisher={Elsevier}
}

@article{kourentzes2014neural,
  title={Neural network ensemble operators for time series forecasting},
  author={Kourentzes, Nikolaos and Barrow, Devon K and Crone, Sven F},
  journal={Expert Systems with Applications},
  volume={41},
  number={9},
  pages={4235--4244},
  year={2014},
  publisher={Elsevier}
}

@article{kuan1995forecasting,
  title={Forecasting exchange rates using feedforward and recurrent neural networks},
  author={Kuan, Chung-Ming and Liu, Tung},
  journal={Journal of applied econometrics},
  volume={10},
  number={4},
  pages={347--364},
  year={1995},
  publisher={Wiley Online Library}
}

@article{labonne2020capturing,
  title={Capturing GDP nowcast uncertainty in real time},
  author={Labonne, Paul},
  journal={arXiv preprint arXiv:2012.02601},
  year={2020}
}

@article{loermann2019nowcasting,
  title={Nowcasting US GDP with artificial neural networks},
  author={Loermann, Julius and Maas, Benedikt},
  year={2019}
}

@article{lundberg2017unified,
  title={A Unified Approach to Interpreting Model Predictions},
  author={Lundberg, Scott M and Lee, Su-In},
  journal={Advances in Neural Information Processing Systems},
  volume={30},
  year={2017}
}

@article{marcellino2010factor,
  title={Factor MIDAS for nowcasting and forecasting with ragged-edge data: A model comparison for German GDP},
  author={Marcellino, Massimiliano and Schumacher, Christian},
  journal={Oxford Bulletin of Economics and Statistics},
  volume={72},
  number={4},
  pages={518--550},
  year={2010},
  publisher={Wiley Online Library}
}

@article{mariano2003new,
  title={A new coincident index of business cycles based on monthly and quarterly series},
  author={Mariano, Roberto S and Murasawa, Yasutomo},
  journal={Journal of applied Econometrics},
  volume={18},
  number={4},
  pages={427--443},
  year={2003},
  publisher={Wiley Online Library}
}

@article{matheson2014new,
  title={New indicators for tracking growth in real time},
  author={Matheson, Troy D},
  journal={OECD Journal: Journal of Business Cycle Measurement and Analysis},
  volume={2013},
  number={2},
  pages={51--71},
  year={2014},
  publisher={OECD}
}

@article{mccracken2016fred,
  title={FRED-MD: A monthly database for macroeconomic research},
  author={McCracken, Michael W and Ng, Serena},
  journal={Journal of Business \& Economic Statistics},
  volume={34},
  number={4},
  pages={574--589},
  year={2016},
  publisher={Taylor \& Francis}
}

@article{richardson2021nowcasting,
  title={Nowcasting GDP using machine-learning algorithms: A real-time assessment},
  author={Richardson, Adam and van Florenstein Mulder, Thomas and Vehbi, Tu{\u{g}}rul},
  journal={International Journal of Forecasting},
  volume={37},
  number={2},
  pages={941--948},
  year={2021},
  publisher={Elsevier}
}

@article{shapley1953value,
  title={A value for n-person games},
  author={Shapley, Lloyd S},
  journal={Contribution to the Theory of Games},
  volume={2},
  year={1953}
}

@article{stock2002forecasting,
  title={Forecasting using principal components from a large number of predictors},
  author={Stock, James H and Watson, Mark W},
  journal={Journal of the American statistical association},
  volume={97},
  number={460},
  pages={1167--1179},
  year={2002},
  publisher={Taylor \& Francis}
}

@article{stock2002macroeconomic,
  title={Macroeconomic forecasting using diffusion indexes},
  author={Stock, James H and Watson, Mark W},
  journal={Journal of Business \& Economic Statistics},
  volume={20},
  number={2},
  pages={147--162},
  year={2002},
  publisher={Taylor \& Francis}
}

@article{tkacz2001neural,
  title={Neural network forecasting of Canadian GDP growth},
  author={Tkacz, Greg},
  journal={International Journal of Forecasting},
  volume={17},
  number={1},
  pages={57--69},
  year={2001},
  publisher={Elsevier}
}

@article{torres2018applying,
  title={Applying recurrent neural networks for multivariate time series forecasting of volatile financial data},
  author={Torres, Douglas Garcia and Qiu, Hongliang},
  journal={KTH Royal Institute of Technology: Stockholm, Sweden},
  year={2018}
}

@article{wallis1986forecasting,
  title={Forecasting with an econometric model: The ‘ragged edge’problem},
  author={Wallis, Kenneth F},
  journal={Journal of Forecasting},
  volume={5},
  number={1},
  pages={1--13},
  year={1986},
  publisher={Wiley Online Library}
}

@article{werbos1990backpropagation,
  title={Backpropagation through time: what it does and how to do it},
  author={Werbos, Paul J},
  journal={Proceedings of the IEEE},
  volume={78},
  number={10},
  pages={1550--1560},
  year={1990},
  publisher={IEEE}
}

@book{ekman2021learning,
  title={Learning Deep Learning: Theory and Practice of Neural Networks, Computer Vision, NLP, and Transformers Using TensorFlow},
  author={Ekman, Magnus},
  year={2021},
  publisher={Addison-Wesley Professional}
}

@book{goodfellow-et-al_2016,
    title={Deep Learning},
    author={Ian Goodfellow and Yoshua Bengio and Aaron Courville},
    publisher={MIT Press},
    note={\url{http://www.deeplearningbook.org}},
    year={2016}
}

@book{greene2003econometric,
  title={Econometric analysis},
  author={Greene, William H},
  year={2017},
  edition={Third},
  publisher={Pearson Education}
}

@book{hastie2009elements,
  title={The Elements of Statistical Learning: Data Mining, Inference, and Prediction},
  author={Hastie, Trevor and Tibshirani, Robert and Friedman, Jerome},
  year={2009},
  publisher={Springer New York, NY}
}

@misc{ganesh2019,
  author = {Ganesh, Prakhar},
  title = {Types of Convolution Kernels: Simplified},
  year = {2019},
  publisher={Towards Data Science},
  URL={https://towardsdatascience.com/types-of-convolution-kernels-simplified-f040cb307c37}
}

@misc{olah2015lstm,
title	= {Understanding LSTM Networks},
author	= {Olah, Christopher},
year	= {2015},
URL	= {http://colah.github.io/posts/2015-08-Understanding-LSTMs/}
}

@misc{fulton2024dfm,
title = {Dynamic factors and coincident indices},
author = {Fulton, Chad and Augspurger, Tom and Sheppard, Kevin and Morton, Jamie},
year = {2024},
URL	= {https://www.statsmodels.org/dev/examples/notebooks/generated/statespace_dfm_coincident.html},
note = {Accessed: 2024-08-15}
}

\newpage

\appendix
\renewcommand\thefigure{\thesection.\arabic{figure}} 
\renewcommand\thetable{\thesection.\arabic{table}}

\section{Data appendix}     \label{app:data}

\setcounter{figure}{0} 
\setcounter{table}{0}

\subsection{Target series}  

\begin{figure}[H]
    \centering
    
    \begin{subfigure}[t]{0.44\textwidth}
        \centering
        \includegraphics[width=\textwidth]{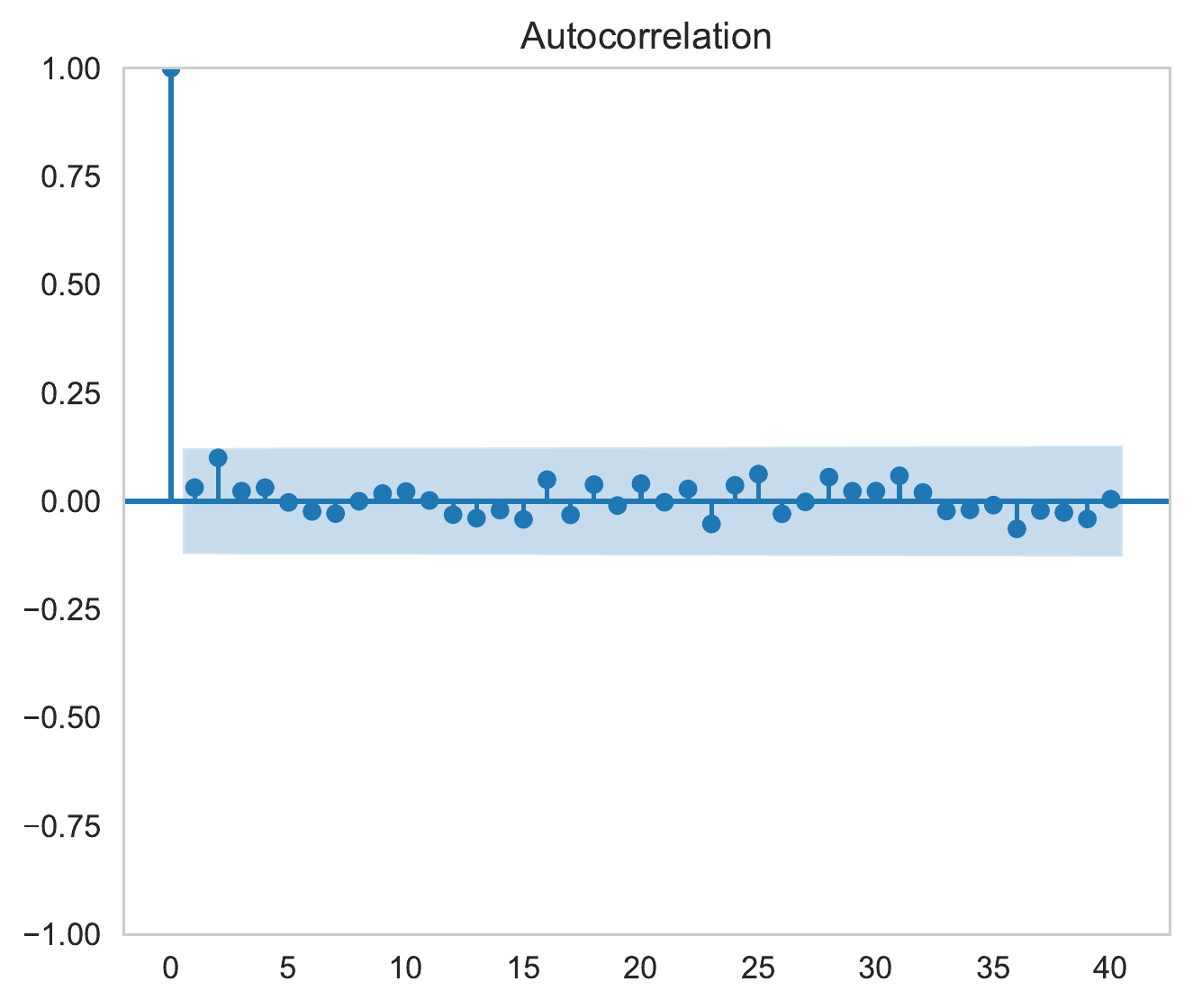} 
        \caption{\footnotesize Sample period: 1960:Q1 -- 2024:Q2. Lags: 40.}
    \end{subfigure}
    \hfill
    \begin{subfigure}[t]{0.44\textwidth}
        \centering
        \includegraphics[width=\textwidth]{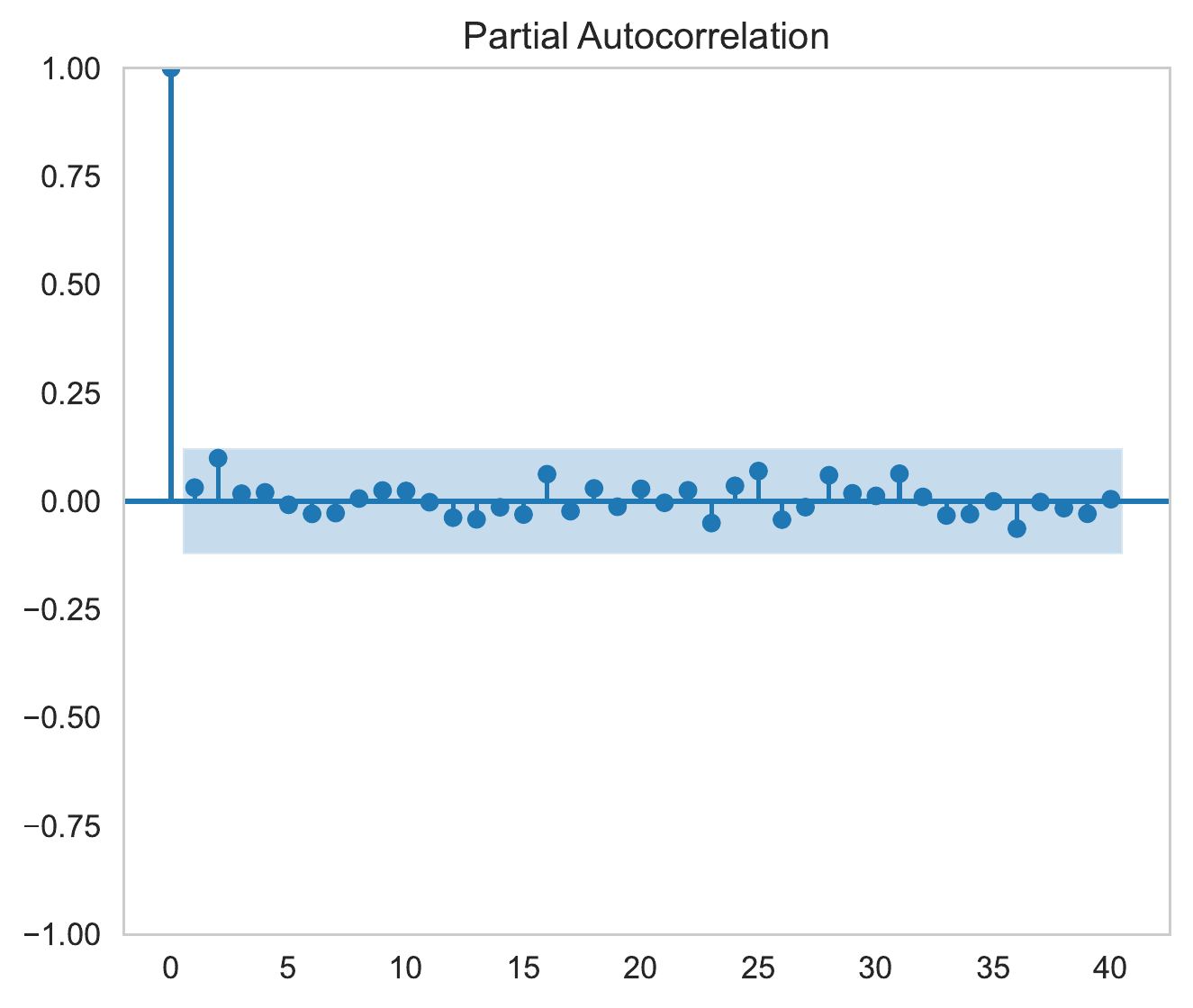} 
        \caption{\footnotesize Sample period: 1960:Q1 -- 2024:Q2. Lags: 40.}
    \end{subfigure}

    \vspace{0.3cm} 
    
    \begin{subfigure}[t]{0.44\textwidth}
        \centering
        \includegraphics[width=\textwidth]{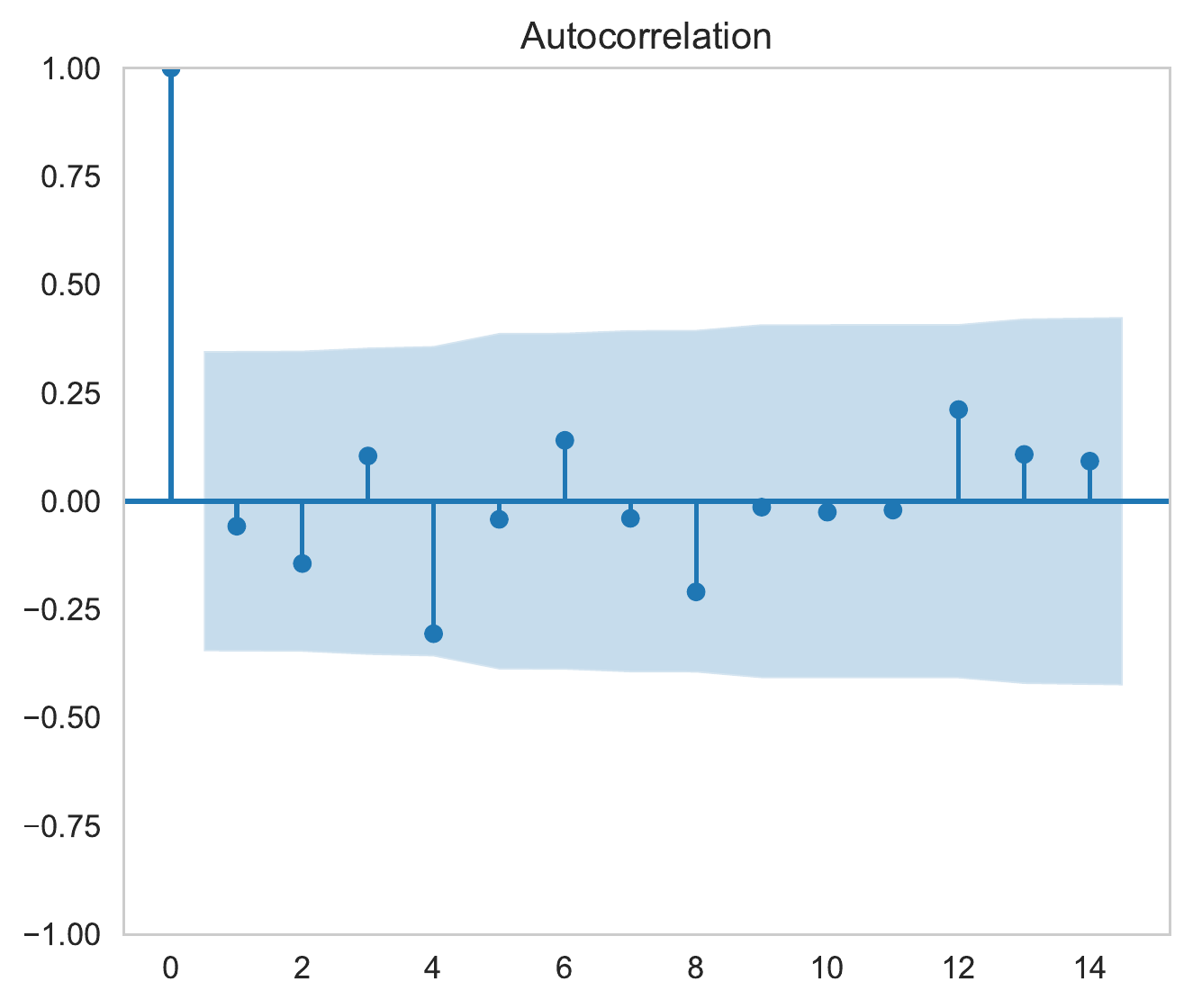} 
        \caption{\footnotesize Sample period: 2012:Q1 -- 2019:Q4. Lags: 14.}
    \end{subfigure}
    \hfill
    \begin{subfigure}[t]{0.44\textwidth}
        \centering
        \includegraphics[width=\textwidth]{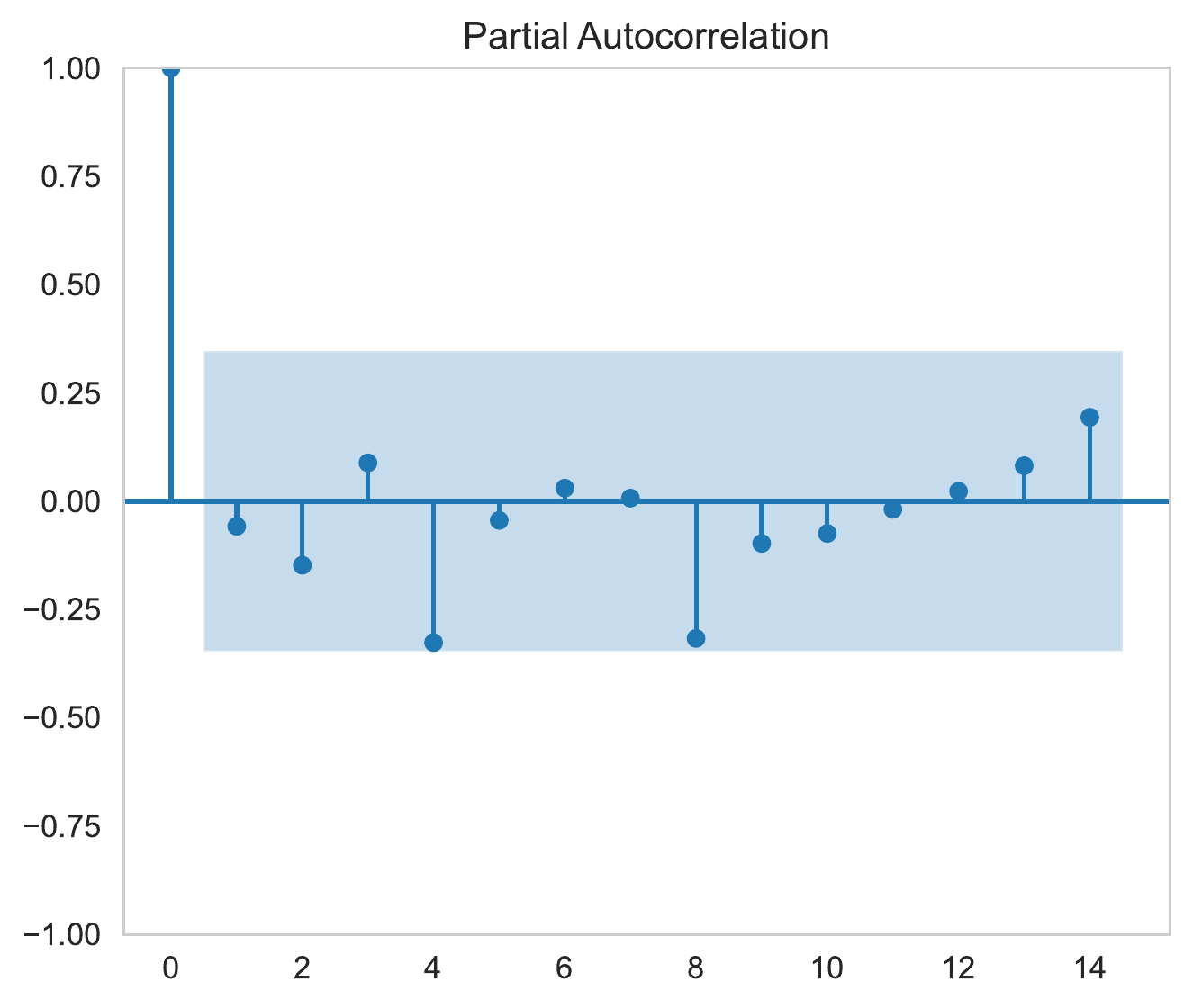} 
        \caption{\footnotesize Sample period: 2012:Q1 -- 2019:Q4. Lags: 14.}
    \end{subfigure}

    \vspace{0.3cm}
    
    \begin{subfigure}[t]{0.44\textwidth}
        \centering
        \includegraphics[width=\textwidth]{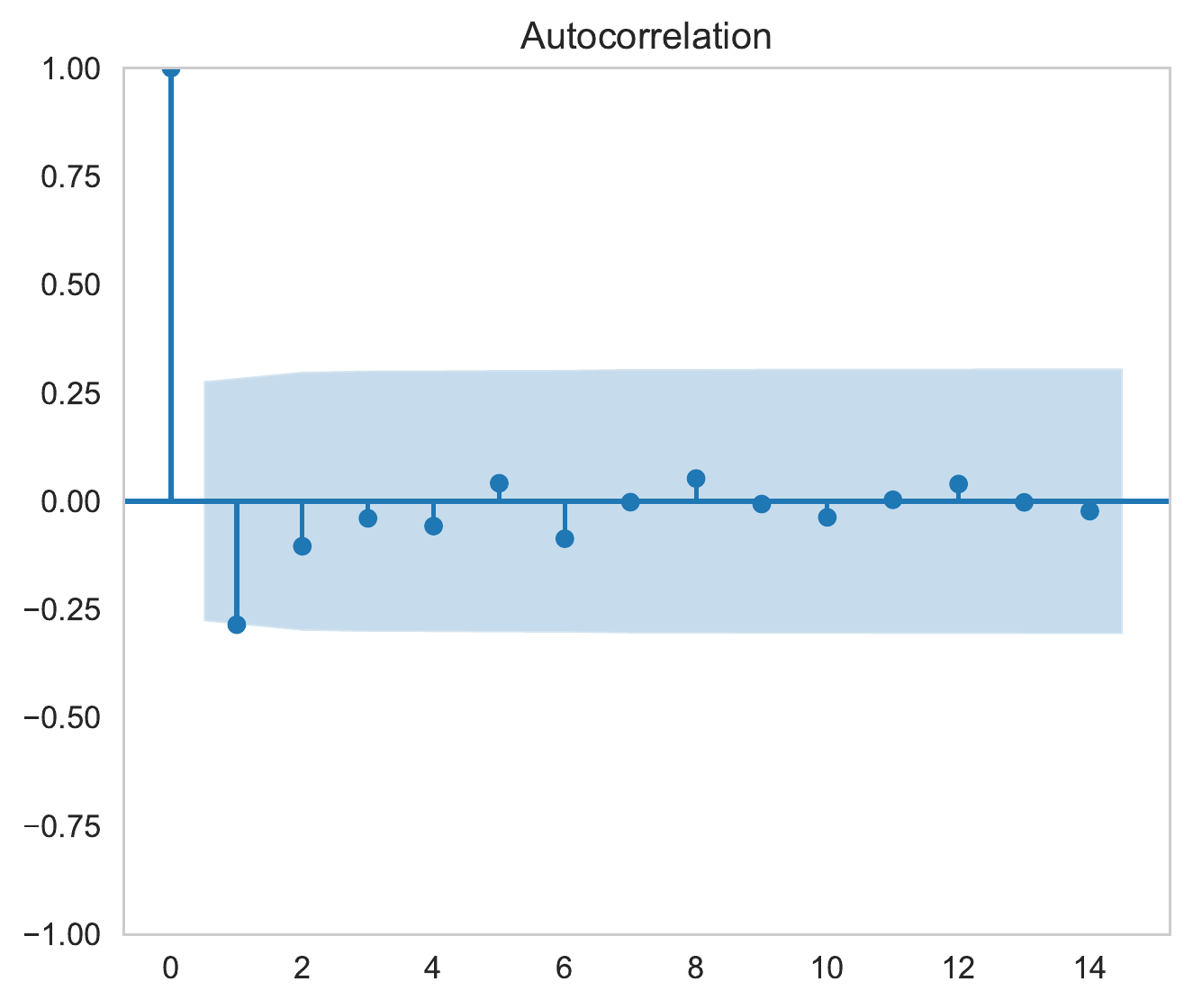} 
        \caption{\footnotesize Sample period: 2012:Q1 -- 2024:Q2. Lags: 14.}
    \end{subfigure}
    \hfill
    \begin{subfigure}[t]{0.44\textwidth}
        \centering
        \includegraphics[width=\textwidth]{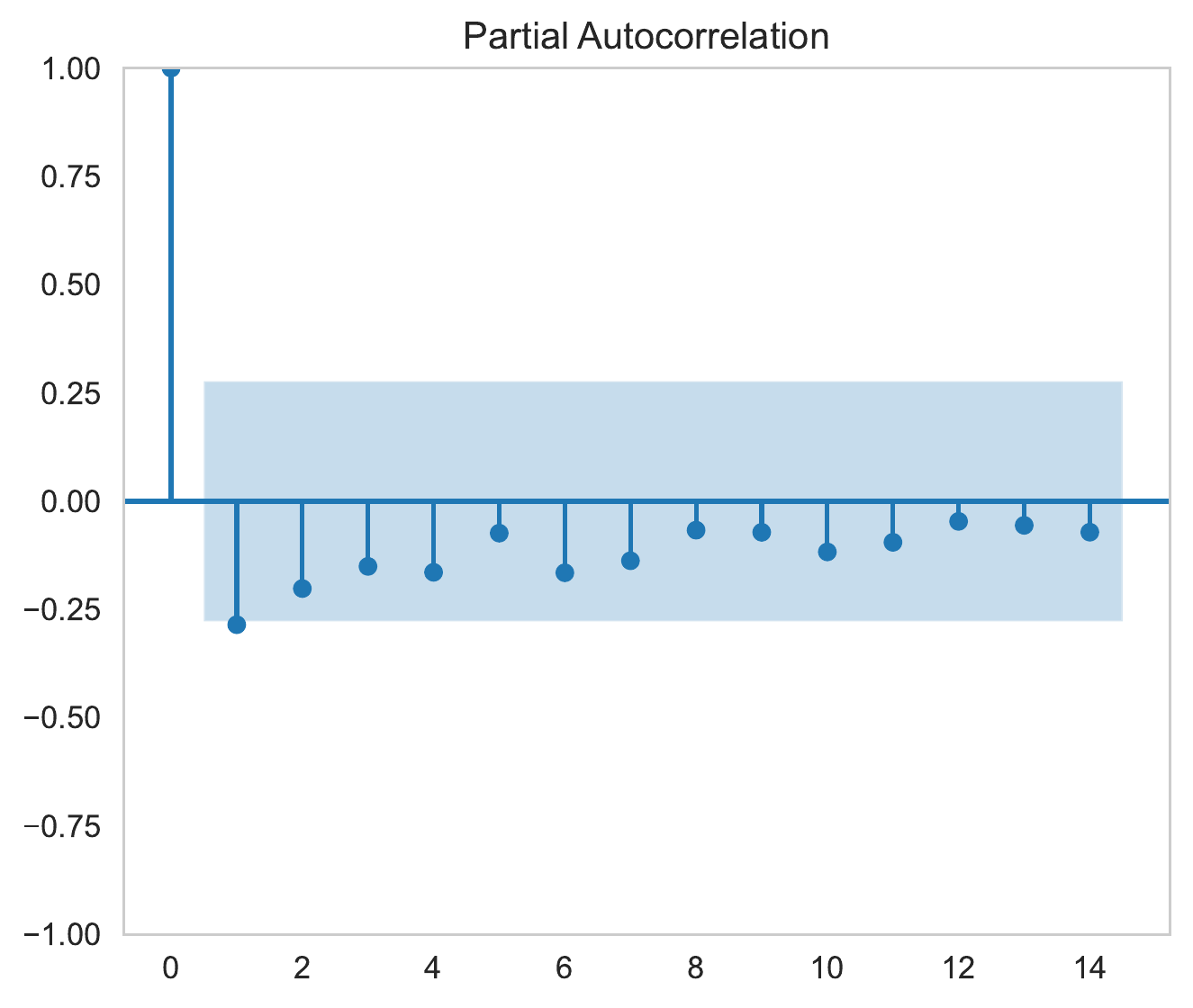} 
        \caption{\footnotesize Sample period: 2012:Q1 -- 2024:Q2. Lags: 14.}
    \end{subfigure}

\caption{ACF (left column) and PACF plots of quarterly GDP growth.}
\label{fig:target_acf_pacf}
\end{figure}

\newpage

\begin{figure}[H]
    \centering
    
    \begin{subfigure}[b]{0.85\textwidth}
        \centering
        \includegraphics[width=\textwidth]{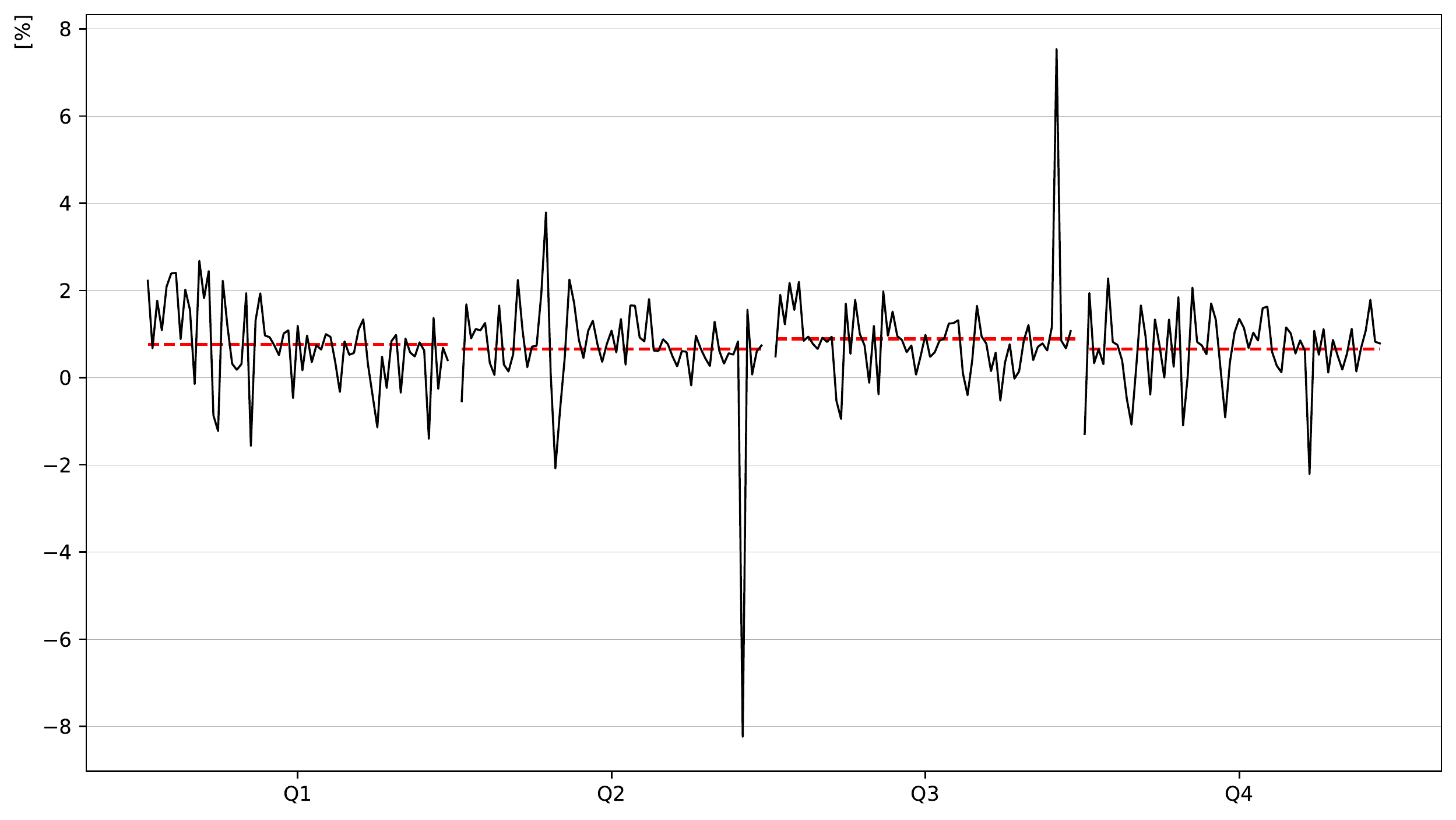}
        \caption{Quarterly grouping: 1960:Q1 -- 2024:Q2.}
    \end{subfigure}
    

    \begin{subfigure}[b]{0.70\textwidth}
        \centering
        \includegraphics[width=\textwidth]{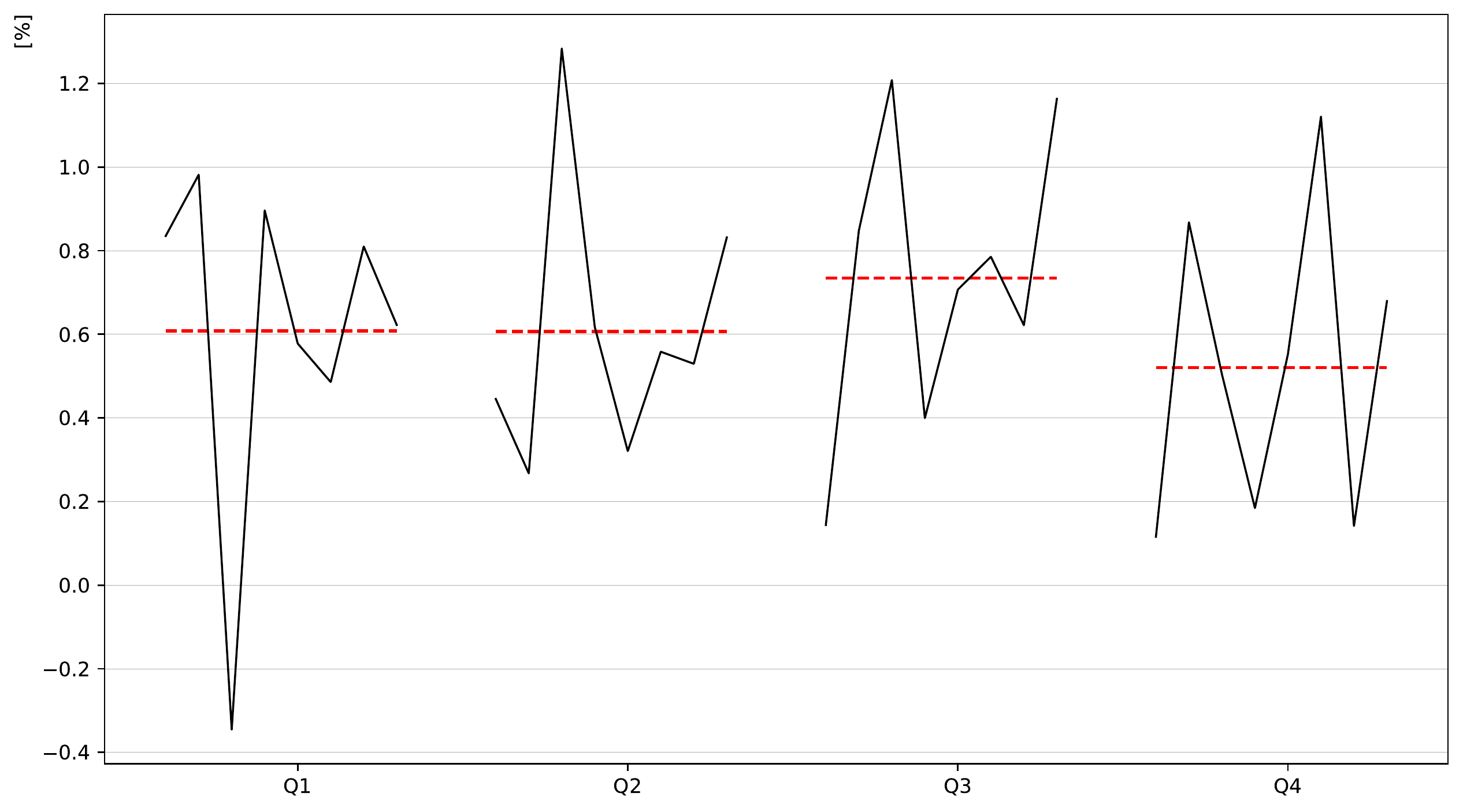} 
        \caption{Quarterly grouping: 2012:Q1 -- 2019:Q4.}
    \end{subfigure}
    

    \begin{subfigure}[b]{0.70\textwidth}
        \centering
        \includegraphics[width=\textwidth]{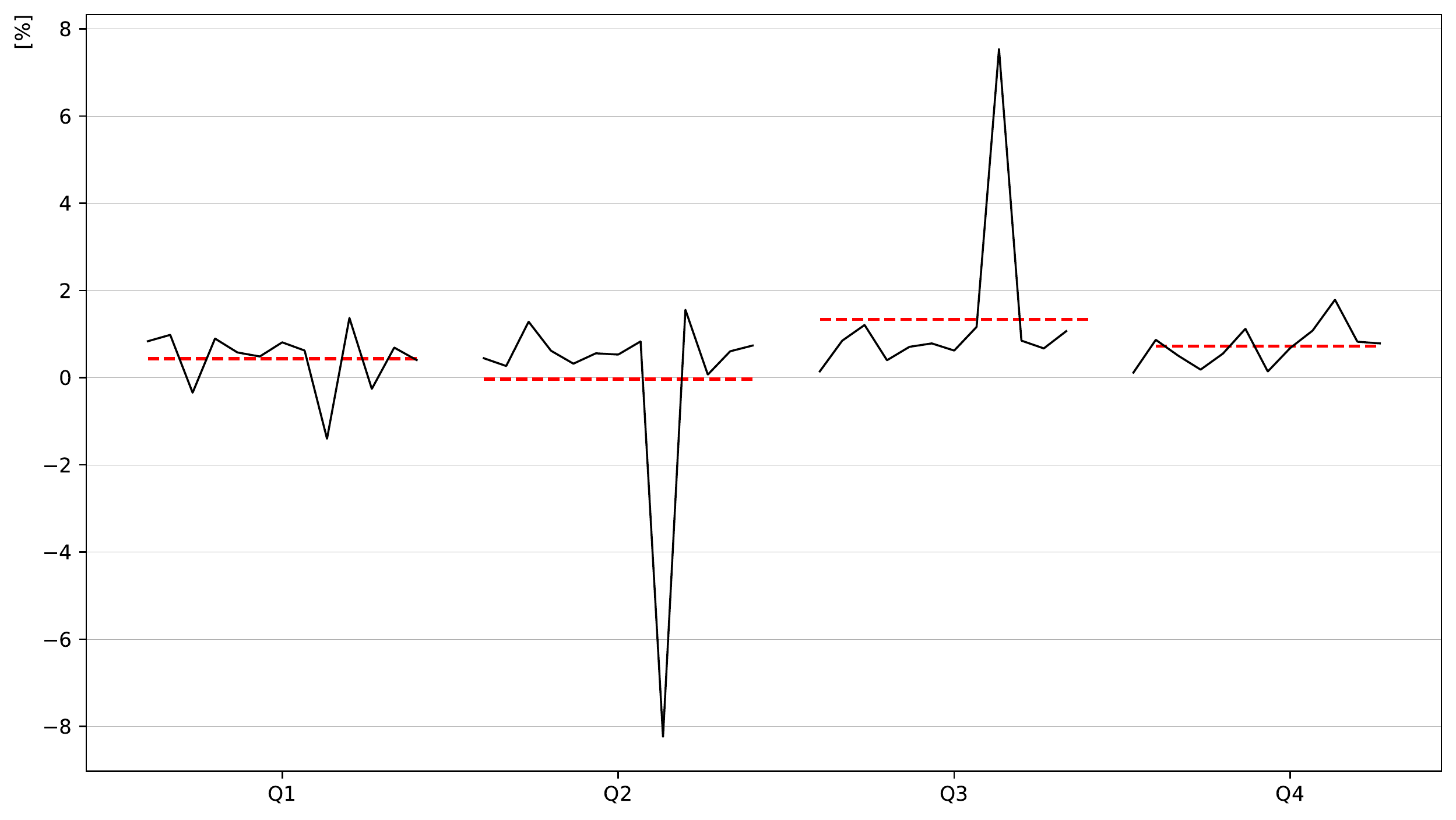} 
        \caption{Quarterly grouping: 2012:Q1 -- 2024:Q2.}
    \end{subfigure}

    \caption{Checking for residual seasonality in GDP growth: Quarterly grouping.}
    \label{fig:target_seasonality}
\end{figure}

\newpage

\subsection{Feature data}  

The TCODE column denotes the following data transformation for a monthly indicator (regressor) $x$:
\begin{enumerate}
\item No transformation
\item $\Delta x_t$
\item $\Delta^{2} x_t$
\item $log(x_t)$
\item $\Delta log(x_t)$
\item $\Delta^{2} log(x_t)$
\item $\Delta \left( \frac{x_t - x_{t-1}}{x_{t-1}} \right)$
\end{enumerate} 

The FRED column gives mnemonics in FRED followed by a short description. 
Some series require adjustments to the raw data available in FRED. 
These variables are tagged by an asterisk to indicate that they have been adjusted and thus differ from the series from the source. 
For a detailed summary of the adjustments see \cite{mccracken2016fred}. 

\vspace{30pt}

\begin{table}[H]
\centering
\caption{Group 1 -- Output and income}
\begin{tabular}{lllll}
\toprule
  & ID & tcode & FRED & Description \\
\midrule
1 & 1 & 5 & RPI & Real Personal Income \\
2 & 2 & 5 & W875RX1 & Real personal income ex transfer receipts \\
3 & 6 & 5 & INDPRO & IP Index \\
4 & 7 & 5 & IPFPNSS & IP: Financial Products and Nonindustrial Supplies \\
5 & 8 & 5 & IPFINAL & IP: Final Products (Market Group) \\
6 & 9 & 5 & IPCONGD & IP: Consumer Goods \\
7 & 10 & 5 & IPDCONGD & IP: Durable Consumer Goods \\
8 & 11 & 5 & IPNCONGD & IP: Nondurable Consumer Goods \\
9 & 12 & 5 & IPBUSEQ & IP: Business Equipment \\
10 & 13 & 5 & IPMAT & IP: Materials \\
11 & 14 & 5 & IPDMAT & IP: Durable Materials \\
12 & 15 & 5 & IPNMAT & IP: Nondurable Materials \\
13 & 16 & 5 & IPMANSICS & IP: Manufacturing (SIC) \\
14 & 17 & 5 & IPB51222s & IP: Residential Utilities \\
15 & 18 & 5 & IPFUELS & IP: Fuels \\
16 & 19 & 1 & NAPMPI & ISM Manufacturing: Production Index \\
17 & 20 & 2 & CUMFNS & Capacity Utilization: Manufacturing \\
\bottomrule
\end{tabular}
\label{tab:fred-md_group_01}
\end{table}

\newpage

\begin{table}[H]
\centering
\caption{Group 2 -- Labor market}
\begin{tabular}{lllll}
\toprule
  & ID & tcode & FRED & Description \\
\midrule
1 & 21* & 2 & HWI & Help-Wanted Index for United States \\
2 & 22* & 2 & HWIURATIO & Ratio of Help Wanted/No. Unemployed \\
3 & 23 & 5 & CLF160OV & Civilian Labor Force \\
4 & 24 & 5 & CE160V & Civilian Employment \\
5 & 25 & 2 & UNRATE & Civilian Unemployment Rate \\
6 & 26 & 2 & UEMPMEAN & Average Duration of Unemployment (Weeks) \\
7 & 27 & 5 & UEMPLT5 & Civilians Unemployed -- Less Than 5 Weeks \\
8 & 28 & 5 & UEMP5TO14 & Civilians Unemployed for 5-14 Weeks \\
9 & 29 & 5 & UEMP15OV & Civilians Unemployed -- 15 Weeks and Over \\
10 & 30 & 5 & UEMP15T26 & Civilians Unemployed for 15 -- 26 Weeks \\
11 & 31 & 5 & UEMP27OV & Civilians Unemployed for 27 Weeks and Over \\
12 & 32* & 5 & CLAIMSx & Initial Claims \\
13 & 33 & 5 & PAYEMS & All Employees: Total nonfarm \\
14 & 34 & 5 & USGOOD & All Employees: Goods-Producing Industries \\
15 & 35 & 5 & CES1021000001 & All Employees: Mining and Logging: Industries \\
16 & 36 & 5 & USCONS & All Employees: Construction \\
17 & 37 & 5 & MANEMP & All Employees: Manufacturing \\
18 & 38 & 5 & DMANEMP & All Employees: Durable Goods \\
19 & 39 & 5 & NDMANEMP & All Employees: Nondurable Goods \\
20 & 40 & 5 & SRVPRD & All Employees: Service-Providing Industries \\
21 & 41 & 5 & USTPU & All Employees: Trade, Transportation and Utilities \\
22 & 42 & 5 & USWTRADE & All Employees: Wholesale Trade \\
23 & 43 & 5 & USTRADE & All Employees: Retail Trade \\
24 & 44 & 5 & USFIRE & All Employees: Financial Activities \\
25 & 45 & 5 & USGOVT & All Employees: Government \\
26 & 46 & 1 & CES0600000007 & Avg Weekly Hours: Goods-Producing \\
27 & 47 & 2 & AWOTMAN & Avg Weekly Overtime Hours: Manufacturing \\
28 & 48 & 1 & AWHMAN & Avg Weekly Hours: Manufacturing \\
29 & 49 & 1 & NAPMEI & ISM Manufacturing: Employment Index \\
30 & 127 & 6 & CES0600000008 & Avg Hourly Earnings: Goods-Producing \\
31 & 128 & 6 & CES2000000008 & Avg Hourly Earnings: Construction \\
32 & 129 & 6 & CES3000000008 & Avg Hourly Earnings: Manufacturing \\
\bottomrule
\end{tabular}
\label{tab:fred-md_group_02}
\end{table}

\newpage

\begin{table}[H]
\centering
\caption{Group 3 -- Housing}
\begin{tabular}{lllll}
\toprule
  & ID & tcode & FRED & Description \\
\midrule
1 & 50 & 4 & HOUST & Housing Starts: Total New Privately Owned \\
2 & 51 & 4 & HOUSTNE & Housing Starts: Northeast \\
3 & 52 & 4 & HOUSTMW & Housing Starts: Midwest \\
4 & 53 & 4 & HOUSTS & Housing Starts: South \\
5 & 54 & 4 & HOUSTW & Housing Starts: West \\
6 & 55 & 4 & PERMIT & New Private Housing Permits (SAAR) \\
7 & 56 & 4 & PERMITNE & New Private Housing Permits: Northeast (SAAR) \\
8 & 57 & 4 & PERMITMW & New Private Housing Permits: Midwest (SAAR) \\
9 & 58 & 4 & PERMITS & New Private Housing Permits: South (SAAR) \\
10 & 59 & 4 & PERMITW & New Private Housing Permits: West (SAAR) \\
\bottomrule
\end{tabular}
\label{tab:fred-md_group_03}
\end{table}

\vspace{30pt}

\begin{table}[H]
\centering
\caption{Group 4 -- Consumption, orders and inventories}
\begin{tabular}{lllll}
\toprule
  & ID & tcode & FRED & Description \\
\midrule
1 & 3 & 5 & DPCERA3M086SBEA & Real personal consumption expenditures \\
2 & 4* & 5 & CMRMTSPLx & Real Manu. and Trade Industries Sales \\
3 & 5* & 5 & RETAILx & Retail and Food Services Sales \\
4 & 60 & 1 & NAPM & ISM: PMI Composite Index \\
5 & 61 & 1 & NAPMNOI & ISM: New Orders Index \\
6 & 62 & 1 & NAPMSDI & ISM: Supplier Deliveries Index \\
7 & 63 & 1 & NAPMII & ISM: Inventories Index \\
8 & 64 & 5 & ACOGNO & New Orders for Consumer Goods \\
9 & 65* & 5 & AMDMNOx & New Orders for Durable Goods \\
10 & 66* & 5 & ANDENOx & New Orders for Nondefense Capital Goods \\
11 & 67* & 5 & AMDMUOx & Unfilled Orders for Durable Goods \\
12 & 68* & 5 & BUSINVx & Total Business Inventories \\
13 & 69* & 2 & ISRATIOx & Total Business: Inventories to Sales Ratio \\
14 & 130* & 2 & UMSCENTx & Consumer Sentiment Index \\
\bottomrule
\end{tabular}
\label{tab:fred-md_group_04}
\end{table}

\newpage

\begin{table}[H]
\centering
\caption{Group 5 -- Money and credit}
\begin{tabular}{lllll}
\toprule
  & ID & tcode & FRED & Description \\
\midrule
1 & 70 & 6 & M1SL & M1 Money Stock \\
2 & 71 & 6 & M2SL & M2 Money Stock \\
3 & 72 & 5 & M2REAL & Real M2 Money Stock \\
4 & 73 & 6 & AMBSL & St. Louis Adjusted Monetary Base \\
5 & 74 & 6 & TOTRESNS & Total Reserves of Depository Institutions \\
6 & 75 & 7 & NONBORRES & Reserves of Depository Institutions \\
7 & 76 & 6 & BUSLOANS & Commercial and Industrial Loans \\
8 & 77 & 6 & REALLN & Real Estate Loans at All Commercial Banks \\
9 & 78 & 6 & NONREVSL & Total Nonrevolving Credit \\
10 & 79* & 2 & CONSPI & Nonrevolving consumer credit to Personal Income \\
11 & 131 & 6 & MZMSL & MZM Money Stock \\
12 & 132 & 6 & DTCOLNVHFNM & Consumer Motor Vehicle Loans Outstanding \\
13 & 133 & 6 & DTCTHFNM & Total Consumer Loans and Leases Outstanding \\
14 & 134 & 6 & INVEST & Securities in Bank Credit at All Commercial Banks \\
\bottomrule
\end{tabular}
\label{tab:fred-md_group_05}
\end{table}

\newpage

\begin{table}[H]
\centering
\caption{Group 6 -- Interest and exchange rates}
\begin{tabular}{lllll}
\toprule
  & ID & tcode & FRED & Description \\
\midrule
1 & 84 & 2 & FEDFUNDS & Effective Federal Funds Rate \\
2 & 85* & 2 & CP3Mx & 3-Month AA Financial Commercial Paper Rate \\
3 & 86 & 2 & TB3MS & 3-Month Treasury Bill \\
4 & 87 & 2 & TB6MS & 6-Month Treasury Bill \\
5 & 88 & 2 & GS1 & 1-Year Treasury Rate \\
6 & 89 & 2 & GS5 & 5-Year Treasury Rate \\
7 & 90 & 2 & GS10 & 10-Year Treasury Rate \\
8 & 91 & 2 & AAA & Moody's Seasoned Aaa Corporate Bond Yield \\
9 & 92 & 2 & BAA & Moody's Seasoned Baa Corporate Bond Yield \\
10 & 93* & 1 & COMPAPFFx & 3-Month Commercial Paper Minus FEDFUNDS \\
11 & 94 & 1 & TB3SMFFM & 3-Month Treasury C Minus FEDFUNDS \\
12 & 95 & 1 & TB6SMFFM & 6-Month Treasury C Minus FEDFUNDS \\
13 & 96 & 1 & T1YFFM & 1-Year Treasury C Minus FEDFUNDS \\
14 & 97 & 1 & T5YFFM & 5-Year Treasury C Minus FEDFUNDS \\
15 & 98 & 1 & T10YFFM & 10-Year Treasury C Minus FEDFUNDS \\
16 & 99 & 1 & AAAFFM & Moody's Aaa Corporate Bond Minus FEDFUNDS \\
17 & 100 & 1 & BAAFFM & Moody's Baa Corporate Bond Minus FEDFUNDS \\
18 & 101 & 5 & TWEXMMTH & Trade Weighted U.S. Dollar Index: Major Currencies \\
19 & 102* & 5 & EXSZUSx & Switzerland / U.S. Foreign Exchange Rate \\
20 & 103* & 5 & EXJPUSx & Japan / U.S. Foreign Exchange Rate \\
21 & 104* & 5 & EXUSUKx & U.S. / U.K. Foreign Exchange Rate \\
22 & 105* & 5 & EXCAUSx & Canada / U.S. Foreign Exchange Rate \\
\bottomrule
\end{tabular}
\label{tab:fred-md_group_06}
\end{table}

\newpage

\begin{table}[H]
\centering
\caption{Group 7 -- Prices}
\begin{tabular}{lllll}
\toprule
  & ID & tcode & FRED & Description \\
\midrule
1 & 106 & 6 & WPSFD49207 & PPI: Finished Goods \\
2 & 107 & 6 & WPSFD49502 & PPI: Finished Consumer Goods \\
3 & 108 & 6 & WPSID61 & PPI: Intermediate Materials \\
4 & 109 & 6 & WPSID62 & PPI: Crude Materials \\
5 & 110* & 6 & OILPRICEx & Crude Oil, spliced WTI and Cushing \\
6 & 111 & 6 & PPICMM & PPI: Metals and metal products \\
7 & 112 & 1 & NAPMPRI & ISM Manufacturing: Prices Index \\
8 & 113 & 6 & CPIAUCSL & CPI: All Items \\
9 & 114 & 6 & CPIAPPSL & CPI: Apparel \\
10 & 115 & 6 & CPITRNSL & CPI: Transportation \\
11 & 116 & 6 & CPIMEDSL & CPI: Medical Care \\
12 & 117 & 6 & CUSR0000SAC & CPI: Commodities \\
13 & 118 & 6 & CUSR0000SAD & CPI: Durables \\
14 & 119 & 6 & CUSR0000SAS & CPI: Service \\
15 & 120 & 6 & CPIULFSL & CPI: All Items less Food \\
16 & 121 & 6 & CUSR0000SA0L2 & CPI: All Items less Shelter \\
17 & 122 & 6 & CUSR0000SA0L5 & CPI: All Items less Medical Care \\
18 & 123 & 6 & PCEPI & Personal Cons. Expend.: Chain Index \\
19 & 124 & 6 & DDURRG3M086SBEA & Personal Cons. Expend.: Durable Goods \\
20 & 125 & 6 & DNDGRG3M086SBEA & Personal Cons. Expend.: Nondurable Goods \\
21 & 126 & 6 & DSERRG3M086SBEA & Personal Cons. Expend.: Services \\
\bottomrule
\end{tabular}
\label{tab:fred-md_group_07}
\end{table}

\vspace{20pt}

\begin{table}[H]
\centering
\caption{Group 8 -- Stock market}
\begin{tabular}{lllll}
\toprule
  & ID & tcode & FRED & Description \\
\midrule
1 & 80* & 5 & S\&P 500 & S\&P's Common Stock Price Index: Composite \\
2 & 81* & 5 & S\&P: indust & S\&P's Common Stock Price Index: Industrials \\
3 & 82* & 2 & S\&P div yield & S\&P's Composite Common Stock: Dividend Yield \\
4 & 83* & 5 & S\&P PE ratio & S\&P's Composite Common Stock: Price-Earnings Ratio \\
\bottomrule
\end{tabular}
\label{tab:fred-md_group_08}
\end{table}

\newpage

\begin{figure}[H]
     \centering
     \begin{subfigure}[b]{0.9\textwidth}
         \centering
         \includegraphics[width=\textwidth]{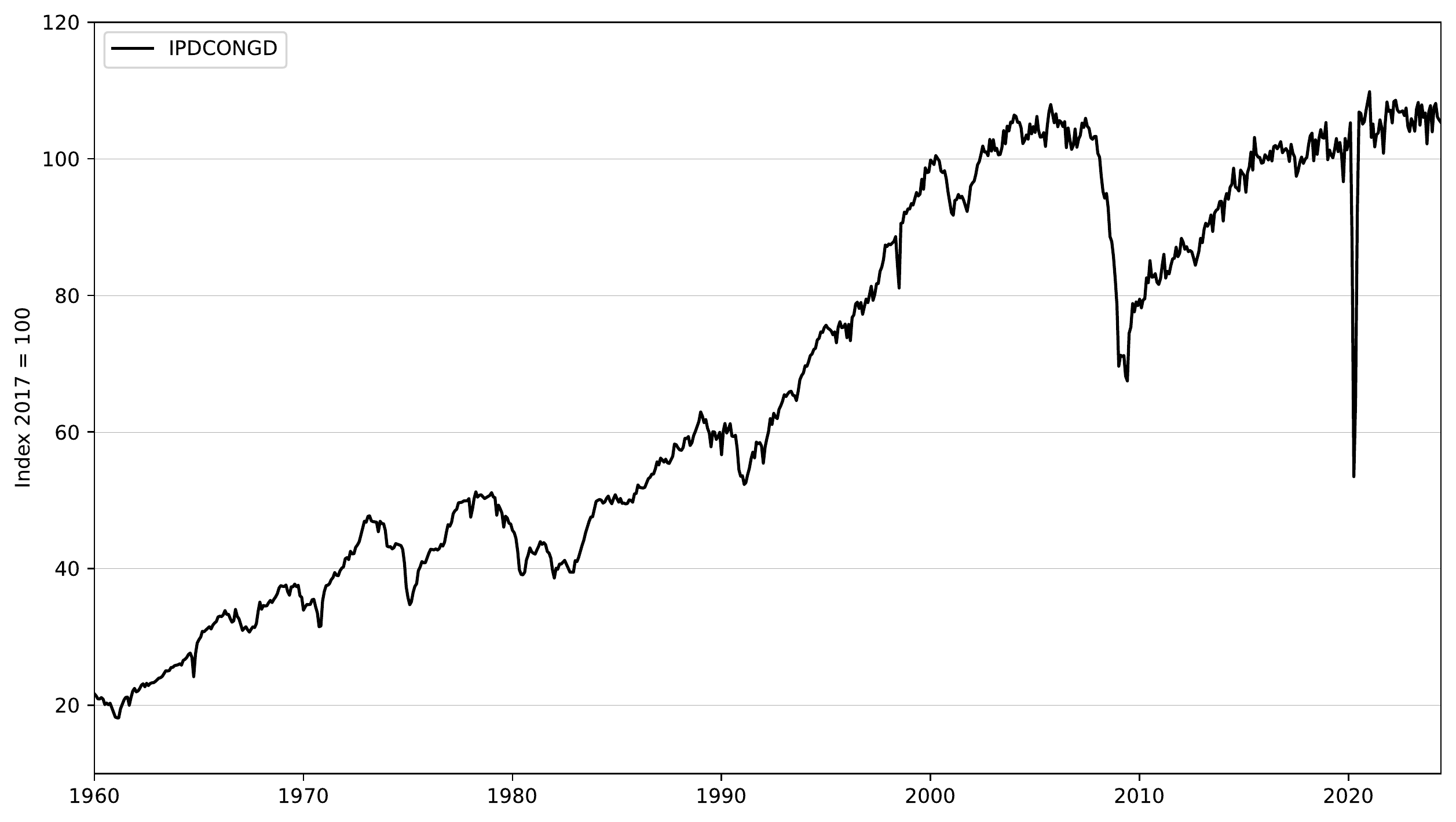}
         \caption{IP: Durable Consumer Goods (IPDCONGD): Original time series with trend.}
     \end{subfigure}
     \hfill
     \begin{subfigure}[b]{0.9\textwidth}
         \centering
         \includegraphics[width=\textwidth]{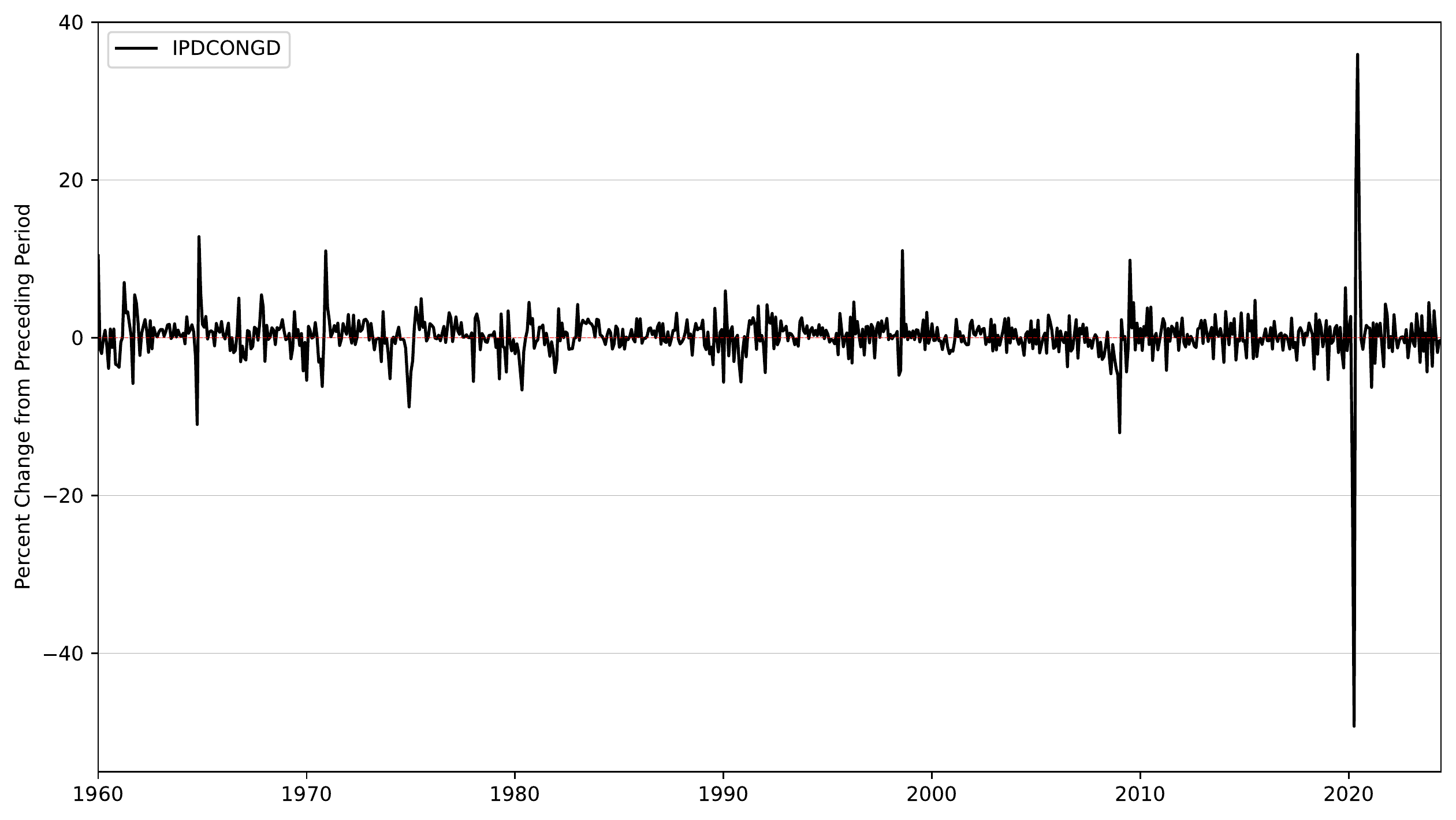}
         \caption{IP: Durable Consumer Goods (IPDCONGD): Log-differenced, detrended time series.}
     \end{subfigure}     

\caption{Industrial Production: Durable Consumer Goods (IPDCONGD, Group 1, \textit{ID}: 10) before and after the corresponding data transformation (\textit{tcode:} 5). Source: FRED-MD.}
\label{fig:ipdcongd}

\end{figure}

\newpage

    \begin{figure}[H]
        \centering
        \begin{subfigure}[h]{0.475\textwidth}
            \centering
            \includegraphics[width=\textwidth]{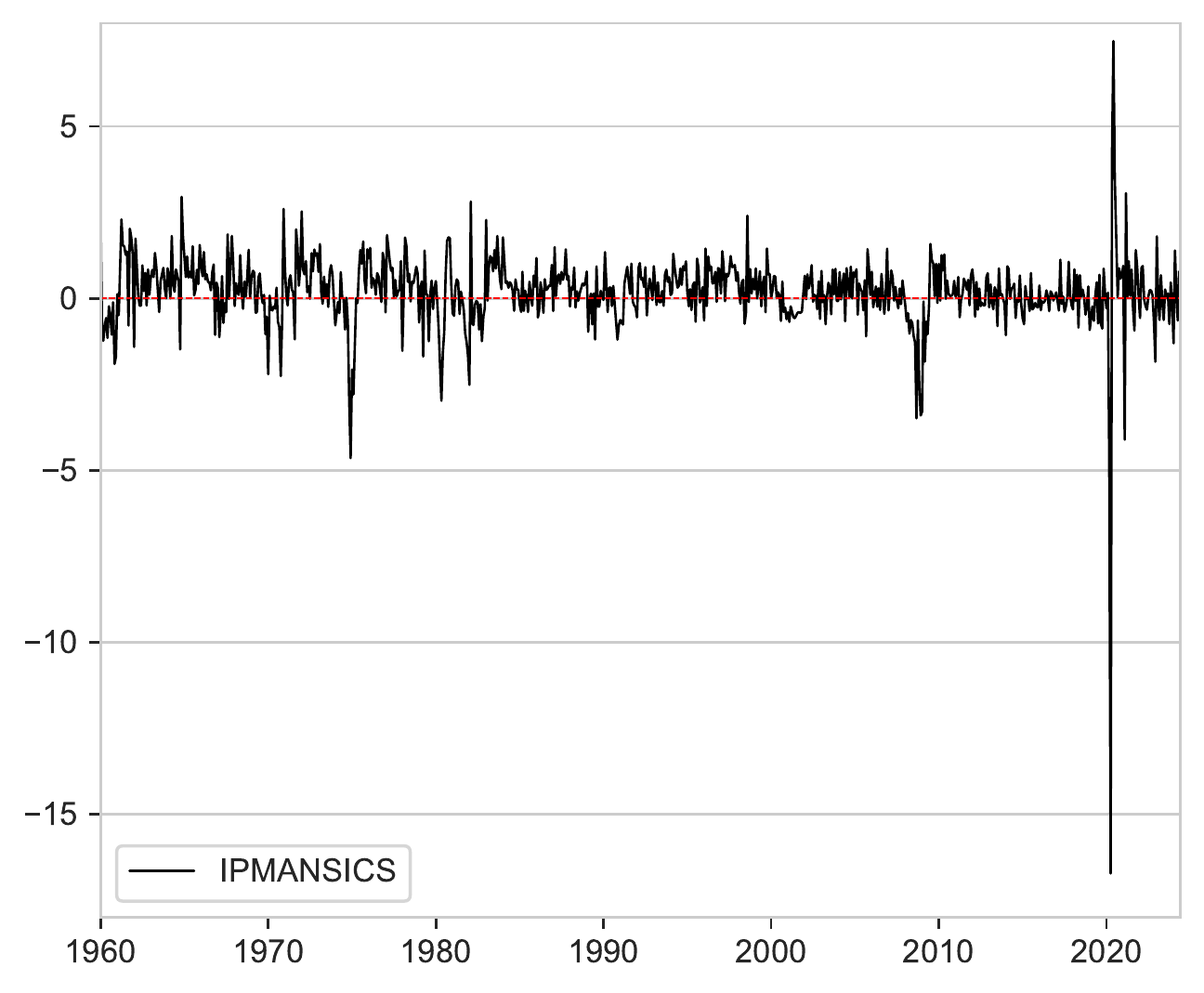}
            \caption[]%
            {{\small IP: Manufacturing (SIC)}}    
        \end{subfigure}
        \hfill  
        \begin{subfigure}[h]{0.475\textwidth}  
            \centering 
            \includegraphics[width=\textwidth]{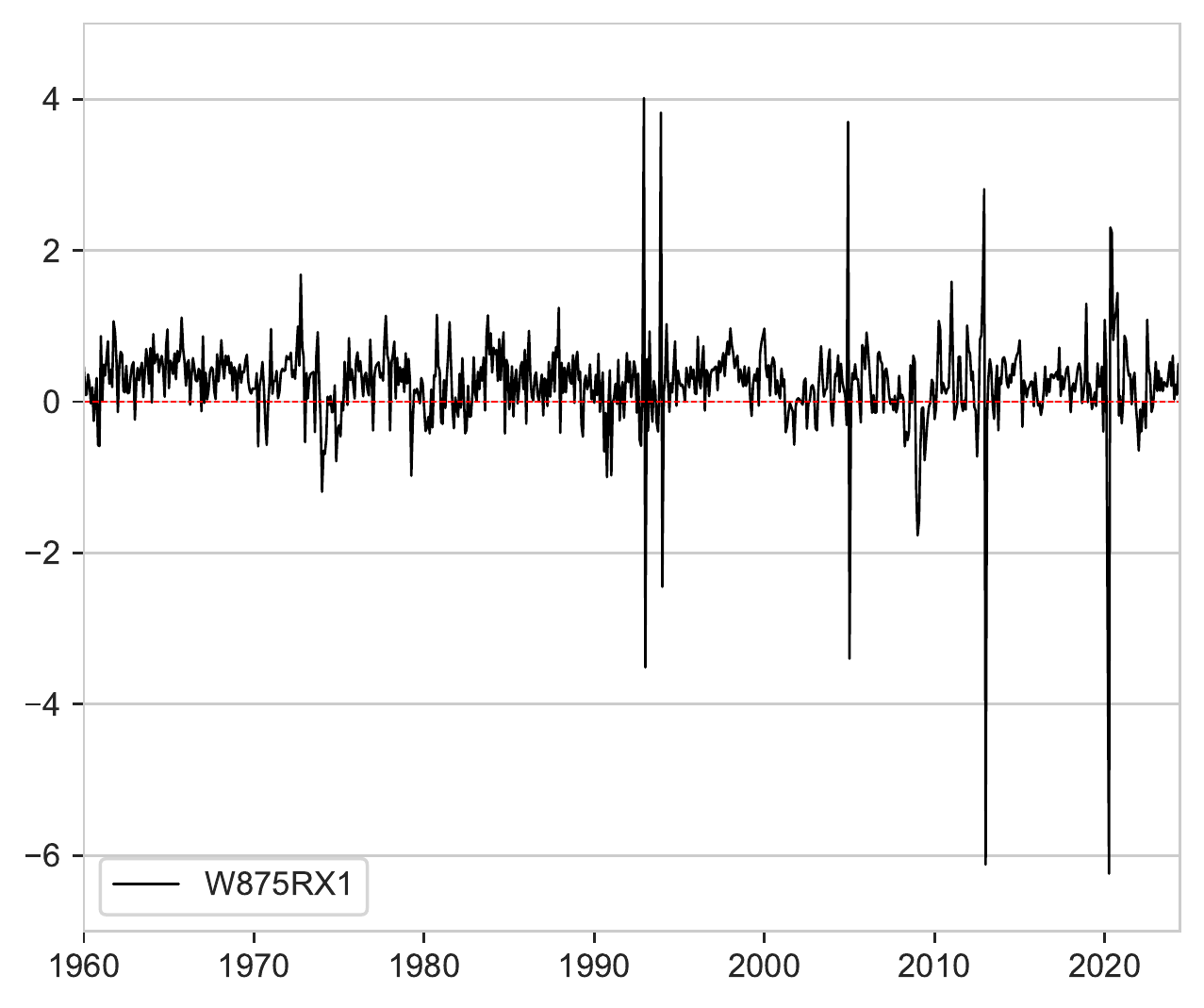}
            \caption[]%
            {{\small Real personal income ex. transfer receipts}}    
        \end{subfigure}
        \vskip\baselineskip
        \begin{subfigure}[h]{0.475\textwidth}   
            \centering 
            \includegraphics[width=\textwidth]{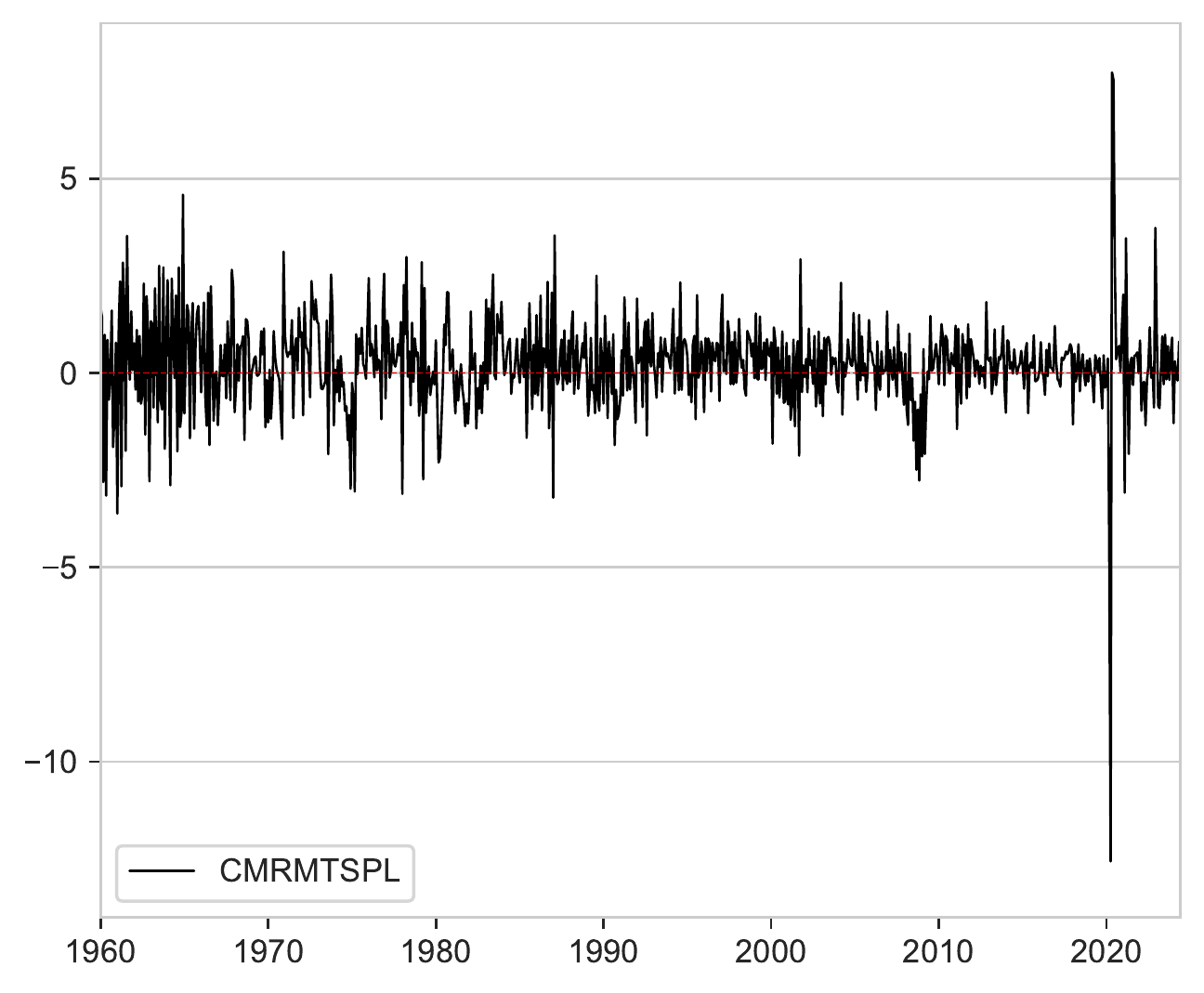}
            \caption[]%
            {{\small Real Manu. and Trade Industries Sales}}    
        \end{subfigure}
        \hfill
        \begin{subfigure}[h]{0.475\textwidth}   
            \centering 
            \includegraphics[width=\textwidth]{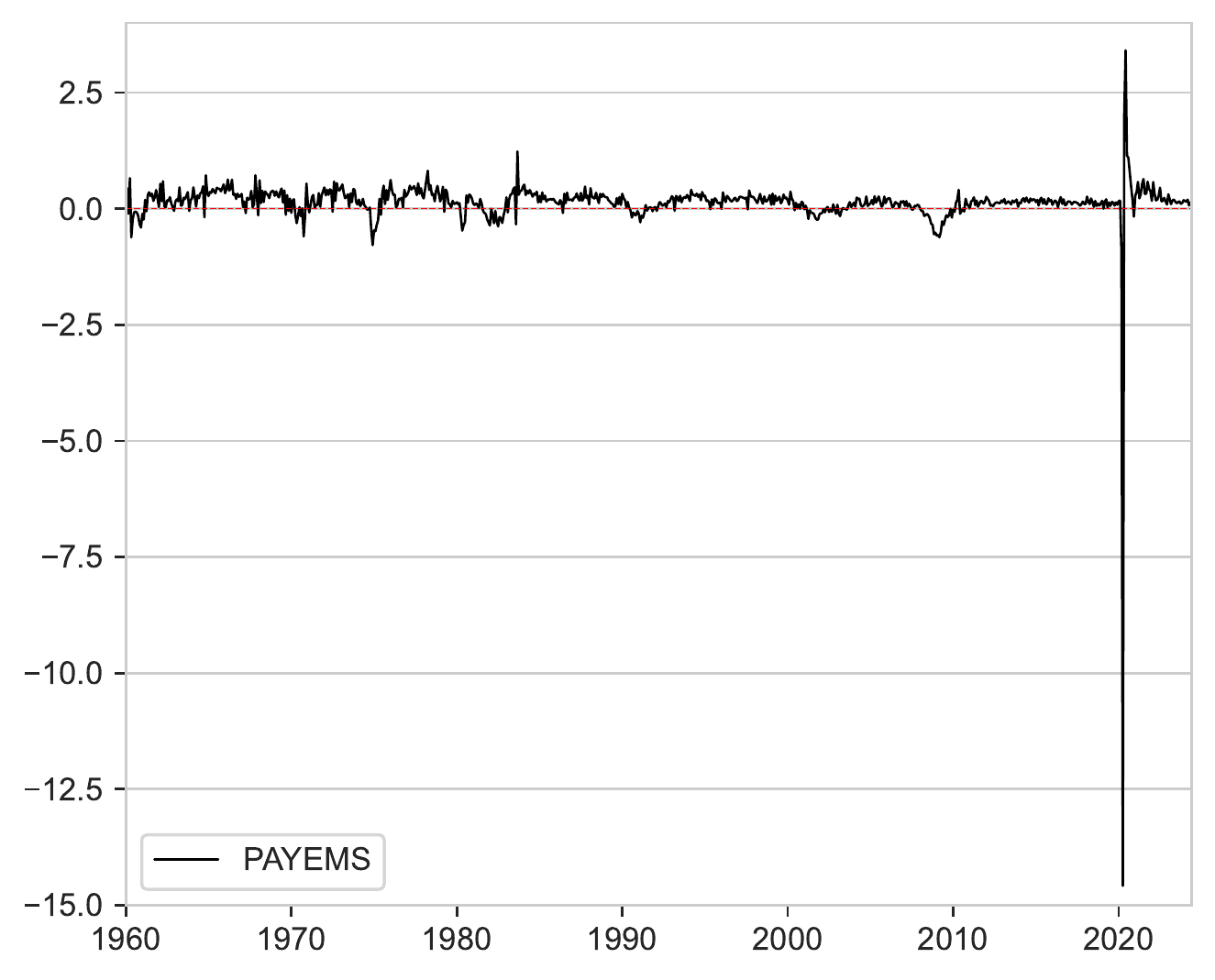}
            \caption[]%
            {{\small All Employees: Total nonfarm}}    
        \end{subfigure}

\caption[The average and standard deviation of critical parameters] {Transformed series for monthly indicators included in the benchmark DFM's measurement vector. Source: FRED-MD.} 
\label{fig:features_dfm}

\end{figure}

\newpage

\begin{table}[H]
\centering
\caption{Test statistics of the KPSS and the ADF stationary test for the untransformed (raw) monthly indicators used in the benchmark DFM.} 
\centering
\def\arraystretch{1.0}
\begin{tabular}{ p{5.0cm} | c | c | c }
\toprule
{} & 1960:M1 -- 2024:M6 & 2012:M1 -- 2019:M12 & 2012:M1 -- 2024:M6 \\
\hline \hline
IPDCONGD - KPSS  &      $4.104^{***}$ &       $1.386^{***}$ &      $1.211^{***}$ \\
IPDCONGD - ADF   &           $-1.141$ &      $-3.551^{***}$ &           $-2.520$ \\
\midrule

IPMANSICS - KPSS &      $4.255^{***}$ &             $0.298$ &            $0.256$ \\
IPMANSICS - ADF  &           $-1.388$ &      $-3.567^{***}$ &           $-2.482$ \\
\midrule

W875RX1 - KPSS   &      $4.331^{***}$ &       $1.672^{***}$ &      $1.758^{***}$ \\
W875RX1 - ADF    &            $2.346$ &             $0.267$ &           $-0.083$ \\
\midrule 

CMRMTSPLx - KPSS &      $4.354^{***}$ &       $1.666^{***}$ &      $1.670^{***}$ \\
CMRMTSPLx - ADF  &            $0.646$ &            $-2.092$ &           $-0.515$ \\
\midrule

PAYEMS - KPSS    &      $4.350^{***}$ &       $1.703^{***}$ &      $1.369^{***}$ \\
PAYEMS - ADF     &           $-0.558$ &            $-1.672$ &           $-1.225$ \\
\bottomrule
\end{tabular}
\label{tab:stationarity_regressors_raw}
\end{table}

\begin{table}[H]
\centering
\caption{Test statistics of the KPSS and the ADF stationary test for the transformed (detrended) monthly indicators used in the benchmark DFM.} 
\centering
\def\arraystretch{1.0}
\begin{tabular}{ p{5.0cm} | c | c | c }
\toprule
{} & 1960:M1 -- 2024:M6 & 2012:M1 -- 2019:M12 & 2012:M1 -- 2024:M6 \\
\hline \hline
$\Delta \log$(IPDCONGD) - KPSS  &            $0.219$ &             $0.283$ &       $0.467^{**}$ \\
$\Delta \log$(IPDCONGD) - ADF   &    $-15.300^{***}$ &       $-3.147^{**}$ &     $-6.912^{***}$ \\
\midrule

$\Delta \log$(IPMANSICS) - KPSS &       $0.497^{**}$ &             $0.227$ &            $0.061$ \\
$\Delta \log$(IPMANSICS) - ADF  &     $-6.828^{***}$ &       $-3.334^{**}$ &    $-10.774^{***}$ \\
\midrule

$\Delta \log$(W875RX1) - KPSS   &       $0.553^{**}$ &             $0.074$ &            $0.042$ \\
$\Delta \log$(W875RX1) - ADF    &     $-5.877^{***}$ &        $-2.719^{*}$ &    $-10.364^{***}$ \\
\midrule

$\Delta \log$(CMRMTSPLx) - KPSS  &            $0.168$ &        $0.479^{**}$ &            $0.085$ \\
$\Delta \log$(CMRMTSPLx) - ADF   &     $-7.512^{***}$ &            $-1.932$ &     $-5.159^{***}$ \\
\midrule

$\Delta \log$(PAYEMS) - KPSS    &       $0.568^{**}$ &        $0.666^{**}$ &            $0.059$ \\
$\Delta \log$(PAYEMS) - ADF     &     $-8.382^{***}$ &            $-2.284$ &    $-10.276^{***}$ \\
\bottomrule
\end{tabular}
\label{tab:stationarity_regressors_transformed}
\end{table}

\newpage

\section{Results appendix}  \label{app:results}

\setcounter{figure}{0}    
\setcounter{table}{0}

\begin{table}[H]
\centering
\caption{Monthly timestamps of the FRED-MD vintages corresponding to each intra-quarterly information set (nowcasting scenario).}
\def\arraystretch{1.0}
\begin{tabular}{l | c | c | c}
\toprule
{} & $\Omega_{m_{1}}$ & $\Omega_{m_{2}}$ & $\Omega_{m_{3}}$ \\
\midrule
FRED-MD vintages & M1, M4, M7, M10 & M2, M5, M8, M11 & M3, M6, M9, M12 \\
\bottomrule
\end{tabular}
\label{tab:fred_md_timestamps}
\end{table}

\vspace{1cm}

\begin{figure}[H] 
\centering
\caption{Formation of monthly input sequences (regressor vectors) corresponding to each intra-quarterly information set.}

\begin{tikzpicture}

\pgfmathsetmacro{\start}{0}     
\pgfmathsetmacro{\fin}{12}      
\pgfmathsetmacro{\tickstep}{1}  
\pgfmathsetmacro{\quarterstep}{3} 

\draw[thick] (\start, 0) -- (\fin, 0);

\foreach \x in {\start, ..., \fin} {
    \draw[thin] (\x, 0.1) -- (\x, -0.1); 
}

\foreach \x in {\start, ..., \fin} {
    \pgfmathtruncatemacro{\xint}{\x} 
    \ifnum\xint>0
        \ifnum\xint<13
            \node[above] at (\x-0.5, +0.2) {M\xint}; 
        \fi
    \fi
    
    \pgfmathtruncatemacro{\modresult}{mod(\xint,\quarterstep)}
    \ifnum\modresult=0
        \ifnum\xint>0
            \node[below] at (\x, -0.3) {Q\the\numexpr\x/\quarterstep\relax}; 
            \draw[thick] (\x, 0.2) -- (\x, -0.2); 
        \fi
    \fi
}

\pgfmathsetmacro{\lineend}{\fin} 
\pgfmathsetmacro{\linestart}{\lineend-8} 

\draw[thick, yellow, <->] (\linestart-2, -1.2) -- (\lineend-2, -1.2); 
\draw[thick, orange, <->] (\linestart-1, -2.2) -- (\lineend-1, -2.2); 
\draw[thick, red, <->] (\linestart, -3.2) -- (\lineend, -3.2); 

\draw[decorate, decoration={brace, mirror, amplitude=7pt}] 
    (\linestart-2, -1.3) -- (\lineend-2, -1.3) node[midway, below=7pt] {$\mathbf{\Omega} = \Omega_{m_{1}}, \ l = 8$};
\draw[decorate, decoration={brace, mirror, amplitude=7pt}] 
    (\linestart-1, -2.3) -- (\lineend-1, -2.3) node[midway, below=7pt] {$\mathbf{\Omega} = \Omega_{m_{2}}, \ l = 8$};
\draw[decorate, decoration={brace, mirror, amplitude=7pt}] 
    (\linestart, -3.3) -- (\lineend, -3.3) node[midway, below=7pt] {$\mathbf{\Omega} = \Omega_{m_{3}}, \ l = 8$};

\end{tikzpicture}
\label{fig:visu_regressor_vectors}

\end{figure}

\newpage

\begin{table}[H]
\centering
\caption{Nowcasting performance of the competitor models relative to a naive constant growth model for GDP and a benchmark DFM specification: MAE evaluation. Evaluation period: 2012:Q1 -- 2019:Q4.} 
\centering
\begin{tabular}{l | l l | l l | l l}
& \multicolumn{2}{c |}{$v = 3t-2$} & \multicolumn{2}{c |}{$v = 3t-1$} & \multicolumn{2}{c}{$v = 3t$} \\
& Naive & DFM & Naive & DFM & Naive & DFM \\
\hline\hline

MLP ($l$ = 8)     &  $\mathbf{0.720^{**}}$ &  $\mathbf{1.070}$ &  $\mathbf{0.743^{**}}$ &  $\mathbf{1.103}$ &  $\mathbf{0.700^{**}}$ &  $\mathbf{1.081}$ \\
MLP ($l$ = 18)    &   $0.753^{*}$ &  $1.119$ &   $0.756^{*}$ &  $1.121$ &   $0.750^{*}$ &  $1.158$ \\
MLP ($l$ = 36)    &   $0.773^{*}$ &  $1.149$ &   $0.756^{*}$ &  $1.122$ &   $0.767^{*}$ &  $1.184$ \\
MLP ($l$ = 48)    &   $0.782^{*}$ &  $1.162$ &   $0.776^{*}$ &  $1.151$ &   $0.771^{*}$ &  $1.191$ \\
\midrule

1D CNN ($l$ = 8)  &   $0.761^{*}$ &  $1.131$ &   $0.745^{*}$ &  $1.106$ &  $0.700^{**}$ &  $1.082$ \\
1D CNN ($l$ = 18) &   $0.759^{*}$ &  $1.127$ &  $0.737^{**}$ &  $1.093$ &  $0.728^{**}$ &  $1.125$ \\
1D CNN ($l$ = 36) &  $\mathbf{0.726^{**}}$ &  $\mathbf{1.079}$ &  $\mathbf{0.684^{**}}$ &  $\mathbf{1.015}$ &  $\mathbf{0.697^{**}}$ &  $\mathbf{1.077}$ \\
1D CNN ($l$ = 48) &   $0.788^{*}$ &  $1.170$ &  $0.725^{**}$ &  $1.076$ &  $0.727^{**}$ &  $1.123$ \\
\midrule

RNN ($l$ = 8)     &  $0.719^{**}$ &  $1.068$ &   $0.745^{*}$ &  $1.105$ &  $\underline{\mathbf{0.646^{**}}}$ &  $\underline{\mathbf{0.999}}$ \\
RNN ($l$ = 18)    &  $0.712^{**}$ &  $1.057$ &  $\mathbf{0.741^{**}}$ &  $\mathbf{1.100}$ &  $0.651^{**}$ &  $1.005$ \\
RNN ($l$ = 36)    &  $0.750^{**}$ &  $1.115$ &  $0.743^{**}$ &  $1.102$ &  $0.658^{**}$ &  $1.016$ \\
RNN ($l$ = 48)    &  $\mathbf{0.711^{**}}$ &  $\mathbf{1.056}$ &   $0.748^{*}$ &  $1.110$ &  $0.659^{**}$ &  $1.019$ \\
\midrule

LSTM ($l$ = 8)    &  $\mathbf{0.725^{**}}$ &  $\mathbf{1.077}$ &   $0.748^{*}$ &  $1.110$ &  $\mathbf{0.666^{**}}$ &  $\mathbf{1.029}$ \\
LSTM ($l$ = 18)   &  $0.735^{**}$ &  $1.092$ &   $0.742^{*}$ &  $1.102$ &  $0.680^{**}$ &  $1.051$ \\
LSTM ($l$ = 36)   &  $0.727^{**}$ &  $1.080$ &   $0.731^{*}$ &  $1.085$ &  $0.719^{**}$ &  $1.110$ \\
LSTM ($l$ = 48)   &  $0.729^{**}$ &  $1.083$ &  $\mathbf{0.726^{**}}$ &  $\mathbf{1.077}$ &  $0.694^{**}$ &  $1.072$ \\
\midrule

GRU ($l$ = 8)     &  $0.712^{**}$ &  $1.057$ &  $\mathbf{0.703^{**}}$ &  $\mathbf{1.043}$ &  $0.657^{**}$ &  $1.015$ \\
GRU ($l$ = 18)    &  $\mathbf{0.704^{**}}$ &  $\mathbf{1.046}$ &  $0.714^{**}$ &  $1.060$ &  $\mathbf{0.655^{**}}$ &  $\mathbf{1.012}$ \\
GRU ($l$ = 36)    &  $0.724^{**}$ &  $1.075$ &  $0.732^{**}$ &  $1.086$ &  $0.680^{**}$ &  $1.050$ \\
GRU ($l$ = 48)    &  $0.721^{**}$ &  $1.071$ &  $0.720^{**}$ &  $1.068$ &  $0.676^{**}$ &  $1.045$ \\
\bottomrule
\end{tabular}
\caption*{\begin{footnotesize}
\textbf{Notes:} This table reports the relative MAE of GDP growth for the competitor ANNs relative to a naive constant growth model and a benchmark DFM specification. 
A value below one indicates that the competitor model beats the naive benchmark model. 
The stars denote statistical significance at 10\%($*$), 5\%($**$) and 1\%($***$) level of the one-sided \textcite{diebold1995comparing} test. 
Columns are related to the individual nowcasting scenarios: e.g. $v = 3t-2$ refers to the 1-month nowcasting scenario. 
Within each nowcasting scenario, best (lowest) relative MAE values are highlighted in bold for the given competitor model. 
The overall best relative MAE values are highlighted in bold and underlined. 
\end{footnotesize}}
\label{tab:eval_no_covid_mae}
\end{table}


\begin{table}[H] 
\centering
\caption{Nowcasting performance of the competitor ANNs relative to the training configuration using the shortest input sequences $(l = 8)$: RMSE and MAE evaluation. Evaluation period: 2012:Q1 -- 2019:Q4.} 
\begin{tabular}{l | l l | l l | l l}

& \multicolumn{2}{c |}{$v = 3t-2$} & \multicolumn{2}{c |}{$v = 3t-1$} & \multicolumn{2}{c}{$v = 3t$} \\

& RMSE & MAE & RMSE & MAE & RMSE & MAE \\
\hline\hline

MLP ($l$ = 8)     &          \multicolumn{1}{c}{$-$} &          \multicolumn{1}{c|}{$-$} &          \multicolumn{1}{c}{$-$} &          \multicolumn{1}{c|}{$-$} &          \multicolumn{1}{c}{$-$} &          \multicolumn{1}{c}{$-$} \\
MLP ($l$ = 18)    &  $1.014$ &  $1.045$ &  $1.041$ &  $1.017$ &  $1.062$ &  $1.072$ \\
MLP ($l$ = 36)    &  $1.040$ &  $1.073$ &  $1.051$ &  $1.017$ &  $1.093$ &  $1.096$ \\
MLP ($l$ = 48)    &  $1.079$ &  $1.086$ &  $1.103$ &  $1.044$ &  $1.126$ &  $1.102$ \\
\midrule

1D CNN ($l$ = 8)  &          \multicolumn{1}{c}{$-$} &          \multicolumn{1}{c|}{$-$} &          \multicolumn{1}{c}{$-$} &          \multicolumn{1}{c|}{$-$} &          \multicolumn{1}{c}{$-$} &          \multicolumn{1}{c}{$-$} \\
1D CNN ($l$ = 18) &  $0.970$ &  $0.997$ &  $0.974$ &  $0.989$ &  $1.021$ &  $1.040$ \\
1D CNN ($l$ = 36) &  $0.936$ &  $0.954$ &  $0.921$ &  $0.918$ &  $0.984$ &  $0.996$ \\
1D CNN ($l$ = 48) &  $1.022$ &  $1.035$ &  $1.020$ &  $0.973$ &  $1.031$ &  $1.039$ \\
\midrule

RNN ($l$ = 8)     &          \multicolumn{1}{c}{$-$} &          \multicolumn{1}{c|}{$-$} &          \multicolumn{1}{c}{$-$} &          \multicolumn{1}{c|}{$-$} &          \multicolumn{1}{c}{$-$} &          \multicolumn{1}{c}{$-$} \\
RNN ($l$ = 18)    &  $0.987$ &  $0.990$ &  $0.999$ &  $0.995$ &  $1.015$ &  $1.006$ \\
RNN ($l$ = 36)    &  $1.024$ &  $1.044$ &  $0.994$ &  $0.997$ &  $1.025$ &  $1.018$ \\
RNN ($l$ = 48)    &  $0.977$ &  $0.989$ &  $1.000$ &  $1.004$ &  $1.035$ &  $1.020$ \\
\midrule

LSTM ($l$ = 8)    &          \multicolumn{1}{c}{$-$} &          \multicolumn{1}{c|}{$-$} &          \multicolumn{1}{c}{$-$} &          \multicolumn{1}{c|}{$-$} &          \multicolumn{1}{c}{$-$} &          \multicolumn{1}{c}{$-$} \\
LSTM ($l$ = 18)   &  $1.007$ &  $1.014$ &  $0.999$ &  $0.992$ &  $1.011$ &  $1.021$ \\
LSTM ($l$ = 36)   &  $0.980$ &  $1.003$ &  $0.989$ &  $0.977$ &  $1.038$ &  $1.079$ \\
LSTM ($l$ = 48)   &  $0.991$ &  $1.006$ &  $0.990$ &  $0.970$ &  $1.025$ &  $1.042$ \\
\midrule

GRU ($l$ = 8)     &          \multicolumn{1}{c}{$-$} &          \multicolumn{1}{c|}{$-$} &          \multicolumn{1}{c}{$-$} &          \multicolumn{1}{c|}{$-$} &          \multicolumn{1}{c}{$-$} &          \multicolumn{1}{c}{$-$} \\
GRU ($l$ = 18)    &  $0.991$ &  $0.989$ &  $1.017$ &  $1.016$ &  $1.000$ &  $0.997$ \\
GRU ($l$ = 36)    &  $1.010$ &  $1.017$ &  $1.044$ &  $1.041$ &  $1.002$ &  $1.034$ \\
GRU ($l$ = 48)    &  $1.009$ &  $1.013$ &  $1.029$ &  $1.024$ &  $1.008$ &  $1.029$ \\
\bottomrule
\end{tabular}
\caption*{\begin{footnotesize}
\textbf{Notes:} This table reports the relative RMSE and MAE of GDP growth for our competitor ANNs relative to the training configuration using the shortest input sequences $(l = 8)$. 
A value below one indicates that the competitor model beats the benchmark training configuration $(l = 8)$. 
The stars denote statistical significance at
10\%($*$), 5\%($**$) and 1\%($***$) level of the one-sided \cite{diebold1995comparing} test. 
Columns are related to the individual nowcasting scenarios: e.g.
$v = 3t-2$ refers to the 1-month nowcasting scenario. 
Relative RMSE and MAE values indicating significant performance advantage over the benchmark training configuration $(l = 8)$ are highlighted in bold. 
\end{footnotesize}}
\label{tab:eval_no_covid_shortest}
\end{table}

\newpage

\begin{table}[H]
\centering
\caption{Nowcasting performance of the competitor models relative to a naive constant growth model for GDP and a benchmark DFM specification: MAE evaluation. Evaluation period: 2012:Q1 -- 2024:Q2.} 
\centering
\begin{tabular}{l | l l | l l | l l}
& \multicolumn{2}{c |}{$v = 3t-2$} & \multicolumn{2}{c |}{$v = 3t-1$} & \multicolumn{2}{c}{$v = 3t$} \\
& Naive & DFM & Naive & DFM & Naive & DFM \\
\hline\hline

MLP ($l$ = 8)     &  $\mathbf{0.569^{*}}$ &  $\mathbf{0.897}$ &  $\mathbf{0.480^{**}}$ &  $\mathbf{0.932}$ &  $\mathbf{0.426^{**}}$ &   $\mathbf{0.787^{*}}$ \\
MLP ($l$ = 18)    &  $0.629^{*}$ &  $0.991$ &  $0.535^{**}$ &  $1.040$ &   $0.545^{*}$ &   $1.007$ \\
MLP ($l$ = 36)    &  $0.686^{*}$ &  $1.081$ &  $0.543^{**}$ &  $1.055$ &  $0.579^{**}$ &   $1.070$ \\
MLP ($l$ = 48)    &  $0.689^{*}$ &  $1.085$ &  $0.584^{**}$ &  $1.134$ &  $0.591^{**}$ &   $1.092$ \\
\midrule 

1D CNN ($l$ = 8)  &  $\mathbf{0.547^{*}}$ &  $\mathbf{0.861}$ &  $\mathbf{0.500^{**}}$ &  $\mathbf{0.972}$ &  $\underline{\mathbf{0.371^{**}}}$ &  $\underline{\mathbf{0.686^{**}}}$ \\
1D CNN ($l$ = 18) &  $0.640^{*}$ &  $1.008$ &   $0.544^{*}$ &  $1.057$ &  $0.418^{**}$ &   $0.773^{*}$ \\
1D CNN ($l$ = 36) &  $0.669^{*}$ &  $1.054$ &   $0.620^{*}$ &  $1.205$ &   $0.606^{*}$ &   $1.119$ \\
1D CNN ($l$ = 48) &  $0.682^{*}$ &  $1.075$ &  $0.602^{**}$ &  $1.169$ &  $0.584^{**}$ &   $1.079$ \\
\midrule 

RNN ($l$ = 8)     &  $\mathbf{0.563^{*}}$ &  $\mathbf{0.887}$ &   $\mathbf{0.556^{*}}$ &  $\mathbf{1.081}$ &  $\mathbf{0.477^{**}}$ &   $\mathbf{0.882}$ \\
RNN ($l$ = 18)    &  $0.568^{*}$ &  $0.895$ &   $0.603^{*}$ &  $1.171$ &  $0.500^{**}$ &   $0.923$ \\
RNN ($l$ = 36)    &  $0.613^{*}$ &  $0.965$ &   $0.639^{*}$ &  $1.242$ &  $0.559^{**}$ &   $1.032$ \\
RNN ($l$ = 48)    &  $0.601^{*}$ &  $0.947$ &   $0.617^{*}$ &  $1.199$ &  $0.570^{**}$ &   $1.053$ \\
\midrule

LSTM ($l$ = 8)    &  $\mathbf{0.601^{*}}$ &  $\mathbf{0.947}$ &   $\mathbf{0.571^{*}}$ &  $\mathbf{1.110}$ &   $0.638^{*}$ &   $1.179$ \\
LSTM ($l$ = 18)   &  $0.628^{*}$ &  $0.989$ &   $0.574^{*}$ &  $1.116$ &   $\mathbf{0.608^{*}}$ &   $\mathbf{1.123}$ \\
LSTM ($l$ = 36)   &  $0.655^{*}$ &  $1.032$ &  $0.593^{**}$ &  $1.152$ &   $0.629^{*}$ &   $1.162$ \\
LSTM ($l$ = 48)   &  $0.643^{*}$ &  $1.013$ &   $0.573^{*}$ &  $1.113$ &   $0.635^{*}$ &   $1.172$ \\
\midrule 

GRU ($l$ = 8)     &  $0.600^{*}$ &  $0.946$ &   $\mathbf{0.568^{*}}$ &  $\mathbf{1.103}$ &   $0.650^{*}$ &   $1.200$ \\
GRU ($l$ = 18)    &  $\mathbf{0.592^{*}}$ &  $\mathbf{0.933}$ &  $0.569^{**}$ &  $1.107$ &   $\mathbf{0.607^{*}}$ &   $\mathbf{1.122}$ \\
GRU ($l$ = 36)    &  $0.644^{*}$ &  $1.015$ &  $0.591^{**}$ &  $1.148$ &   $0.673^{*}$ &   $1.243$ \\
GRU ($l$ = 48)    &  $0.627^{*}$ &  $0.988$ &  $0.554^{**}$ &  $1.077$ &   $0.665^{*}$ &   $1.230$ \\
\bottomrule
\end{tabular}
\caption*{\begin{footnotesize}
\textbf{Notes:} This table reports the relative MAE of GDP growth for the competitor ANNs relative to a naive constant growth model and a benchmark DFM specification. 
A value below one indicates that the competitor model beats the naive benchmark model. 
The stars denote statistical significance at 10\%($*$), 5\%($**$) and 1\%($***$) level of the one-sided \textcite{diebold1995comparing} test. 
Columns are related to the individual nowcasting scenarios: e.g. $v = 3t-2$ refers to the 1-month nowcasting scenario. 
Within each nowcasting scenario, best (lowest) relative MAE values are highlighted in bold for the given competitor model. 
The overall best relative MAE values are highlighted in bold and underlined. 
\end{footnotesize}}
\label{tab:eval_with_covid_mae}
\end{table}

\newpage

\begin{table}[H] 
\centering
\caption{Nowcasting performance of the competitor ANNs relative to the training configuration using the shortest input sequences $(l = 8)$: RMSE and MAE evaluation. Evaluation period: 2012:Q1 -- 2024:Q2.} 
\begin{tabular}{l | l l | l l | l l}

& \multicolumn{2}{c |}{$v = 3t-2$} & \multicolumn{2}{c |}{$v = 3t-1$} & \multicolumn{2}{c}{$v = 3t$} \\

& RMSE & MAE & RMSE & MAE & RMSE & MAE \\
\hline\hline

MLP ($l$ = 8)     &          \multicolumn{1}{c}{$-$} &          \multicolumn{1}{c|}{$-$} &          \multicolumn{1}{c}{$-$} &          \multicolumn{1}{c|}{$-$} &           \multicolumn{1}{c}{$-$} &           \multicolumn{1}{c}{$-$} \\
MLP ($l$ = 18)    &  $1.080$ &  $1.105$ &  $1.135$ &  $1.116$ &   $1.276$ &   $1.278$ \\
MLP ($l$ = 36)    &  $1.183$ &  $1.205$ &  $1.256$ &  $1.131$ &   $1.619$ &   $1.359$ \\
MLP ($l$ = 48)    &  $1.189$ &  $1.210$ &  $1.372$ &  $1.217$ &   $1.654$ &   $1.387$ \\
\midrule

1D CNN ($l$ = 8)  &          \multicolumn{1}{c}{$-$} &          \multicolumn{1}{c|}{$-$} &          \multicolumn{1}{c}{$-$} &          \multicolumn{1}{c|}{$-$} &           \multicolumn{1}{c}{$-$} &           \multicolumn{1}{c}{$-$} \\
1D CNN ($l$ = 18) &  $1.213$ &  $1.171$ &  $1.152$ &  $1.087$ &   $1.140$ &   $1.127$ \\
1D CNN ($l$ = 36) &  $1.354$ &  $1.223$ &  $1.476$ &  $1.239$ &   $2.253$ &   $1.632$ \\
1D CNN ($l$ = 48) &  $1.333$ &  $1.248$ &  $1.526$ &  $1.202$ &   $2.335$ &   $1.574$ \\
\midrule

RNN ($l$ = 8)     &          \multicolumn{1}{c}{$-$} &          \multicolumn{1}{c|}{$-$} &          \multicolumn{1}{c}{$-$} &          \multicolumn{1}{c|}{$-$} &           \multicolumn{1}{c}{$-$} &           \multicolumn{1}{c}{$-$} \\
RNN ($l$ = 18)    &  $0.997$ &  $1.009$ &  $1.169$ &  $1.083$ &   $1.030$ &   $1.047$ \\
RNN ($l$ = 36)    &  $1.065$ &  $1.088$ &  $1.157$ &  $1.149$ &   $1.336$ &   $1.171$ \\
RNN ($l$ = 48)    &  $1.063$ &  $1.067$ &  $1.129$ &  $1.109$ &   $1.391$ &   $1.194$ \\
\midrule

LSTM ($l$ = 8)    &          \multicolumn{1}{c}{$-$} &          \multicolumn{1}{c|}{$-$} &          \multicolumn{1}{c}{$-$} &          \multicolumn{1}{c|}{$-$} &           \multicolumn{1}{c}{$-$} &           \multicolumn{1}{c}{$-$} \\
LSTM ($l$ = 18)   &  $1.029$ &  $1.044$ &  $1.019$ &  $1.005$ &   $0.953^{*}$ &   $0.952^{*}$ \\
LSTM ($l$ = 36)   &  $1.094$ &  $1.089$ &  $1.060$ &  $1.038$ &   $1.000$ &   $0.986$ \\
LSTM ($l$ = 48)   &  $1.070$ &  $1.069$ &  $1.046$ &  $1.003$ &   $1.022$ &   $0.995$ \\
\midrule

GRU ($l$ = 8)     &          \multicolumn{1}{c}{$-$} &          \multicolumn{1}{c|}{$-$} &          \multicolumn{1}{c}{$-$} &          \multicolumn{1}{c|}{$-$} &           \multicolumn{1}{c}{$-$} &           \multicolumn{1}{c}{$-$} \\
GRU ($l$ = 18)    &  $0.998$ &  $0.987$ &  $1.038$ &  $1.003$ &  $0.961^{**}$ &  $0.935^{**}$ \\
GRU ($l$ = 36)    &  $1.088$ &  $1.073$ &  $1.126$ &  $1.041$ &   $1.115$ &   $1.036$ \\
GRU ($l$ = 48)    &  $1.047$ &  $1.044$ &  $1.006$ &  $0.977$ &   $1.094$ &   $1.024$ \\
\bottomrule
\end{tabular}
\caption*{\begin{footnotesize}
\textbf{Notes:} This table reports the relative RMSE and MAE of GDP growth for our competitor ANNs relative to the training configuration using the shortest input sequences $(l = 8)$. 
A value below one indicates that the competitor model beats the benchmark training configuration $(l = 8)$. 
The stars denote statistical significance at
10\%($*$), 5\%($**$) and 1\%($***$) level of the one-sided \cite{diebold1995comparing} test. 
Columns are related to the individual nowcasting scenarios: e.g.
$v = 3t-2$ refers to the 1-month nowcasting scenario. 
Relative RMSE and MAE values indicating significant performance advantage over the benchmark training configuration $(l = 8)$ are highlighted in bold. 
\end{footnotesize}}
\label{tab:eval_with_covid_shortest}
\end{table}

\end{document}